\documentclass[runningheads,a4paper]{llncs}

\usepackage{amssymb}
\usepackage{hyperref}
\usepackage{orcidlink}
\usepackage{graphicx}
\usepackage{soul}
\usepackage{xcolor}
\usepackage{rotating}
\sethlcolor{yellow}
\usepackage{cite}   
\usepackage[utf8]{inputenc}

\usepackage{url}
\usepackage{doi}
\urldef{\mailsa}\path|{alfred.hofmann, ursula.barth, ingrid.haas, frank.holzwarth,|
\urldef{\mailsb}\path|anna.kramer, leonie.kunz, christine.reiss, nicole.sator,|
\urldef{\mailsc}\path|erika.siebert-cole, peter.strasser, lncs}@springer.com|    
\newcommand{\keywords}[1]{\par\addvspace\baselineskip\noindent\keywordname\enspace\ignorespaces#1}

\begin{document}

\mainmatter

\title{Blockchain and AI: Securing Intelligent Networks for the Future}

\titlerunning{Blockchain and AI for Intelligent Network Security}

\author{Joy Dutta\thanks{Corresponding author. Email: \email{joy.dutta.in@ieee.org}}\orcidlink{0000-0003-4862-1589}$^1$%
\and Hossien~B.~Eldeeb\orcidlink{0000-0001-7560-1124}$^{2}$ 
\and Tu Dac Ho\orcidlink{0000-0001-7215-0479}$^3$}
\authorrunning{J. Dutta et al.}

\institute{$^1$Department of Electrical Engineering, Khalifa University of Science and Technology, Abu Dhabi, UAE \\
$^2$Department of Information Systems and Security,College of Information Technology, United Arab Emirates University, Al Ain 15551, UAE\\
$^3$Department of Information Security and Communication Technology, Norwegian University of Science and Technology (NTNU), Trondheim, Norway \\
Emails: joy.dutta.in@ieee.org, hossien@uaeu.ac.ae, tu.d.ho@ntnu.no
}

%
%

\toctitle{Blockchain and AI for Intelligent Network Security: A Taxonomy, Integration Framework, and Benchmarking Blueprint}

\tocauthor{Joy Dutta, Hossien B. Eldeeb, and Tu Dac Ho}
\maketitle

\begin{abstract}
Blockchain and artificial intelligence (AI) are increasingly proposed together for securing intelligent networks, but the literature remains fragmented across ledger design, AI-driven detection, cyber-physical applications, and emerging agentic workflows. This paper synthesizes the area through three reusable contributions: (i) a taxonomy of blockchain-AI security for intelligent networks, (ii) integration patterns for verifiable and adaptive security workflows, and (iii) the Blockchain-AI Security Evaluation Blueprint (BASE), a reporting checklist spanning AI quality, ledger behavior, end-to-end service levels, privacy, energy, and reproducibility. The paper also maps the evidence landscape across IoT, critical infrastructure, smart grids, transportation, and healthcare, showing that the conceptual fit is strong but real-world evidence remains uneven and often prototype-heavy. The synthesis clarifies where blockchain contributes provenance, trust, and auditability, where AI contributes detection, adaptation, and orchestration, and where future work should focus on interoperable interfaces, privacy-preserving analytics, bounded agentic automation, and open cross-domain benchmarks. The paper is intended as a reference for researchers and practitioners designing secure, transparent, and resilient intelligent networks.

\end{abstract}

\section{Introduction} \label{sec:introduction}

The digital landscape is undergoing a profound transformation, characterized by increasingly interconnected systems, exponential growth in data generation, and sophisticated cyber threats that evolve at unprecedented rates. As our critical infrastructure, financial systems, healthcare networks, and transportation grids become more intelligent and automated, they simultaneously become more vulnerable to complex attacks that can have devastating consequences \cite{H4}. Traditional security approaches, primarily reactive in nature, are proving inadequate against the sophisticated threat landscape of today's intelligent networks. This inadequacy stems from their inability to anticipate novel attack vectors, their slow response times, and their limited capacity to adapt to rapidly evolving threats \cite{Ofusori2024Artificial, Dutta2025Empowering}.

In this challenging environment, two revolutionary technologies have emerged as potential game-changers for network security: blockchain and artificial intelligence (AI). Blockchain technology, with its decentralized architecture, immutable record-keeping, and consensus mechanisms, offers a new paradigm for establishing trust and integrity in distributed systems\cite{H1}. Meanwhile, AI, with its ability to analyze vast amounts of data, recognize patterns, and make predictive decisions, provides unprecedented capabilities for threat detection, anomaly identification, and automated response \cite{H2,Salah2019Blockchain, dutta2024advanced}. While each technology individually brings significant advantages to the security domain, their integration creates a synergistic effect that could fundamentally transform how we secure intelligent networks\cite{H3}.

The integration of blockchain and AI for security purposes represents a convergence of complementary strengths. Blockchain provides a tamper-resistant foundation for data integrity and transparent operations, while AI contributes adaptive intelligence and predictive capabilities. Together, they address complementary challenges in traditional security systems: blockchain ensures tamper-proof logging and auditability, which enhances trust in AI outputs, while AI introduces adaptability and context-aware intelligence that complements blockchain's deterministic and rule-based structure \cite{Hussain2021Artificial, dutta2023next}. Recent advances in large language models and emerging agentic AI systems further extend this landscape by enabling natural-language threat intelligence, evidence synthesis, and bounded workflow automation, while also introducing new requirements for action safety, policy compliance, and auditable tool use. This integration enables a security paradigm that is simultaneously proactive and verifiable, adaptive and trustworthy.

The potential applications of this integrated approach span numerous domains, from securing Internet of Things (IoT) ecosystems to protecting critical infrastructure, from enhancing financial system integrity to safeguarding healthcare \cite{H5} and digital twin networks \cite{H20}. In each context, the blockchain-AI integration offers unique advantages that address domain-specific security challenges while providing a common foundation of trust, transparency, and intelligence \cite{Zuo2025Exploring}.

However, this integration is not without challenges. Technical hurdles related to scalability, interoperability, and energy efficiency must be addressed. Ethical considerations regarding privacy, autonomy, and accountability require careful attention. Regulatory frameworks need to evolve to accommodate these new technological approaches while ensuring appropriate governance and oversight \cite{RESSI2024AI}.

\begin{figure}[t]
\centering
\includegraphics[width=0.80 \textwidth]{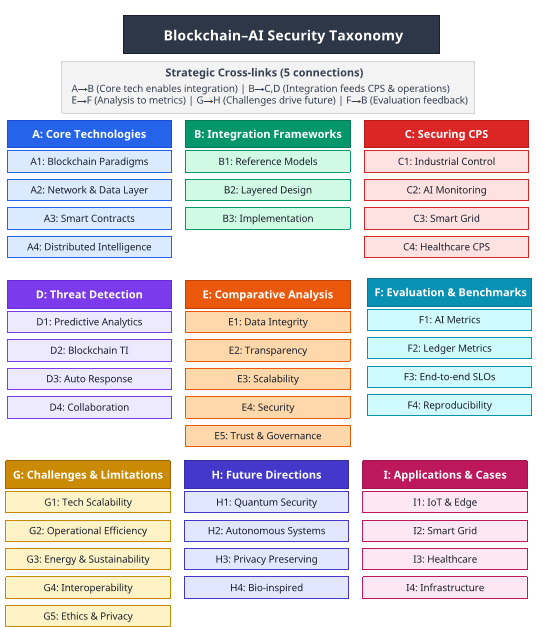}
\caption{Taxonomy of blockchain-AI security and paper roadmap: Each category maps to the main sections of the paper}
\label{fig:chapter_overview}
\end{figure}

This paper examines how blockchain and AI can be co-designed to secure intelligent networks. We begin with the technological foundations of each stack in the security context, then present integration frameworks and implementation strategies that make these technologies work together in verifiable, accountable workflows. We apply these patterns to cyber-physical systems and proactive threat detection, and we include a comparative analysis that clarifies where each technology contributes most. An accompanying evaluation blueprint standardizes how to report AI metrics, ledger performance, end-to-end reliability, energy, and reproducibility. We also analyze challenges, limitations, and ethical considerations, and we close with future directions that include quantum-resilient cryptography, autonomous security systems, privacy-preserving analytics, and bio-inspired paradigms. Representative evidence across domains illustrates early adoption and uneven maturity across use cases. Readers can follow the structure in Figure~\ref{fig:chapter_overview}, which provides an icon-based map of the major sections and their connections.


To make the paper's contributions explicit, we organize the survey around four reusable contributions. First, we synthesize the security roles of blockchain and AI across intelligent-network contexts. Second, we structure their co-design into integration patterns for verifiable and adaptive security workflows. Third, we introduce the \textit{Blockchain-AI Security Evaluation Blueprint (BASE)}, a reporting template spanning AI performance, ledger behavior, end-to-end service metrics, energy, and reproducibility. Fourth, we identify open research priorities across quantum-resilient cryptography, privacy-preserving analytics, autonomous response, and socio-technical validation.

Several prior reviews discuss blockchain for AI, AI for blockchain, or broad convergence across IoT applications \cite{Salah2019Blockchain,Hussain2021Artificial,RESSI2024AI,Zuo2025Exploring}. In contrast, this paper focuses specifically on security workflows for intelligent networks. It combines four elements that are rarely presented together in prior reviews: a security-centric taxonomy, integration patterns for verifiable and adaptive workflows, the BASE evaluation blueprint, and an evidence-and-maturity synthesis across application domains. Table~\ref{tab:prior_surveys} positions this paper relative to representative prior surveys already cited in the manuscript.

\begin{table}[t]
\centering
\caption{Positioning this paper relative to representative prior surveys already cited in the manuscript}
\label{tab:prior_surveys}
\scriptsize
\renewcommand{\arraystretch}{1.12}
\resizebox{\textwidth}{!}{%
\begin{tabular}{|p{2.7cm}|p{3.4cm}|p{2.8cm}|p{1.7cm}|p{1.8cm}|p{3.8cm}|}
\hline
\textbf{Reference} & \textbf{Primary scope} & \textbf{Intelligent-network security focus} & \textbf{LLM / agentic focus} & \textbf{Evaluation blueprint} & \textbf{What this paper adds beyond it} \\
\hline
Salah et al. (2019) \cite{Salah2019Blockchain} & Blockchain for AI: applications, platforms, protocols, and open research challenges & Indirect; not centered on intelligent-network security workflows & No & No & Reframes the area around security workflow design, cross-domain evidence synthesis, and BASE \\
\hline
Hussain and Al-Turjman (2021) \cite{Hussain2021Artificial} & Broad review of AI and blockchain synergies, applications, and challenges & Broad and non-specific; not organized around intelligent-network security operations & No & No & Adds a security taxonomy, integration patterns, and explicit evaluation/reporting criteria \\
\hline
Ressi et al. (2024) \cite{RESSI2024AI} & AI enhancing blockchain technology, including consensus, smart contracts, data privacy, and efficiency & Focuses mainly on AI $\rightarrow$ blockchain improvement rather than end-to-end security workflows & No explicit focus & No & Adds the complementary blockchain-AI security view, CPS and threat-operations emphasis, and BASE \\
\hline
Zuo (2025) \cite{Zuo2025Exploring} & Bidirectional AI-blockchain synergy and convergence across IoT applications and beyond & Closer to secure IoT settings, but still broader than intelligent-network security workflow design & No explicit focus & No & Adds explicit security workflow framing, LLM/agentic considerations, and an evidence/maturity map across domains \\
\hline
\textbf{This paper} & Taxonomy, integration framework, BASE, and evidence landscape for blockchain-AI security in intelligent networks & Yes; centered on verifiable, adaptive, and auditable security workflows & Yes & Yes & Unifies taxonomy, workflow design, benchmarking, and cross-domain maturity assessment in one reference paper \\
\hline
\end{tabular}%
}
\end{table}


\section{Advanced Blockchain Technology for Securing Intelligent Networks} \label{sec:Advanced-Blockchain-Technology}

The evolution of blockchain technology has transcended its initial application in cryptocurrencies to become a foundational element in securing intelligent networks. This section explores advanced blockchain technologies specifically designed to address the complex security requirements of modern distributed systems, with particular emphasis on consensus mechanisms, network architectures, smart contracts, and distributed identity frameworks.

\subsection{Evolution of Blockchain Security Paradigms}

The security paradigms in blockchain technology have undergone significant evolution, moving from simple proof-of-work (PoW) mechanisms to sophisticated Byzantine Fault Tolerance (BFT) variants optimized for intelligent network security \cite{Xu2023Survey}. Traditional PoW mechanisms, while effective for cryptocurrency applications, present limitations in security-critical environments due to their resource-intensive nature and potential for 51\% attacks. Modern BFT variants such as Practical Byzantine Fault Tolerance (PBFT), Delegated Byzantine Fault Tolerance (dBFT), and Federated Byzantine Agreement (FBA) offer improved security guarantees with significantly reduced computational overhead, making them more suitable for intelligent network security applications.

Istanbul Byzantine Fault Tolerance (IBFT) consensus mechanism, for instance, provides immediate transaction finality and malicious node tolerance up to one-third of the network, under partial synchrony with fewer than one-third Byzantine faults, making it particularly valuable for securing critical infrastructure systems where transaction certainty is paramount. Similarly, Proof of Authority (PoA) consensus mechanisms leverage identity verification rather than computational work, offering a security model that aligns well with regulated environments where participant accountability is essential. The recently proposed PoAh 2.0 consensus further advances this landscape by introducing AI-driven dynamic authentication tailored to IoMT-edge workflows, enabling adaptive security based on data sensitivity and establishing a new benchmark for trust management in eHealth blockchain systems \cite{dutta2024poah}.

Zero-knowledge proofs represent another significant advancement in blockchain security paradigms, enabling verification of information without revealing the underlying data. This capability is particularly valuable in intelligent networks where sensitive data must be protected while still enabling security verification \cite{Sun2021Survey}. Zero-knowledge Succinct Non-interactive Arguments of Knowledge (zk-SNARKs) and Zero-Knowledge Scalable Transparent Arguments of Knowledge (zk-STARKs) allow for privacy-preserving verification of transactions and computations, addressing the often-conflicting requirements of transparency and confidentiality in secure networks.

Quantum-resistant approaches to blockchain security have also emerged as a critical area of development, addressing the potential threat that quantum computing poses to current cryptographic methods. Lattice-based cryptography, hash-based signatures, and multivariate polynomial cryptography are being integrated into blockchain protocols to ensure long-term security against quantum attacks \cite{Allende2023Quantum}. These quantum-resistant approaches are particularly important for intelligent networks that must maintain security integrity over extended periods, such as critical infrastructure systems and long-term data storage networks.

\begin{table}[t]
\centering
\caption{Consensus trade-offs for security-critical intelligent-network deployments \cite{Xu2023Survey,dutta2024poah}}
\label{tab:consensus_cps}
\scriptsize
\renewcommand{\arraystretch}{1.15}
\resizebox{\textwidth}{!}{%
\begin{tabular}{|p{2.1cm}|p{1.7cm}|p{2.8cm}|p{1.5cm}|p{1.5cm}|p{3.1cm}|}
\hline
\textbf{Consensus} & \textbf{Finality} & \textbf{Fault / attack tolerance} & \textbf{Latency} & \textbf{Energy profile} & \textbf{Typical fit for intelligent-network security} \\
\hline
PoW & Probabilistic & Strong Sybil resistance, but vulnerable to majority-hash-power concentration and expensive to operate & High & High & Useful when open participation is required, but generally unsuitable for low-latency CPS and operational security pipelines \\
\hline
PBFT / IBFT & Deterministic & Tolerates up to one-third Byzantine nodes under standard BFT assumptions & Low & Low--moderate & Strong fit for permissioned critical infrastructure, industrial coordination, and audit trails where finality and bounded latency matter \\
\hline
PoA & Deterministic or near-deterministic, depending on implementation & Relies on vetted validators and governance accountability rather than open participation & Low & Low & Strong fit for regulated environments, consortium deployments, and cross-organizational security workflows with known participants \\
\hline
PoAh 2.0 & Deterministic within its permissioned workflow setting & Extends authority-based operation with AI-assisted dynamic authentication and sensitivity-aware trust decisions & Low & Low & Best suited to domain-specific settings such as IoMT-edge and adaptive trust management where authentication context changes over time \\
\hline
\end{tabular}%
}
\end{table}

Table~\ref{tab:consensus_cps} condenses the main trade-offs; in security-critical intelligent networks, the practical center of gravity is moving toward permissioned BFT-like and authority-based designs because they provide stronger control over finality, governance, and energy use than open PoW systems.

\subsection{Specialized Network Architectures for Intelligent Systems}

The evolution of blockchain network architectures has produced specialized designs optimized for the unique requirements of intelligent systems. Sharding techniques, which partition blockchain networks into smaller, more manageable segments, address the scalability challenges that have traditionally limited blockchain application in high-throughput intelligent networks \cite{Quan2025Sharding}. By allowing parallel processing of transactions across multiple shards, these architectures enable the performance levels required for real-time security applications in intelligent networks.

Cross-chain interoperability frameworks represent another architectural advancement critical for securing heterogeneous intelligent networks. Solutions such as Polkadot's relay chain architecture, Cosmos's Inter-Blockchain Communication protocol, and Chainlink's oracle networks enable secure information exchange and coordinated security operations across different blockchain networks \cite{LI2025CrossChain}. This interoperability is essential for comprehensive security in complex environments where multiple systems with different trust models must interact securely.

Layer-2 scaling solutions, including state channels, sidechains, and rollups, provide additional architectural approaches to address the performance requirements of intelligent network security. These solutions move much of the transaction processing off the main blockchain while still leveraging its security guarantees, enabling high-throughput security monitoring and response capabilities \cite{Chiedu2025On-Chain-Off-Chain-Scalability}. Optimistic rollups, for instance, can increase transaction throughput by orders of magnitude while maintaining security assurances, making them suitable for real-time security applications in intelligent networks.

Private and permissioned blockchain architectures offer specialized security models for controlled environments, providing the benefits of blockchain technology while maintaining governance over network participants. Hyperledger Fabric's channels and private data collections, Quorum's private transactions, and R3 Corda's notary-based consensus all provide mechanisms for securing sensitive information within intelligent networks while still enabling necessary information sharing \cite{Liang2024Permissioned-and-PermissionlessSurvey}. These architectures are particularly valuable in regulated industries and critical infrastructure protection, where participant verification and information access control are essential security requirements.
\subsection{Advanced Smart Contracts for Security Orchestration}

Smart contracts have evolved from simple transaction automation tools to sophisticated security orchestration mechanisms capable of coordinating complex security operations across intelligent networks \cite{alneyadi2023role, HUSAIN2025Context}. Security-focused smart contract frameworks such as NuCypher for cryptographic access control, Quantstamp for automated security verification, and OpenZeppelin for standardized security implementations provide specialized capabilities for intelligent network protection \cite{Azimi2025SmartContracts}.

Formal verification methods for smart contracts represent a significant advancement in ensuring their security integrity. Techniques such as symbolic execution, model checking, and theorem proving enable mathematical verification of smart contract behavior, reducing the risk of vulnerabilities that could compromise intelligent network security \cite{Fernando2024Vulnerability}. Tools like KEVM, which provides a formal semantics of the Ethereum Virtual Machine, allow for rigorous verification of smart contract security properties before deployment.

Self-sovereign identity frameworks implemented through smart contracts provide advanced authentication and authorization capabilities for intelligent networks. Solutions such as Sovrin, uPort, and Microsoft's ION enable decentralized identity verification without reliance on central authorities, addressing a critical security requirement in distributed intelligent systems \cite{Schardong2022SSI}. These frameworks support fine-grained access control, credential verification, and revocation mechanisms that enhance the security posture of intelligent networks.

Chaincode technology, as implemented in Hyperledger Fabric, extends smart contract capabilities with enhanced privacy and performance features particularly valuable for security applications. The ability to execute chaincode in isolated environments, combined with flexible endorsement policies, enables sophisticated security orchestration across organizational boundaries \cite{MOHANM2023ChainCode}. This capability is especially important in collaborative security scenarios where multiple entities must coordinate their defense mechanisms while maintaining operational independence.
\subsection{Blockchain for Securing Distributed Intelligence}

Blockchain technology provides unique capabilities for securing distributed intelligence systems, where computational resources, data, and decision-making are distributed across multiple nodes. Decentralized machine learning frameworks such as Ocean Protocol, Fetch.ai, and SingularityNET leverage blockchain to enable secure, transparent, and accountable AI operations across distributed environments \cite{Charles2023critical}. These frameworks address critical security challenges in distributed intelligence, including model integrity, data provenance, and result verification.

Secure multi-party computation (MPC) integrated with blockchain enables collaborative intelligence operations while protecting sensitive inputs. This integration allows multiple parties to jointly compute functions over their inputs while keeping those inputs private, a capability essential for collaborative threat intelligence and coordinated defense mechanisms \cite{Ohize2024Blockchain}. The combination of MPC with blockchain's immutable record-keeping creates auditable yet privacy-preserving distributed intelligence systems.

Verifiable computation frameworks built on blockchain technology enable the validation of computational results from untrusted sources, addressing a fundamental security challenge in distributed intelligence systems. Solutions such as TrueBit and Arbitrum provide mechanisms for verifying the correctness of computations performed by external parties, ensuring the integrity of distributed intelligence operations \cite{alrayes2024intrusion}. This capability is particularly valuable in scenarios where security-critical decisions rely on computations performed across distributed nodes with varying trust levels.

Blockchain-based federated learning architectures represent another significant advancement in securing distributed intelligence. These architectures enable collaborative model training across multiple organizations without sharing raw data, addressing both privacy concerns and data silos that hamper effective security analytics \cite{Issa2023Blockchain-Based-Federated-Learning, alqubaisi2023privacy}. By recording model updates on a blockchain, these architectures provide transparency and accountability in the federated learning process, ensuring that all participants adhere to agreed-upon protocols and that the resulting models maintain their integrity.


\section{AI-Driven Techniques for Enhanced Cybersecurity} \label{sec:AI-Driven-Cybersecurity}

\begin{figure*}[!t]
    \centering
    \includegraphics[width=\textwidth]{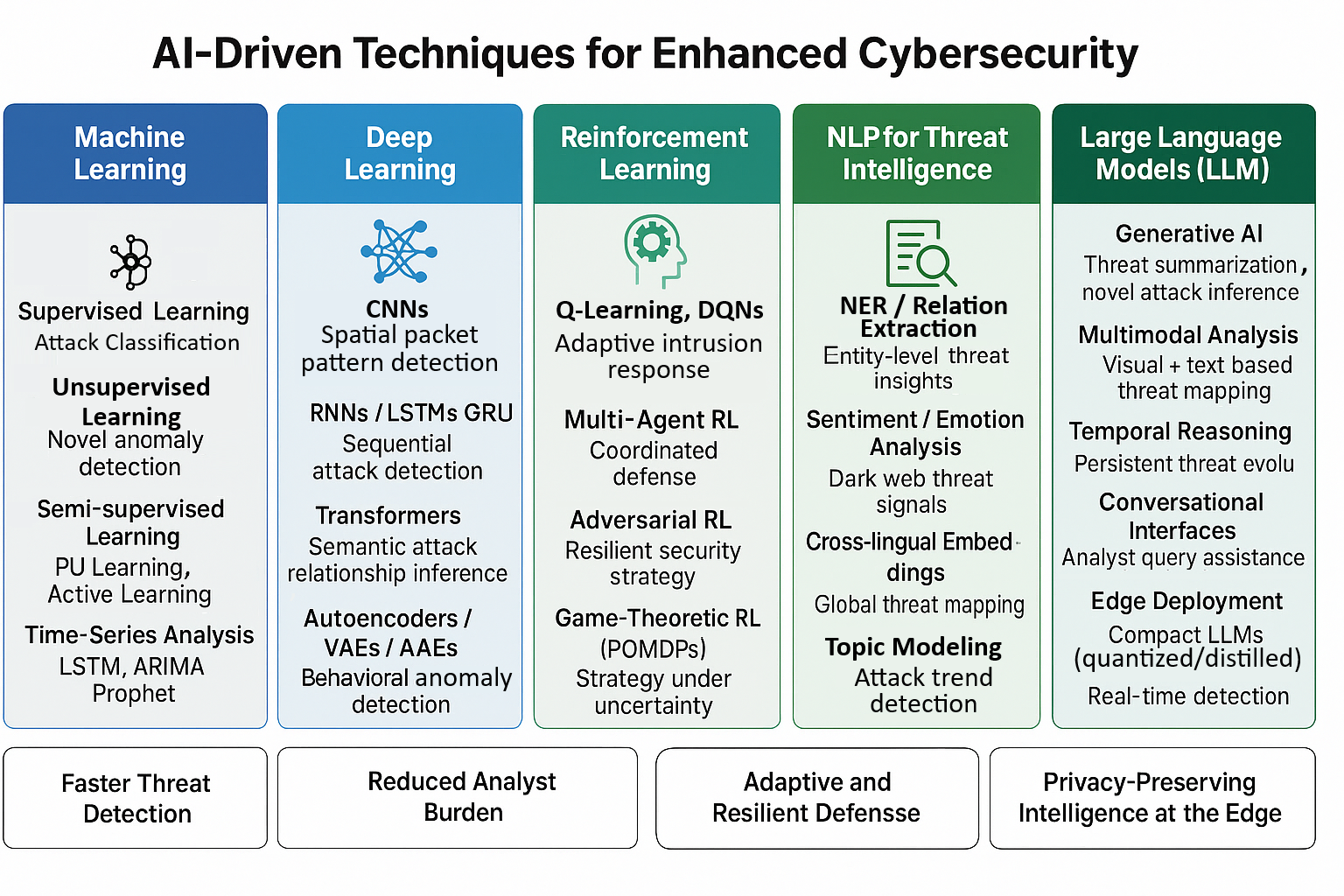}
    \caption{Overview of AI-driven techniques for enhanced cybersecurity}
    \label{fig:ai_cybersecurity_overview}
\end{figure*}

Artificial intelligence has revolutionized network security by introducing capabilities that far exceed traditional rule-based approaches. This section explores the advanced AI technologies being applied to network security challenges, with particular focus on machine learning for anomaly detection, deep learning for pattern recognition, reinforcement learning for adaptive defense, and natural language processing for threat intelligence. Figure~\ref{fig:ai_cybersecurity_overview} provides a visual summary of these AI paradigms and their respective contributions to cybersecurity, helping readers contextualize the techniques discussed in this section.

\subsection{Machine Learning for Anomaly Detection}
Supervised learning approaches have demonstrated remarkable effectiveness in identifying known attack patterns in network traffic. Support Vector Machines (SVMs), Random Forests, and Gradient Boosting algorithms can classify malicious activities with high precision when trained on labeled datasets of normal and attack traffic \cite{Ajagbe2024Intrusion}. These approaches excel at recognizing variations of known attacks but require extensive labeled training data, which can be challenging to obtain in rapidly evolving threat landscapes.

Unsupervised learning techniques address this limitation by identifying anomalies without prior examples of attacks. Clustering algorithms such as K-means, DBSCAN, and Gaussian Mixture Models can identify unusual patterns in network behavior that deviate from normal operations \cite{Liu2023Anomaly}. Isolation Forests and One-Class SVMs have proven particularly effective at detecting outliers in high-dimensional network data, enabling the identification of novel attack vectors that would evade signature-based detection systems.

Semi-supervised approaches bridge these methodologies by leveraging limited labeled data alongside larger unlabeled datasets. Positive-Unlabeled (PU) learning and active learning techniques have shown promise in security contexts where obtaining comprehensive attack labels is impractical \cite{Gang2024PUNet}. These approaches can achieve detection rates comparable to fully supervised methods while requiring significantly less labeled data, making them practical for operational security environments.

Time series analysis using machine learning has emerged as a powerful technique for detecting temporal anomalies in network behavior. Long Short-Term Memory (LSTM) networks, Autoregressive Integrated Moving Average (ARIMA) models, and Prophet forecasting have demonstrated effectiveness in identifying unusual patterns in network traffic over time \cite{Priyadarshini2022TimeSeries}. These approaches can detect subtle changes in behavior that might indicate advanced persistent threats or slowly evolving attack campaigns.
\subsection{Deep Learning for Pattern Recognition}
Convolutional Neural Networks (CNNs) have been successfully applied to network traffic analysis, treating packet data as multi-dimensional arrays similar to images. This approach enables the identification of spatial patterns in network data that might indicate malicious activity \cite{WASIF2025CNN}. CNNs can automatically extract relevant features from raw packet data, reducing the need for manual feature engineering and enabling the detection of subtle attack patterns that might be missed by traditional approaches\cite{H14}.

Recurrent Neural Networks (RNNs) and their variants, particularly Long Short-Term Memory (LSTM) and Gated Recurrent Unit (GRU) architectures, excel at analyzing sequential network data. These architectures can capture temporal dependencies in network traffic, enabling the detection of multi-stage attacks that evolve over time \cite{Dash2025LSTM}. The ability to maintain context across long sequences makes these approaches particularly valuable for identifying sophisticated attack campaigns that develop gradually.

Transformer models have recently demonstrated superior performance in analyzing complex network security data. Their self-attention mechanisms can capture relationships between events regardless of their temporal distance, addressing a limitation of traditional RNN architectures \cite{KHEDDAR2025Transformers}. Models such as BERT and GPT, adapted for security contexts, have shown remarkable capabilities in understanding the semantics of security events and identifying subtle connections between seemingly unrelated activities.

Autoencoders represent another powerful deep learning approach for network security, particularly for anomaly detection. By learning to reconstruct normal network behavior, these models can identify deviations that may indicate security breaches \cite{Torabi2023Autoencoder}. Variational Autoencoders (VAEs) and Adversarial Autoencoders (AAEs) have demonstrated particular effectiveness in generating robust representations of normal behavior that enable high-precision anomaly detection with manageable false positive rates.
\subsection{Reinforcement Learning for Adaptive Defense}
Reinforcement learning (RL) enables security systems to learn optimal defense strategies through interaction with adversarial environments. Q-learning and Deep Q-Networks (DQNs) have been applied to security orchestration, automatically learning optimal response strategies to different attack scenarios \cite{Alavizadeh2022DeepQLearning}. These approaches enable security systems to adapt their responses based on the effectiveness of previous actions, creating increasingly sophisticated defense mechanisms over time.

Multi-agent reinforcement learning extends this capability to coordinated defense across distributed systems. Approaches such as Multi-Agent Deep Deterministic Policy Gradient (MADDPG) and Counterfactual Multi-Agent Policy Gradients enable collaborative defense strategies across multiple security components \cite{FINISTRELLA2025MultiAgent}. These techniques are particularly valuable in complex network environments where coordinated responses across different security domains are necessary to effectively counter sophisticated attacks.

Adversarial reinforcement learning introduces explicit modeling of attacker behavior to develop more robust defense strategies. By training defensive agents against simulated attackers that also learn and adapt, these approaches develop security strategies that are resilient against evolving threats \cite{PURVES2024Causally}. This adversarial training creates defense mechanisms that anticipate and counter novel attack strategies rather than simply responding to known patterns.

Game-theoretic approaches to reinforcement learning provide a framework for modeling the strategic interactions between attackers and defenders. Stochastic games and partially observable Markov decision processes (POMDPs) enable the development of optimal defense strategies under uncertainty about attacker capabilities and objectives \cite{HU2019gametheory}. These approaches are particularly valuable in high-stakes security environments where understanding the strategic implications of defense decisions is critical.
\subsection{Natural Language Processing for Threat Intelligence}
Natural Language Processing (NLP) has transformed threat intelligence by enabling the automated extraction of security-relevant information from unstructured text sources. Named Entity Recognition (NER) and Relationship Extraction techniques can identify threat actors, attack techniques, vulnerabilities, and affected systems from security reports, blogs, and social media \cite{Marinho2023Automated}. These capabilities enable security teams to process vastly more threat intelligence than would be possible through manual analysis.

Sentiment analysis and emotion detection applied to dark web forums and underground marketplaces provide early warning of emerging threats. By analyzing the emotional content and sentiment trends in these communities, security systems can identify growing interest in particular targets or attack techniques before they materialize into actual attacks \cite{TANG2024Cyber}. This proactive intelligence enables organizations to strengthen defenses against anticipated threats rather than reacting to attacks in progress.

Cross-lingual threat intelligence correlation addresses the global nature of cyber threats by connecting information across language barriers. Neural machine translation and cross-lingual embeddings enable the identification of connections between threat activities discussed in different languages, providing a more comprehensive view of global threat landscapes \cite{chen2021cross}. This capability is increasingly important as threat actors operate across international boundaries and communicate in multiple languages.

Topic modeling and trend analysis applied to security discussions enable the identification of emerging threat vectors and attack methodologies. Latent Dirichlet Allocation (LDA), Non-negative Matrix Factorization (NMF), and BERTopic can identify evolving themes in security discussions that may indicate new attack trends \cite{ALQURASHI2024TopicModelling, khuzaie2021survey}. By monitoring these trends, security teams can anticipate novel attack vectors and develop appropriate countermeasures before these attacks become widespread.

\subsection{Large Language Models for Threat Intelligence}

The emergence of Large Language Models (LLMs) represents a paradigm shift in natural language processing capabilities, with significant implications for threat intelligence. Unlike traditional NLP approaches, LLMs leverage massive parameter counts and extensive pre-training to develop sophisticated language understanding and generation capabilities that can transform threat intelligence operations. This section explores how these advanced models are revolutionizing security analytics, with particular focus on their application in threat detection, intelligence synthesis, and edge deployment scenarios.

\subsubsection{Generative AI for Advanced Threat Analysis: }

Generative AI models, particularly transformer-based architectures like GPT (Generative Pre-trained Transformer) and its variants, have demonstrated unprecedented capabilities in understanding and contextualizing security information. These models can process unstructured security data at scale, extracting meaningful insights from diverse sources including security bulletins, incident reports, and technical documentation \cite{Hasanov2024LLM}. The ability to understand semantic relationships and contextual nuances enables these models to identify subtle connections between seemingly unrelated security events that might indicate coordinated attack campaigns.

Zero-shot and few-shot learning capabilities in modern LLMs enable threat analysis without extensive labeled training data, addressing a critical limitation in traditional supervised approaches. These models can identify potential threats based on general language  rather than specific examples, making them particularly valuable for detecting novel attack vectors \cite{Shahriar2025Zeroshot, Meshkin2024ZeroShot}. Reports on zero-shot and few-shot LLMs indicate meaningful accuracy on previously unseen attack patterns without task-specific training.

Multimodal threat analysis leveraging LLMs capable of processing both text and visual data enables comprehensive security analytics across different information formats. These models can analyze security-relevant information from screenshots, network visualizations, and textual reports simultaneously, creating integrated threat perspectives \cite{Fahad2025Sentiment}. Multimodal threat analysis with large language models that fuse text with code or network telemetry has been reported to improve attack-campaign identification compared with text-only baselines.

Temporal reasoning capabilities in advanced LLMs enable the tracking of threat evolution over time, identifying how attack methodologies develop and adapt. These models can maintain context across extended sequences of security events, recognizing patterns that emerge gradually over weeks or months \cite{Zhang2025LLMs}. LLM-based temporal reasoning that maintains context across extended event sequences can reduce detection time for slowly evolving threats.

\subsubsection{LLM-Enhanced Threat Intelligence Operations: }

Automated threat report generation using LLMs transforms raw security data into comprehensive, human-readable intelligence reports. These models can synthesize information from multiple sources, identify key findings, and generate actionable reports with appropriate context and recommendations \cite{Perrina2023AGIR}. Template-guided threat-report generation with large language models reduces analyst effort while preserving the structure and clarity required for dissemination.

Conversational interfaces for threat intelligence powered by LLMs enable security analysts to interact with complex security data through natural language queries. These interfaces allow analysts to explore threat data, request specific analyses, and receive explanations without requiring specialized query languages or technical interfaces \cite{FERRAG2025Generative}. In comparative evaluations, analysts reached actionable findings faster when using LLM-powered conversational interfaces than with conventional security tools.

Threat intelligence enrichment using LLMs enhances raw security data with additional context, historical information, and relevant connections. These models can automatically supplement threat indicators with information about associated tactics, techniques, procedures, threat actors, and affected systems \cite{Fieblinger2024Actionable}. In controlled evaluations, LLM-based contextual enrichment has been reported to improve analysts’ comprehension of complex threats by linking indicators to tactics, techniques, procedures, actors, and affected systems

Counterfactual threat analysis leveraging LLMs' reasoning capabilities enables security teams to explore potential attack scenarios and their implications. These models can generate plausible attack variations, helping security teams identify defensive blind spots and develop more comprehensive protection strategies \cite{Li2024PromptingLLMs}. In controlled evaluations, LLM-based counterfactual generators have been reported to surface previously overlooked attack paths in test environments, enabling proactive hardening before exploitation.

\subsubsection{Edge LLM Deployment for Real-Time Threat Intelligence: }

Edge-optimized LLM architectures enable sophisticated language processing capabilities on resource-constrained devices at the network edge. Techniques such as quantization, pruning, and knowledge distillation reduce model size and computational requirements while preserving core capabilities \cite{Dantas2025review}. Edge-optimized LLMs can retain much of the detection performance of larger models while fitting resource-constrained devices.

Real-time threat detection at the edge using optimized LLMs enables immediate identification of potential security incidents without cloud dependencies. These models can analyze local security telemetry, network traffic, and system logs in real-time, identifying suspicious patterns that warrant further investigation. Edge-deployed detection models show lower latency than cloud-dependent approaches, which narrows the adversarial window \cite{Shankar2020Benchmarking}.

Federated learning for edge LLM improvement enables continuous enhancement of threat detection capabilities without centralizing sensitive data. Organizations can contribute to model improvements by sharing model updates rather than raw security data, preserving privacy while enabling collective intelligence \cite{cai2025blockchain, PICCIALLI2025Federated}. Federated learning programs for edge LLM improvement report faster personalization and reduced data exposure by sharing model updates rather than raw logs across participating organizations \cite{Alshehri2024Deep}.

LLM-specific risks must be addressed explicitly in threat-intelligence pipelines. Prompt injection and tool-use abuse can coerce unintended actions or data exfiltration when the model is connected to external tools. Hallucinations and ungrounded attributions can fabricate indicators or misidentify threat actors, which inflates false positives and propagates bad intelligence. Secrets may leak through the context window and associated logs or caches when retrieved evidence, credentials, or private tickets are included. Practical defenses include retrieval grounding from authenticated sources, strict allow-lists and least-privilege for tool invocation, output verification against typed schemas and policy checkers, and routine red-team evaluations using curated attack libraries. These controls should be instrumented with failure-rate tracking, incident postmortems, and dataset versioning to provide auditability and continuous hardening.

\noindent These developments also point toward a broader shift from LLM-assisted analysis to agentic AI. Unlike prompt-driven LLM use, agentic systems can combine language reasoning with memory, tool access, and multi-step planning so they can gather evidence, query external systems, draft actions, and revise decisions across longer security workflows \cite{Ferrag2025AutonomousAgents,Lazer2026AgenticCyber}. In security operations, this makes them useful for bounded tasks such as alert triage, evidence collection, playbook selection, and ticket preparation. At the same time, once a model can act rather than only interpret, the design focus must expand from answer quality to action safety, policy compliance, rollback, and auditable tool use. For that reason, agentic AI is best understood here as a controlled extension of the LLM trend rather than a replacement for human judgment.


\section{Blockchain-AI Integration Frameworks for Network Security} \label{sec:Blockchain-AI-Integration}

The integration of blockchain and AI technologies creates powerful new paradigms for securing intelligent networks, combining the trustworthiness and transparency of blockchain with the adaptive intelligence of AI \cite{dutta2022ai}. This section explores comprehensive frameworks for this integration, examining reference architectures, formal models, layered approaches, and implementation strategies that enable these technologies to work together effectively for enhanced security.

\subsection{Security-Oriented Reference Architectures and Formal Models}
Research on blockchain-AI integration is beginning to converge on a set of recurring architectural patterns rather than a single canonical reference model. Recent studies describe modular designs that separate data acquisition, cryptographic verification, AI analytics, policy enforcement, and orchestration, while defining explicit interfaces between blockchain services and AI components \cite{Ghadi2025hybrid, Elkhodr2025Framework}. These proposals are useful because they expose trust boundaries, data lineage, and control points, even though the field has not yet adopted a universally accepted architecture.

The formalization of such integrated systems remains comparatively immature. Existing work is strongest in domain-specific trust, compliance, and verification workflows rather than in broadly accepted end-to-end formal models \cite{KUMAR2025DynamicTrust, Alevizos2025Automated}. In practice, current systems benefit more from modular and verifiable security workflows than from attempting to present the whole pipeline as fully formalized.

Trust management is a more mature thread within this literature. Hybrid trust frameworks use blockchain to anchor identities, evidence, and policy states, while AI estimates behavioural trust, risk, or anomaly scores from dynamic observations \cite{KUMAR2025DynamicTrust}. This division of labour is especially useful in environments where static credentials alone are insufficient.

Security verification frameworks for integrated blockchain-AI systems are also beginning to emerge. Rather than claiming full formal closure, recent work focuses on verifiable logging, compliance checking, and reproducible validation of model and ledger behaviour across implementations \cite{Alevizos2025Automated}. This is a practical direction for current deployments because it supports accountability even when complete end-to-end formal verification is not yet feasible.

\subsection{Layered Security Architectures for Integrated Systems}

\begin{figure*}[!t]
    \centering
    \includegraphics[width=\textwidth]{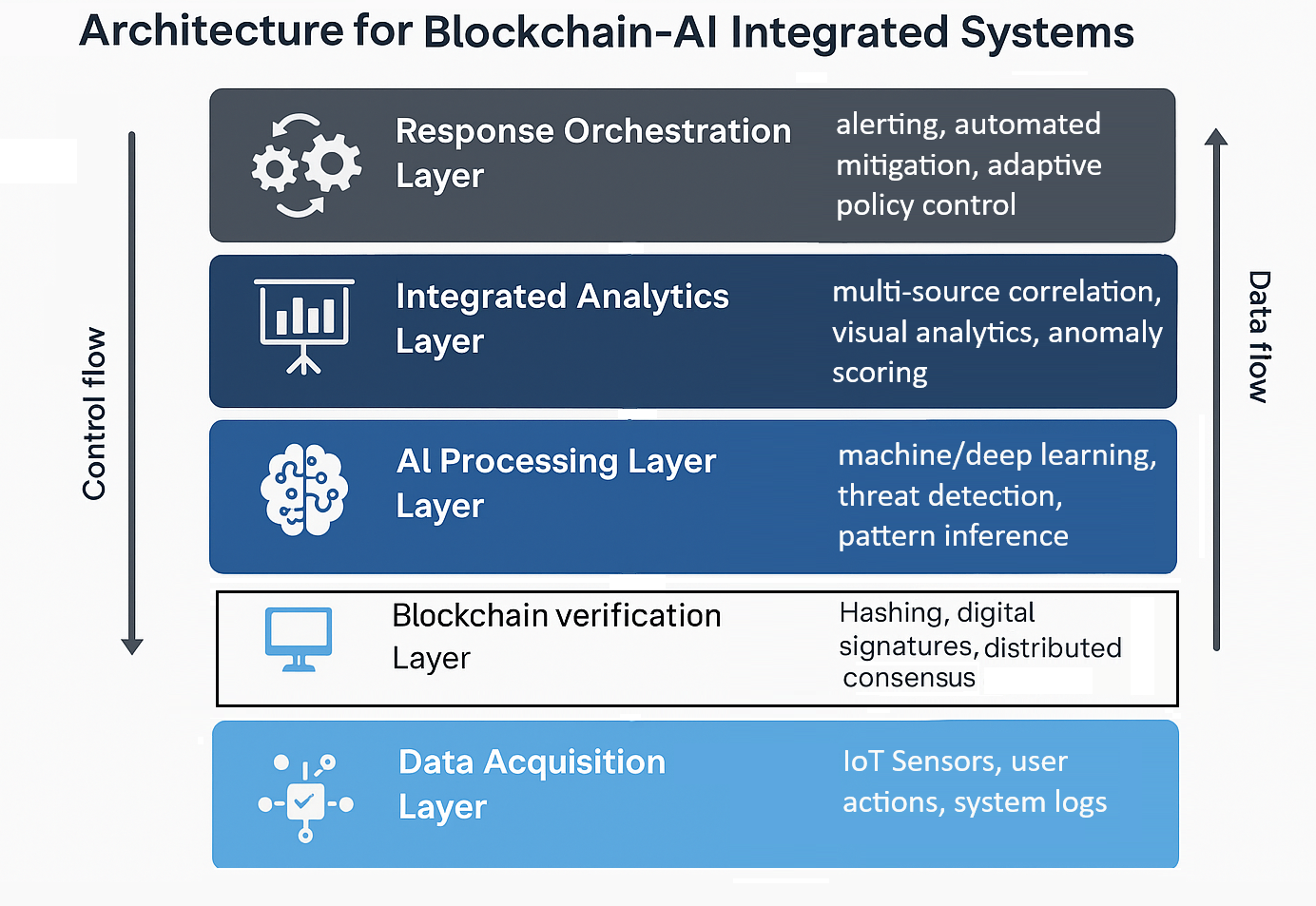} 
    \caption{Layered Security Architecture for Blockchain-AI Integrated Systems}
    \label{fig:SILM}
\end{figure*}

Layered architecture approaches to blockchain-AI integration define structured relationships between components, creating clear security boundaries and responsibilities. Layered security models have been suggested, typically structured into data acquisition, blockchain verification, AI processing, analytics, and response orchestration layers \cite{Elkhodr2025Framework, Ghadi2025hybrid}. Figure~\ref{fig:SILM} visually illustrates this multi-layered security architecture, providing readers with a clear understanding of how each layer interacts and contributes to the overall security framework. This layered approach enables modular security designs where each layer has specific security responsibilities and interfaces, facilitating both implementation and security analysis.

Data integrity layers ensure that information flowing between blockchain and AI components maintains its trustworthiness throughout the integration. These layers implement cryptographic verification, provenance tracking, and tamper detection to prevent the compromise of data as it moves between systems \cite{Hossain2024Enhancing}. Recent approaches propose data integrity pipelines that continuously verify information flow from source to consumption, ensuring that AI components operate on trustworthy data from blockchain sources.

Intelligence verification layers provide mechanisms for validating the operations and outputs of AI components within integrated systems. These layers implement techniques such as zero-knowledge proofs of AI operations, reproducible training verification, and decision auditing to create transparency in AI processes \cite{Scaramuzza2025Trustworthy}. Researchers have proposed transparent protocols for recording AI operations on blockchain in ways that enable verification while protecting sensitive aspects of AI models.

Cross-technology orchestration layers coordinate the operations of blockchain and AI components to create cohesive security capabilities. These layers implement event-driven architectures, state synchronization, and coordinated decision processes that enable blockchain and AI to work together effectively \cite{Elkhodr2025Framework}. Security orchestration approaches have been studied to coordinate blockchain verification with AI-driven security responses, enabling responsive yet trustworthy operations \cite{Alevizos2025Automated}.
\subsection{Security Paradigm Transformation Through Integration}

The integration of blockchain and AI transforms security paradigms from reactive to predictive-preventive models. Traditional security approaches focus on detecting and responding to attacks after they occur, while integrated approaches leverage AI's predictive capabilities with blockchain's verification mechanisms to anticipate and prevent attacks before they succeed. This shift has led to proposals for predictive security frameworks that combine AI’s forecasting with blockchain-based verification to enable proactive defenses.

Continuous authentication paradigms replace point-in-time verification with ongoing behavioral validation secured by blockchain. These approaches use AI to continuously analyze entity behavior while recording authentication evidence on blockchain, creating tamper-proof authentication histories that enable detection of compromised credentials \cite{yao2023blockchain}. Blockchain-based continuous authentication protocols have been explored to validate user and device behavior across domains, creating tamper-proof authentication histories.

Trust-but-verify approaches to security operations leverage blockchain's verification capabilities to enable trustworthy automation through AI. These approaches allow AI systems to make autonomous security decisions while maintaining accountability through blockchain verification, addressing a fundamental challenge in security automation \cite{Kulothungan2025Blockchain}. Autonomous security concepts have been proposed where AI systems make security decisions while blockchain ensures accountability through verifiable logging.

Collaborative-competitive security models leverage game-theoretic aspects of both blockchain and AI to create security systems that continuously evolve against adversarial threats. These models implement competitive dynamics between defensive AI systems while using blockchain to ensure fair play and record outcomes, creating security capabilities that improve through controlled adversarial processes \cite{Avarikioti2025ComposableGame}. Evolutionary security strategies using game-theoretic principles have been investigated, where blockchain records ensure fairness and AI adapts defenses against adversarial threats.
\subsection{Implementation Strategies for Security-Focused Integration}

Event-driven integration architectures provide flexible, loosely-coupled connections between blockchain and AI components. These architectures use event streams and message queues to coordinate operations between technologies without creating tight dependencies, enabling modular security implementations that can evolve over time \cite{ROSABILBAO2023CEPEDALoCo}. Event-driven integration frameworks have been proposed to standardize how blockchain and AI components interact in common security operations.

Microservices approaches to blockchain-AI integration decompose security functions into independent services with well-defined interfaces. This approach enables specialized security services that combine blockchain and AI capabilities for specific functions, such as identity verification, threat detection, or access control. Microservices-based integration patterns define how to implement and orchestrate independent blockchain and AI security services so they can be composed into larger capabilities \cite{Akhdar2024Exploring}.

API-driven integration strategies define standardized interfaces between blockchain and AI components, enabling consistent integration patterns across different implementations. These strategies define clear contracts for data exchange, operation invocation, and result verification between technologies \cite{Alevizos2025Automated}. API-driven integration guidelines standardize interfaces for common patterns such as identity, attestation, and inference, which improves interoperability across heterogeneous implementations.

DevSecOps practices for integrated systems address the unique challenges of developing, deploying, and maintaining blockchain-AI security solutions. These practices extend traditional DevSecOps approaches with specific considerations for blockchain governance, AI model management, and integrated testing \cite{Fu2024AIforDevSecOps}. DevSecOps practices tailored to blockchain–AI systems specify processes and checkpoints that surface security requirements early and verify them continuously throughout the development lifecycle.


\section{Securing Cyber-Physical Systems with Blockchain and AI} \label{sec:SecuringCPS}

Cyber-physical systems (CPS) represent a critical frontier in security, where digital vulnerabilities can have direct physical consequences \cite{H13}. The integration of blockchain and AI technologies offers powerful new approaches to securing these systems, addressing the unique challenges presented by the convergence of digital control and physical operations. This section explores how this integration can enhance security in industrial control systems, smart grids, healthcare systems, and other critical cyber-physical environments.
\subsection{Security Challenges in Cyber-Physical Environments}

Cyber-physical systems face unique security challenges stemming from their hybrid nature and critical operational requirements. The convergence of IT and OT (Operational Technology) creates complex security boundaries where traditional approaches often prove inadequate \cite{H7,H6,H19}. The long operational lifecycles of many physical components, sometimes measured in decades, create security challenges when integrating with rapidly evolving digital systems. Additionally, the real-time requirements and safety-critical nature of many CPS operations impose strict constraints on security mechanisms, requiring solutions that ensure security without compromising operational performance or safety.

Resource constraints represent another significant challenge in CPS security. Many edge devices in cyber-physical systems have limited computational capabilities, memory, and power resources, restricting the security mechanisms they can support \cite{H9,H12}. These constraints are particularly challenging when implementing sophisticated security approaches like AI-based authentication\cite{H11,H10}, anomaly detection\cite{H15,H16} or blockchain verification\cite{Zuo2025Exploring,RESSI2024AI}, which traditionally require significant computational resources.

Multi-stakeholder governance in cyber-physical environments creates additional security complexities. Many CPS environments, such as industrial control systems or smart cities, involve multiple organizations with different security requirements, capabilities, and responsibilities \cite{H8}. Coordinating security across these organizational boundaries while maintaining appropriate separation of concerns requires sophisticated governance frameworks that can be difficult to implement with traditional security approaches.

The physical impact potential of security breaches in cyber-physical systems dramatically raises the stakes of security failures. Unlike purely digital systems where breaches might compromise data or services, CPS breaches can result in physical damage, environmental harm, or even loss of life \cite{H17,H18}. This reality necessitates security approaches with higher assurance levels and more comprehensive protection than might be required in traditional IT environments.
\subsection{Blockchain-Enhanced Security for Industrial Control Systems}

Decentralized authentication frameworks based on blockchain technology address critical security vulnerabilities in industrial control systems. By implementing device identity and authentication on blockchain, these frameworks eliminate central points of failure that could compromise entire control systems \cite{R1}. Blockchain-based identity protocols for industrial control systems have been suggested, combining device registration with physical attestation to detect compromised components.

Secure command verification using blockchain ensures the integrity and authorization of control commands in industrial environments. By recording command provenance, authorization, and verification on an immutable ledger, these approaches prevent unauthorized or malicious control operations \cite{R2}. Blockchain-based command verification methods have been developed to support multi-factor verification of critical control commands, ensuring provenance and auditability.

Immutable audit trails for industrial operations provide tamper-proof records of control activities, enabling both security monitoring and regulatory compliance. These blockchain-based audit systems record critical operations, configuration changes, and security events in ways that prevent modification even by privileged insiders \cite{R3}. Blockchain-based industrial audit ledgers have been introduced to provide tamper-resistant records of operational activities with role-based visibility.

Blockchain-secured firmware management addresses the critical challenge of ensuring software integrity in industrial control components. These approaches use blockchain to verify firmware provenance, validate update integrity, and record deployment history, preventing malicious code injection through the software supply chain \cite{R4}. Blockchain-enabled firmware update mechanisms have been proposed to verify update integrity and provenance, using hardware-based attestation for additional assurance.
\subsection{AI-Driven Monitoring for Cyber-Physical Security}
Process-aware anomaly detection leverages AI to identify suspicious activities in cyber-physical operations by understanding the underlying physical processes being controlled. Unlike traditional IT security monitoring, these approaches model the physical processes and their expected behaviors, enabling the detection of attacks that might appear legitimate from a purely digital perspective \cite{R5}. AI-based process behavior modeling techniques, such as physics-informed neural networks, are increasingly applied to learn normal operational patterns and detect anomalies.

Multi-sensor correlation analysis uses AI to identify coordinated attacks across different subsystems in cyber-physical environments. By analyzing relationships between seemingly unrelated anomalies across different sensors and control systems, these approaches can detect sophisticated attacks that might evade system-specific monitoring \cite{R6}. Recent approaches employ tensor-based deep learning to correlate anomalies across operational domains, enabling detection of coordinated attack campaigns.

Predictive maintenance security leverages AI to distinguish between legitimate equipment failures and security-induced malfunctions. By modeling expected degradation patterns and failure modes, these approaches can identify abnormal failures that might indicate cyber attacks targeting physical equipment \cite{R7,Amato2024Detecting}. Ensemble learning approaches have been applied for predictive maintenance security, combining operational data with telemetry to distinguish between normal wear, failures, and malicious activity.

Behavioral baselining for control operations uses AI to establish normal operational patterns for both human operators and automated systems. By learning these normal patterns, security systems can identify unusual behaviors that might indicate compromised credentials or systems \cite{R8,Gauthama2021Machine}. Unsupervised learning techniques have been explored to model operator actions and detect anomalous command patterns that can indicate insider threats.

\subsection{Integrated Approaches for Smart Grid Security}
Blockchain-based energy trading security addresses the unique challenges of securing decentralized energy markets and peer-to-peer trading platforms. These approaches implement secure transaction verification, fraud prevention, and dispute resolution mechanisms specifically designed for energy trading environments \cite{R9}. Blockchain-based energy trading mechanisms implement on-chain transaction verification with AI-driven fraud detection, supporting trustworthy and efficient peer-to-peer markets at grid scale.

Grid stability protection through secure demand response leverages both blockchain and AI to ensure that demand response programs enhance rather than compromise grid stability. Blockchain provides verification of demand response commitments and actions, while AI optimizes response patterns to maintain grid stability under varying conditions \cite{R10}. Verified demand-response schemes record commitments and actions on a distributed ledger and use AI to schedule responses that respect grid-stability constraints.

Distributed energy resource (DER) integration security addresses the challenges of securely incorporating numerous small-scale energy resources into the grid. Blockchain provides secure identity and transaction verification for DERs, while AI optimizes their integration to maintain grid security and efficiency \cite{R11}. Research prototypes for distributed energy resource integration register and authenticate resources on a distributed ledger and use AI to coordinate secure interconnection and dispatch across heterogeneous devices.

Advanced metering infrastructure (AMI) security leverages blockchain for data integrity and AI for anomaly detection in smart meter deployments. Blockchain ensures the immutability of meter data for billing and operations, while AI identifies unusual consumption patterns that might indicate tampering or fraud \cite{R12}. AMI security combines ledger-backed meter data with AI-based consumption analytics to detect theft and billing anomalies while preserving auditability.

\subsection{Healthcare Cyber-Physical Systems Security}
Medical device authentication frameworks leverage blockchain to establish and verify device identity in healthcare environments. These frameworks create tamper-evident device registries with cryptographic identity verification, preventing the introduction of unauthorized or counterfeit devices into clinical settings \cite{R13}. Medical device identity and access control can register devices on a distributed ledger and issue verifiable credentials so that only authenticated devices connect to clinical networks and access patient data.

Secure telemetry for patient monitoring combines blockchain data integrity with AI-based anomaly detection to ensure the security and privacy of patient monitoring systems. Blockchain provides verifiable records of telemetry data, while AI identifies unusual patterns that might indicate device tampering or patient emergencies \cite{R14}. Patient-monitoring pipelines combine blockchain-based provenance with privacy-preserving analytics to secure data flows while supporting timely clinical response.

Safety-preserving security for medical devices addresses the critical challenge of implementing security measures without compromising patient safety. These approaches use formal verification and safety-aware design to ensure that security mechanisms never interfere with essential medical functions, even during security incidents \cite{R15}. In safety-critical medical devices, layered security designs incorporate explicit safety overrides so that protective controls uphold essential clinical functions and enhance, rather than compromise, patient safety.

Patient-controlled health data sharing leverages blockchain for access control and AI for privacy protection in health information exchange. These approaches give patients granular control over their health data while ensuring that sharing occurs securely and in compliance with regulatory requirements \cite{R16}. Consent-management protocols record patient authorization on-chain and use AI policy engines to enforce share-with constraints during data exchange.


\section{Proactive Threat Detection and Response with Blockchain and AI} \label{sec:ProactiveThreatDetection}
\begin{figure*}[!t]
    \centering
    \includegraphics[width=.75 \textwidth]{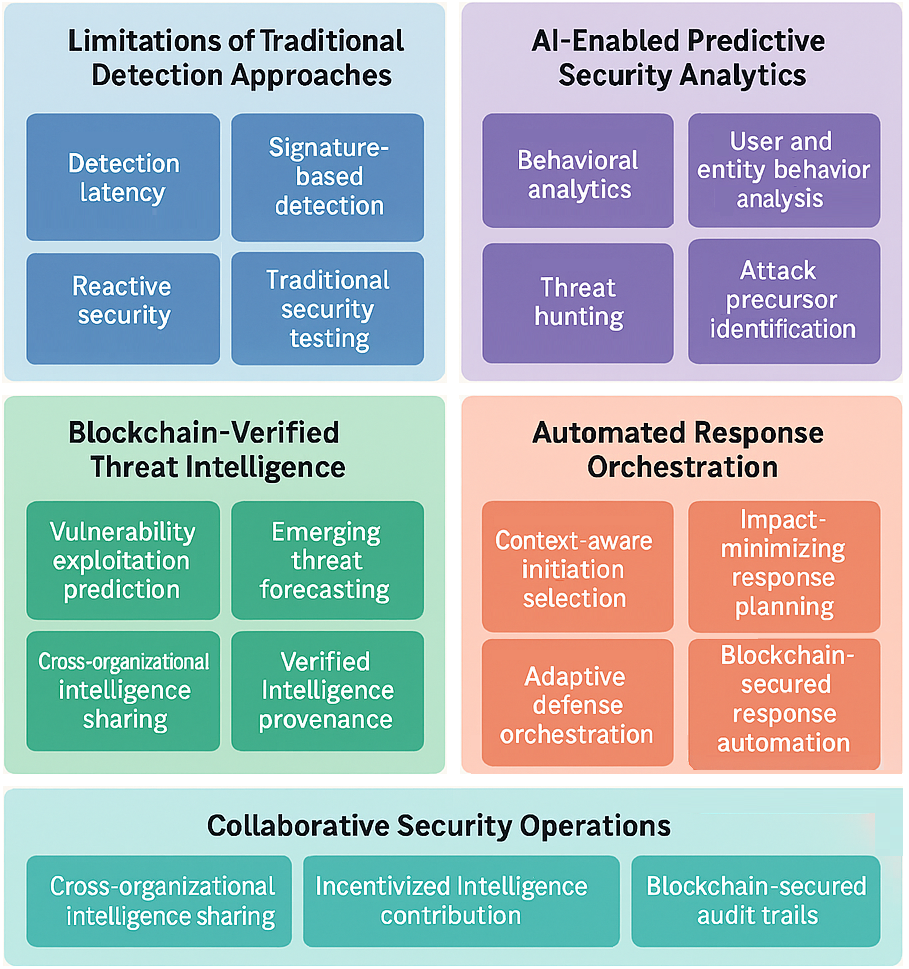}
    \caption{Overview of proactive threat detection and response strategies enabled by the integration of blockchain and AI}
    \label{fig:proactive_security}
\end{figure*}

Traditional security approaches have predominantly been reactive, responding to threats after they have been detected. The integration of blockchain and AI enables a paradigm shift toward proactive security, where threats can be anticipated and mitigated before they cause significant damage. This section explores advanced approaches to proactive threat detection and mitigation enabled by this technological integration. Figure~\ref{fig:proactive_security} provides a structured visualization of the key capabilities across five focus areas: limitations of traditional approaches, AI-powered predictive analytics, blockchain-verified threat intelligence, automated response orchestration, and collaborative security operations.

\subsection{Limitations of Traditional Detection Approaches}
Traditional security monitoring faces significant limitations that reduce its effectiveness against sophisticated threats. Detection latency in conventional systems often allows attackers to achieve their objectives before defensive responses can be implemented \cite{R17}. Industry breach-lifecycle studies continue to indicate prolonged attacker presence and response timelines; IBM’s 2025 Cost of a Data Breach Report reports a global average breach lifecycle of 241 days, indicating that many organizations still require months to identify and contain breaches\cite{IBM2025DataBreach}.

Signature-based detection, while effective against known threats, proves inadequate against novel attack vectors and zero-day exploits. These approaches can only identify threats that match pre-defined patterns, leaving organizations vulnerable to new attack methodologies \cite{R18}. The increasing sophistication of attackers, who deliberately design their techniques to evade known signatures, further reduces the effectiveness of these traditional approaches.

Reactive security frameworks that focus on incident response rather than prevention create fundamental limitations in security effectiveness. These approaches accept that breaches will occur and focus on minimizing damage rather than preventing initial compromise \cite{R19}. While incident response capabilities remain essential, the exclusive focus on reaction rather than prevention creates security postures that are perpetually disadvantaged against determined attackers.

The limitations of traditional security testing further compound these challenges. Point-in-time vulnerability assessments and periodic penetration tests provide only snapshots of security posture rather than continuous evaluation \cite{R20}. The rapidly evolving threat landscape means that new vulnerabilities may emerge immediately after testing is completed, creating security blind spots until the next assessment cycle.
\subsection{AI-Enabled Predictive Security Analytics}
Behavioral analytics leverages AI to establish baseline patterns of normal activity and identify deviations that may indicate security threats. Unlike rule-based approaches, behavioral analytics can detect novel attack patterns that don't match known signatures \cite{R21}. Advanced implementations use unsupervised learning techniques to model normal behavior across multiple dimensions, enabling the detection of subtle anomalies that might indicate sophisticated attacks in their early stages.

User and entity behavior analytics (UEBA) extends behavioral analysis to focus specifically on the actions of users and entities within networks. These approaches build comprehensive behavioral profiles and identify deviations that may indicate account compromise or insider threats \cite{R22}. By analyzing patterns across authentication, access, data movement, and system interactions, UEBA can detect malicious activities that might appear legitimate when viewed through traditional security lenses.

Threat hunting leverages AI to proactively search for indicators of compromise that may have evaded automated detection. Unlike traditional monitoring, which waits for alerts, threat hunting actively searches for subtle signs of malicious activity \cite{R23}. AI-enabled hunting platforms use machine learning to identify potential areas of investigation and guide human analysts toward suspicious patterns that warrant deeper examination, creating a proactive approach to threat discovery.

Attack precursor identification uses AI to recognize patterns that typically precede specific types of attacks, enabling preventive action before attacks fully materialize. These approaches analyze subtle indicators across network traffic, authentication attempts, and system behaviors to identify the early stages of attack campaigns \cite{R24}. By recognizing these precursors, security teams can implement targeted countermeasures that disrupt attack progressions before significant damage occurs.

\subsection{Blockchain-Verified Threat Intelligence} \label{subsec:BVTI}
Vulnerability exploitation prediction uses machine learning to forecast which vulnerabilities are most likely to be exploited in the near future. These approaches analyze factors such as vulnerability characteristics, exploit availability, and threat actor behaviors to prioritize patching efforts \cite{R25}. By focusing remediation on vulnerabilities with the highest exploitation probability rather than just technical severity, organizations can optimize their security resources for maximum risk reduction.

Emerging threat forecasting extends predictive analytics to identify new attack methodologies before they become widespread. These approaches analyze trends in threat actor communications, tool development, and early attack instances to forecast the emergence of new threat vectors \cite{R26}. This forecasting enables security teams to develop countermeasures for emerging threats before they face active exploitation attempts, shifting security posture from reactive to anticipatory.

Cross-organizational intelligence sharing leverages blockchain to create trusted mechanisms for exchanging threat information across organizational boundaries. Unlike traditional threat sharing, which relies on centralized platforms or trust relationships, blockchain-based approaches create decentralized yet verifiable intelligence exchange \cite{R27, alqubaisi2023privacy}. These systems implement cryptographic verification of intelligence sources and content, enabling organizations to make informed trust decisions about shared intelligence.

Verified intelligence provenance uses blockchain to establish and maintain the chain of custody for threat intelligence. These approaches record the origin, handling, and modification of intelligence through cryptographically secured transactions, preventing tampering or misattribution \cite{R28}. By maintaining verifiable records of intelligence provenance, these systems enable consumers to evaluate the reliability and potential biases of intelligence before acting upon it.
\subsection{Automated Response Orchestration}
Context-aware mitigation selection uses AI to determine the most appropriate response to security incidents based on comprehensive contextual analysis. Unlike static playbooks, these approaches consider factors such as attack characteristics, system criticality, operational impact, and available resources to optimize response strategies \cite{R29}. By tailoring responses to specific contexts, these systems maximize security effectiveness while minimizing operational disruption.

Impact-minimizing response planning leverages AI to develop response strategies that contain threats while minimizing negative effects on legitimate operations. These approaches model the potential consequences of different response actions and identify strategies that achieve security objectives with minimal collateral impact \cite{Dutta2025Empowering}. This capability is particularly valuable in operational technology environments where security responses must be balanced against safety and continuity requirements.

Adaptive defense orchestration implements dynamic security controls that evolve based on the current threat landscape and system state. Unlike static security configurations, adaptive orchestration continuously adjusts security posture based on risk assessments and threat intelligence \cite{dutta2024advanced}. These systems implement graduated security measures that intensify in response to increased threat levels, creating efficient security operations that concentrate resources where they are most needed.

Blockchain-secured response automation addresses the trust challenges in automated security responses by creating verifiable records of response decisions and actions. These approaches implement cryptographic verification of response authorization and execution, preventing unauthorized or malicious automation \cite{dutta2023next}. By maintaining immutable audit trails of automated responses, these systems create accountability in security automation while still enabling the speed and scale benefits of automated operations.
\subsection{Collaborative Security Operations}
As summarized in \autoref{subsec:BVTI}, cross-organizational intelligence sharing relies on blockchain-verified provenance; here we focus on incentives, auditability, and distributed monitoring. Incentivized intelligence contribution addresses the "free-rider" problem in threat intelligence sharing by creating verifiable records of contributions and establishing reward mechanisms. These approaches use blockchain to track intelligence contributions and implement token-based or reputation-based incentive systems \cite{alneyadi2023role}. By creating tangible benefits for intelligence sharing while maintaining cryptographic verification of contributions, these systems foster sustainable intelligence ecosystems.

Blockchain-secured audit trails provide tamper-proof records of security events, creating trustworthy forensic evidence and compliance documentation. Unlike traditional logging systems, which can be manipulated by attackers or insiders, blockchain-based audit trails implement cryptographic verification and distributed storage \cite{almansoori2023machine}. These characteristics make them particularly valuable for security operations that span organizational boundaries or require regulatory compliance verification.

Distributed security monitoring implements sensor networks that maintain effectiveness even when parts of the network are compromised. These approaches use blockchain to coordinate monitoring activities and verify the integrity of security telemetry across distributed environments \cite{Islam2025Decentralized}. By eliminating central points of failure and implementing cryptographic verification of monitoring data, these systems create resilient visibility that can withstand targeted attacks against the monitoring infrastructure itself.


\section{Challenges and Limitations in Blockchain-AI Integration} \label{sec:challenges}

\begin{figure*}[!t]
    \centering
    \includegraphics[width=\textwidth]{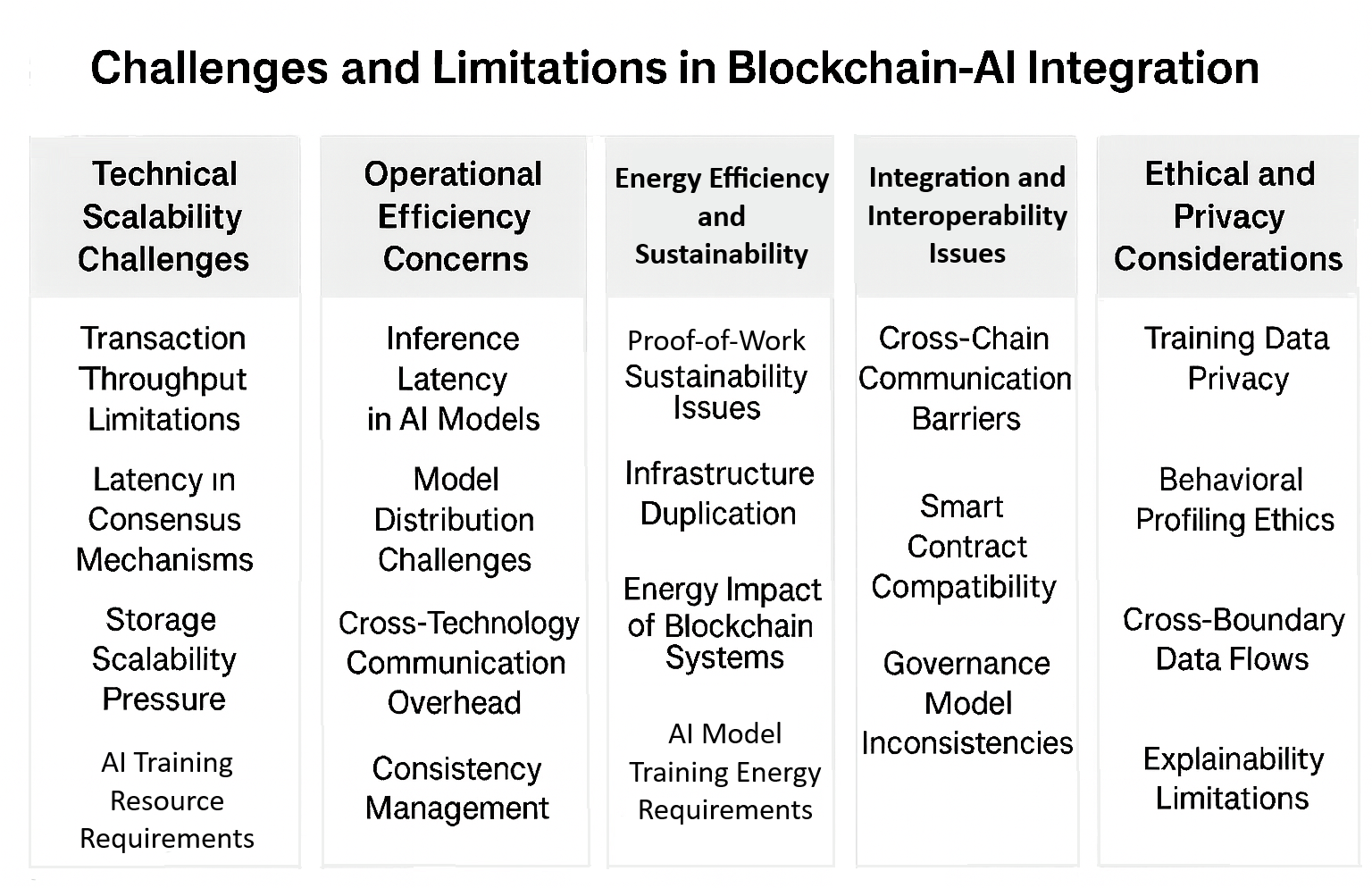}
    \caption{Visual breakdown of key challenges in integrating blockchain and AI for intelligent network security}
    \label{fig:blockchain_ai_challenges}
\end{figure*}

While the integration of blockchain and AI offers powerful new capabilities for securing intelligent networks, this integration also presents significant challenges and limitations that must be addressed. This section examines the technical, operational, and ethical challenges associated with blockchain-AI integration for security applications, providing a balanced view of both the potential and the constraints of these technologies. Figure~\ref{fig:blockchain_ai_challenges} presents a categorized overview of these challenges, helping readers quickly grasp the scope and structure of limitations discussed throughout the section.

\subsection{Technical Scalability Challenges}
Transaction throughput limitations in blockchain systems create significant challenges for security applications that require high-volume data processing. Traditional public blockchains such as Bitcoin and Ethereum mainnet can process only a limited number of transactions per second at the protocol level, typically ranging from single-digit to low-tens transactions per second, which remains far below the demands of many security monitoring workloads \cite{Maurya2024Blockchain}. While newer blockchain architectures implement various scaling approaches, fundamental trade-offs between throughput, decentralization, and security create ongoing challenges for high-volume security applications.

Latency challenges in blockchain consensus mechanisms can conflict with the real-time requirements of many security operations. The time required to achieve consensus and finality in blockchain networks, ranging from seconds to minutes depending on the protocol, may be incompatible with security functions that require immediate response \cite{Alharby2022Transaction}. This latency creates particular challenges for time-sensitive security operations such as access control, attack mitigation, and critical infrastructure protection.

Storage scalability issues emerge as blockchain-based security systems accumulate data over time. The requirement to maintain complete historical records, while valuable for security auditing and forensics, creates ever-growing storage demands that can become unsustainable \cite{Eren2025Security}. While various pruning and archiving strategies have been proposed, the fundamental tension between comprehensive record-keeping and storage efficiency remains a significant challenge.

AI training resource requirements present another scalability challenge for integrated systems. Advanced AI models for security applications often require substantial computational resources for training, with state-of-the-art models demanding GPU clusters operating for days or weeks \cite{Thompson2020TheCL}. These resource requirements can make sophisticated AI capabilities inaccessible for many security implementations, particularly in resource-constrained environments like IoT networks or industrial control systems.
\subsection{Operational Efficiency Concerns}
Inference latency in complex AI models can create operational bottlenecks in security systems that require real-time decision-making. While training happens offline, inference must occur in real-time for many security applications, and complex models may introduce latency that impacts operational effectiveness \cite{Deng2020Edge}. This challenge is particularly acute in resource-constrained environments where computational capabilities for inference may be limited.

Model distribution challenges emerge when deploying AI capabilities across distributed security environments. The size of sophisticated AI models, sometimes reaching billions of parameters, creates significant challenges for deployment to edge devices or resource-constrained environments \cite{Lim2020Federated}. While various model compression and edge optimization techniques exist, fundamental trade-offs between model capability and deployment flexibility remain.

Cross-technology communication overhead in integrated systems can create performance bottlenecks and operational complexity. The interfaces between blockchain and AI components introduce additional processing layers that can impact system performance and reliability \cite{SANKA2021Review}. This overhead is particularly challenging in high-throughput security applications where efficient processing is essential for operational effectiveness.

Consistency management in distributed blockchain-AI systems presents significant operational challenges. Maintaining consistent state across blockchain networks while simultaneously updating AI models creates complex synchronization requirements \cite{Martínez2023Decentralized}. These challenges are particularly acute in security applications where inconsistencies could create vulnerabilities or lead to incorrect security decisions.


\subsection{Energy Efficiency and Sustainability}
Proof-of-work sustainability issues in blockchain security applications create significant environmental concerns \cite{STOLL2019PoW}. The energy consumption of proof-of-work consensus mechanisms, particularly in large-scale blockchain networks, raises questions about the environmental impact of blockchain-based security solutions. While alternative consensus mechanisms like proof-of-stake dramatically reduce energy requirements, they introduce different security trade-offs that must be carefully evaluated for each application context.

Infrastructure duplication in blockchain networks, where multiple nodes maintain identical copies of the ledger, creates inherent inefficiencies in resource utilization \cite{Huang2021Survey}. This redundancy, while valuable for security and resilience, multiplies the resource requirements of blockchain-based security systems. The tension between redundancy for security and efficiency in resource utilization represents an ongoing challenge for sustainable blockchain implementations.

The energy impact of blockchain security systems extends beyond consensus mechanisms to include network communication, storage, and general operation. The distributed nature of blockchain systems, with multiple nodes performing similar operations, creates multiplicative effects on energy consumption \cite{SANKA2021Review}. These effects must be carefully considered when evaluating the sustainability of blockchain-based security approaches, particularly for large-scale or long-term implementations.

AI model training energy requirements present another sustainability challenge for integrated security systems. The computational resources required for training sophisticated AI models translate directly to significant energy consumption, with environmental impacts that must be considered \cite{Patterson2022Carbon}. While various techniques for efficient training exist, the fundamental relationship between model sophistication and training energy requirements creates ongoing sustainability challenges.
\subsection{Integration and Interoperability Issues}
Cross-chain communication barriers create significant challenges for security implementations that span multiple blockchain networks. The lack of standardized cross-chain protocols limits the ability to create comprehensive security solutions that operate across different blockchain environments \cite{Wang2023Exploring}. While various cross-chain technologies are emerging, fundamental challenges related to security guarantees, transaction finality, and protocol compatibility remain.

Smart contract compatibility issues across different blockchain platforms create integration challenges for security implementations. Variations in smart contract languages, execution environments, and capabilities limit the portability of security logic across different blockchain platforms \cite{Ren2023Interoperability}. These limitations can create significant development overhead and operational complexity for security implementations that must span multiple blockchain environments.

Governance model inconsistencies between different blockchain networks create challenges for integrated security operations. Variations in how different blockchains handle protocol updates, dispute resolution, and community governance can create unpredictable behaviors in cross-chain security implementations \cite{Wang2023Exploring}. These inconsistencies are particularly challenging for security applications where governance predictability is essential for risk management.

AI model interface inconsistencies create integration challenges when combining different AI capabilities within security systems. Variations in input formats, output structures, and operational parameters create complexity when building integrated security capabilities from multiple AI components. These inconsistencies can lead to development overhead, operational complexity, and potential security vulnerabilities at component interfaces.
\subsection{Ethical and Privacy Considerations}
Training data privacy concerns emerge when developing AI security capabilities, particularly when sensitive data is used for model training. The potential for privacy violations, data misuse, or inadvertent information leakage creates significant ethical challenges \cite{Liu2021When}. These concerns are particularly acute in security contexts where training data may contain sensitive information about systems, vulnerabilities, or user behaviors.

Behavioral profiling ethics in AI security systems raise questions about surveillance, privacy, and potential discrimination. The detailed behavioral models used for anomaly detection and threat identification may capture sensitive information about individuals and organizations \cite{Hu2022Membership}. The tension between security effectiveness and privacy preservation represents an ongoing ethical challenge in AI-based security approaches.

Cross-boundary data flows in integrated security systems create regulatory compliance challenges related to data sovereignty and privacy regulations. The global nature of many security threats conflicts with increasingly stringent data localization requirements in different jurisdictions \cite{Liu2021When}. These conflicts create particular challenges for blockchain-based approaches where data is inherently distributed across multiple nodes that may span jurisdictional boundaries.

Explainability limitations in AI security decisions raise significant ethical and operational concerns. The ``black-box'' nature of many advanced AI models makes it difficult to justify, verify, and audit security decisions, creating challenges for accountability and operator trust \cite{Li2023Trustworthy, Dutta2026Explainable}. These limitations are particularly problematic in security contexts where decisions may have significant consequences and may require justification to stakeholders or regulators.


\section{Comparative Analysis of Blockchain-AI Synergy} \label{sec:ComparativeAnalysis}

The interaction between blockchain and artificial intelligence (AI) is bidirectional, with each technology reinforcing the strengths of the other. Blockchain contributes primarily to the \textit{trust, transparency, and integrity} of AI-driven systems, while AI improves the \textit{efficiency, scalability, and adaptability} of blockchain operations. Table~\ref{tab:synergy} summarizes this relationship across key dimensions.

\begin{table}[h!]
\centering
\caption{Comparative perspectives on Blockchain-AI synergy}
\label{tab:synergy}
\begin{tabular}{|p{3.5cm}|p{5.5cm}|p{5.5cm}|}
\hline
\textbf{Dimension} & \textbf{Blockchain $\rightarrow$ AI} & \textbf{AI $\rightarrow$ Blockchain} \\
\hline
Data Integrity \& Provenance \cite{Huang2021Survey, SANKA2021Review, Wang2023Exploring}& Provides tamper-proof training datasets, verifiable model updates, and immutable provenance trails that ensure AI operates on trustworthy data. & AI cannot change immutability guarantees; it improves upstream data quality via preprocessing and anomaly detection, which benefits on-chain provenance.
 \\
\hline
Transparency \& Explainability \cite{Li2023Trustworthy, Dutta2026Explainable}& Offers immutable audit logs of AI model decisions, enabling verifiable accountability. & Supports interpretability of blockchain records by applying explainable AI techniques to smart contracts and transaction analysis. \\
\hline
Scalability \& Efficiency \cite{SANKA2021Review, Huang2021Survey, Lim2020Federated} & No direct effect on model scalability; blockchain can indirectly support scalable AI through trusted data sharing and provenance. & Optimizes blockchain consensus (e.g., resource-aware scheduling, reinforcement learning for adaptive consensus) and reduces computational overhead. \\
\hline
Security \cite{Huang2021Survey, Zhou2019EdgeIntelligence, Ivanov2023Security} & Ensures AI pipelines are resilient to data poisoning and tampering by anchoring evidence on-chain. & Detects anomalies, adversarial behavior, or malicious nodes within blockchain networks using machine learning models. \\
\hline
Trust \cite{SANKA2021Review, GUO2025auditing, Darshan2023Architecture} \& Governance & Smart contracts enforce provenance, fairness, and policy compliance for AI models (e.g., in federated learning). & AI enables automated auditing and vulnerability detection in smart contracts, improving blockchain governance. \\
\hline
\end{tabular}
\end{table}

\noindent As Table~\ref{tab:synergy} shows, certain dimensions exhibit a stronger one-way contribution. For example, blockchain uniquely provides immutable provenance (hence the AI $\rightarrow$ Blockchain column remains limited under \textit{Data Integrity}), while AI uniquely enhances scalability of consensus (hence blockchain has no direct role under \textit{Scalability}). This asymmetry is expected, since the two technologies operate at different layers of the computational stack: blockchain at the \textit{trust and data integrity layer}, and AI at the \textit{analytics and decision-making layer}. 

Overall, the comparative view highlights that blockchain and AI are not simply complementary but mutually reinforcing. Blockchain introduces trust and accountability into AI-driven systems, whereas AI enhances the adaptability and efficiency of blockchain infrastructures. Together, they create a foundation for secure, transparent, and resilient intelligent networks. Next, to move from comparative claims to evidence, we operationalize the dimensions in Table~\ref{tab:synergy} into testable criteria. The next subsection specifies metrics and reproducibility artifacts that allow blockchain-AI implementations to be evaluated under common workloads and fault conditions. This makes integrity, transparency, scalability, security, and governance comparable across systems rather than merely asserted.
\subsection{BASE: Evaluation and Benchmarking Checklist} \label{sec:EvaluationBenchmarks}

We formalize the discussion in Table~\ref{tab:synergy} into a reusable reporting template, which we term the \textit{Blockchain-AI Security Evaluation Blueprint (BASE)}. BASE is intended to make integrated security systems comparable across domains by requiring joint reporting of AI performance, ledger behavior, closed-loop operational impact, energy, and reproducibility.

\paragraph{AI detection:} Report class-imbalance aware metrics such as AUROC, AUPRC, and Matthews Correlation Coefficient (MCC). Include alert-volume reduction relative to a vetted ruleset or SIEM baseline, and time-to-detect measured from the first malicious event to the first correct alert (median and 95th percentile). When using LLMs, also report calibration error or Brier score for probabilistic outputs \cite{Li2023Trustworthy, Zhang2025LLMs}.

\paragraph{Blockchain components:} Report throughput in transactions per second at the target finality setting, confirmation latency to probabilistic finality (public) or block commit (permissioned), and reorganization resilience as the frequency of orphaned blocks and the probability of a reorganization deeper than $k$ under bursty and adversarial workloads \cite{Alharby2022Transaction, SANKA2021Review}. Include audit latency for provenance queries over $N$ blocks and resource overhead covering CPU, memory, storage, and energy \cite{R28, Huang2021Survey}.

\paragraph{End-to-end systems:} Define service-level objectives for detection-to-mitigation and audit-to-action under mixed load, then measure them with closed-loop tests. Provide failure-injection results covering node crashes, network partitions, clock skew, stale features, and model poisoning. If differential privacy is used, report the privacy loss budget $(\varepsilon,\delta)$ consumed per release and its effect on utility \cite{Maurya2024Blockchain, Lim2020Federated, Liu2021When}.

\paragraph{Artifacts and protocol:} Release a reproducibility package containing fixed data splits or synthetic workload generators, attack and fault scripts, configuration files, random seeds, environment manifests, and analysis notebooks \cite{Fu2024AIforDevSecOps}. Document the test harness and dataset licensing, and version all inputs so third parties can rerun the full pipeline \cite{Li2023Trustworthy}.

\begin{table}[t]
\centering
\caption{BASE checklist for reporting blockchain-AI security systems \cite{Li2023Trustworthy,Zhang2025LLMs,Alharby2022Transaction,SANKA2021Review,R28,Huang2021Survey,Maurya2024Blockchain,Lim2020Federated,Liu2021When,Fu2024AIforDevSecOps}}
\label{tab:base_checklist}
\scriptsize
\renewcommand{\arraystretch}{1.15}
\resizebox{\textwidth}{!}{%
\begin{tabular}{|p{2.1cm}|p{4.3cm}|p{4.8cm}|}
\hline
\textbf{Layer} & \textbf{Minimum items to report} & \textbf{Representative measures} \\
\hline
AI quality & Detection quality, false-alarm burden, calibration, drift sensitivity & AUROC, AUPRC, MCC, false positives/day, Brier score, time-to-detect \\
\hline
Ledger behavior & Finality-aware throughput, commit latency, audit latency, resilience & TPS at finality, commit time, orphan/reorg statistics, provenance query latency \\
\hline
Closed-loop security & Operational service levels under attack and failure & Detection-to-mitigation time, audit-to-action time, failure injection, poisoning tolerance \\
\hline
Privacy and governance & Privacy guarantees, policy traceability, operator accountability & $(\varepsilon,\delta)$, consent traceability, decision logs, approval checkpoints \\
\hline
Energy and cost & Resource footprint of the full pipeline & Per-inference energy, per-transaction energy, storage growth, GPU-hours, carbon-aware scheduling notes \\
\hline
Reproducibility & Inputs, scripts, environments, and workload definitions & Data splits, seeds, manifests, attack scripts, licenses, notebooks \\
\hline
\end{tabular}%
}
\end{table}


\section{Emerging Trends and Future Directions} \label{sec:EmergingTrends}
\begin{figure*}[!t]
    \centering
    \includegraphics[width=\textwidth]{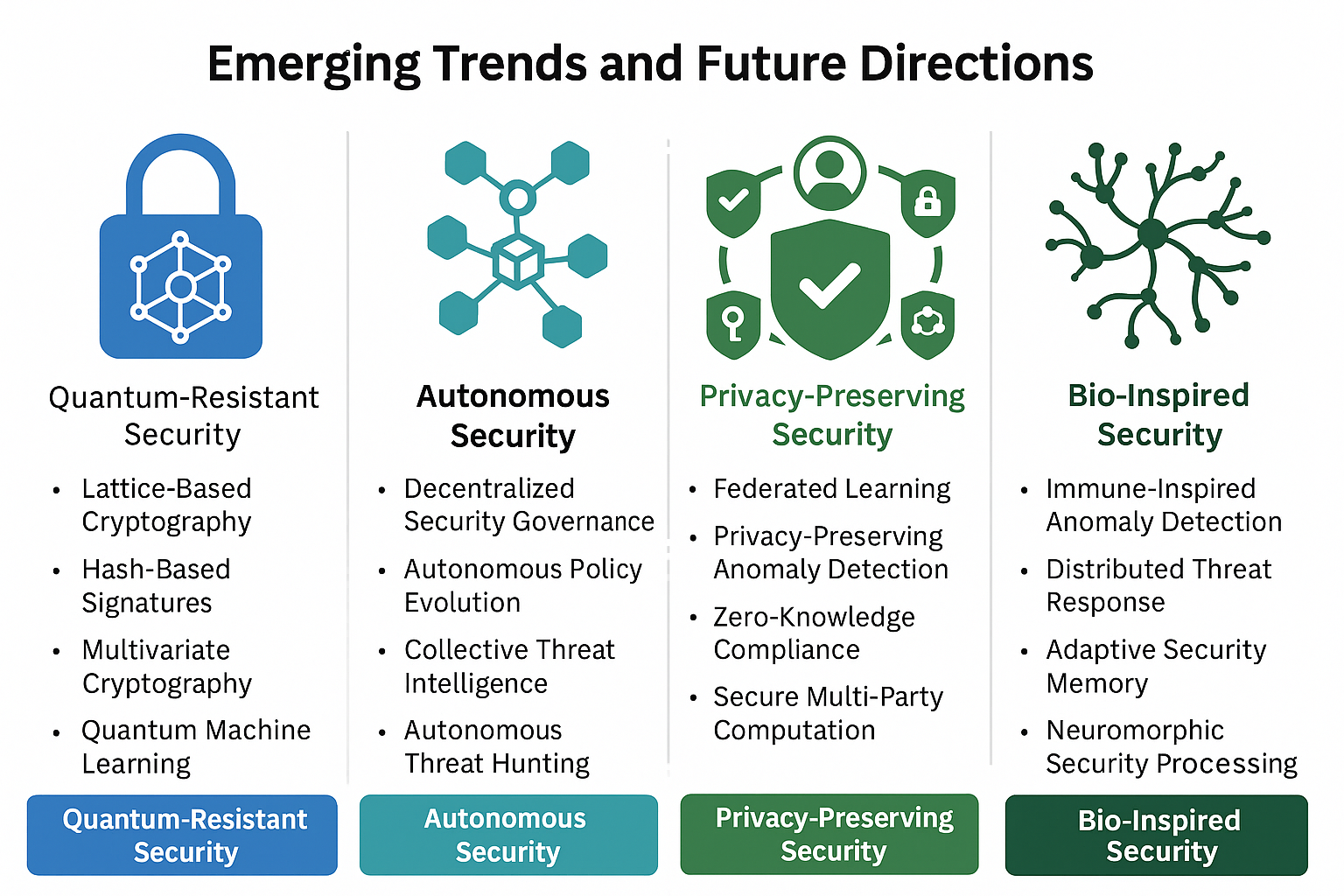}
    \caption{Emerging trends and future directions in securing intelligent networks through blockchain and AI integration}
    \label{fig:Emerging-Trends}
\end{figure*}

The integration of blockchain and AI for securing intelligent networks continues to evolve rapidly, with emerging technologies and approaches promising to address current limitations and enable new capabilities. This evolution is expected to become even more consequential in next-generation 6G environments, where ultra-dense connectivity, terahertz communication, native AI support, and highly distributed service architectures expand the attack surface and intensify the need for security mechanisms that are adaptive, privacy-aware, and verifiable \cite{Kumar2026Next}. This section explores the future directions and trends in this field, examining quantum-resistant approaches, autonomous and agentic security systems, privacy-preserving techniques, and bio-inspired security paradigms. Figure~\ref{fig:Emerging-Trends} presents a visual overview of these four categories, offering readers a structured understanding of how these innovations collectively shape the future landscape of secure intelligent systems.

\subsection{Quantum-Resistant Security Approaches}
Lattice-based cryptography is a leading approach for quantum-resistant blockchain security. Unlike traditional cryptographic methods that are vulnerable to quantum attacks, lattice-based approaches rely on mathematical problems that remain hard even for quantum computers \cite{Allende2023Quantum}. The integration of lattice-based signatures and key establishment into blockchain protocols creates quantum-resistant security foundations that can withstand future advances in quantum computing.

Hash-based signature schemes provide another quantum-resistant approach for blockchain security applications. These schemes rely solely on the security of cryptographic hash functions, which are believed to maintain significant security margins even against quantum attacks \cite{LIU2025hash}. The integration of hash-based signatures into blockchain protocols, particularly for long-term security applications, provides protection against future quantum threats while maintaining compatibility with existing blockchain architectures. Stateless schemes such as SPHINCS+ avoid state-management risks but have larger signatures, so plan for storage and verification costs.

Multivariate cryptography offers additional quantum-resistant options for blockchain security. These approaches use systems of multivariate polynomial equations, which present computational challenges even for quantum algorithms \cite{Fernández2020Towards}. While more computationally intensive than traditional cryptography, multivariate approaches provide strong security guarantees against quantum attacks, making them valuable for high-security blockchain applications. These schemes remain under active cryptanalysis and standardization, so adopt cautiously and track updates from standards bodies.

Quantum machine learning does not provide post-quantum cryptographic protection; it targets analytics performance and detection quality. By leveraging quantum computing principles for pattern recognition and anomaly detection, these approaches promise advances in threat identification capabilities \cite{CORLI2025Quantum}. While practical implementations remain limited by current quantum hardware capabilities, the theoretical foundations for quantum-enhanced security analytics are rapidly developing.

For near-term migrations, prioritize the NIST-standardized primitives ML-KEM for key establishment and ML-DSA or SLH-DSA for signatures \cite{NIST2024PQC}, while also tracking compact signature candidates such as Falcon-derived schemes as they progress toward standard publication as FN-DSA. Pilot deployments should benchmark signature size, verification cost, key and certificate overhead, and on-chain storage impacts under realistic block and transaction loads before broad rollout. Hybrid certificate chains that combine classical and post-quantum signatures can smooth migration without requiring abrupt replacement of existing trust infrastructures.

\subsection{Autonomous Security Systems}
Decentralized security governance frameworks enable collaborative security management without central control points \cite{Fu2024AIforDevSecOps}. These frameworks use blockchain to implement distributed decision-making for security policies, incident response, and resource allocation. By eliminating central governance authorities while maintaining coordinated security operations, these approaches create resilient security ecosystems that can withstand targeted attacks against governance mechanisms.

Autonomous security policy evolution leverages AI to continuously adapt security rules and configurations based on emerging threats and changing environments \cite{PURVES2024Causally}. Unlike traditional approaches where policies are manually updated, these systems implement self-modifying security frameworks that evolve through AI-driven learning \cite{HU2019gametheory}. The integration of blockchain for policy verification and audit creates trustworthy autonomous policy systems that maintain accountability despite their self-modifying nature.

Collective threat intelligence systems implement distributed mechanisms for gathering, validating, and acting upon security intelligence \cite{Marinho2023Automated,Perrina2023AGIR}. These systems use blockchain to create verifiable intelligence repositories while leveraging AI for intelligence analysis and correlation. By distributing both intelligence collection and analysis across multiple participants, these systems create comprehensive threat awareness that no single organization could achieve independently.

Autonomous threat hunting systems proactively search for indicators of compromise without human direction. These systems implement AI-driven hunting methodologies that continuously evolve based on discovered threats and attack patterns \cite{Hasanov2024LLM,Shahriar2025Zeroshot}. The integration of blockchain for hunt verification and findings documentation creates trustworthy autonomous hunting that maintains human oversight without requiring constant human direction.

A natural next step beyond autonomous analysis is agentic orchestration, where AI components do not only interpret events but can also plan actions, call approved tools, maintain working memory, and coordinate bounded workflows across security platforms \cite{Ferrag2025AutonomousAgents,Lazer2026AgenticCyber}. In this setting, blockchain becomes even more valuable because agentic systems need trustworthy state transitions, policy-aware action logs, signed tool invocations, and verifiable coordination across multiple actors. This opens the door to practical workflows such as evidence gathering, alert prioritization, automated response orchestration, and cross-organizational response support. The safer direction, however, is not unrestricted autonomy. Agentic security systems should operate under explicit guardrails, including least-privilege tool access, policy-as-code enforcement, human approval gates for high-impact actions, and rollback-capable execution trails. This keeps automation useful while preserving accountability, explainability, and operational control.

\subsection{Privacy-Preserving Security Techniques}
Federated learning for cross-organizational threat detection enables collaborative security analytics without sharing sensitive data \cite{Lim2020Federated,Martínez2023Decentralized}. Organizations train models locally on their own data and share only model updates rather than raw data, preserving privacy while enabling collective intelligence. The integration of blockchain for verifying model update integrity and participant compliance creates trustworthy federated learning that maintains privacy guarantees even with untrusted participants \cite{cai2025blockchain}.

Privacy-preserving anomaly detection implements techniques that identify security threats without exposing sensitive details of the underlying data \cite{Liu2021When,Hu2022Membership}. Beyond detection alone, privacy-preserving security analytics increasingly require explainable query and decision mechanisms so that sensitive blockchain-backed data can be analyzed without exposing raw records while still providing transparent justification for system outputs, especially in domains such as eHealth and regulated data-sharing environments \cite{Dutta2026Explainable}. These approaches use methods such as differential privacy, homomorphic encryption, and secure multi-party computation to maintain data confidentiality throughout the detection process. The integration of blockchain for verification and audit creates privacy-preserving detection systems that maintain accountability without compromising confidentiality.

Zero-knowledge compliance verification enables regulatory validation without exposing sensitive security details. These approaches use zero-knowledge proofs to demonstrate compliance with security requirements while revealing minimal information about the underlying systems and data \cite{Sun2021Survey}. The integration of blockchain for recording compliance verification creates auditable security assurance that satisfies regulatory requirements while maintaining operational security through minimal information disclosure \cite{Liang2024Permissioned-and-PermissionlessSurvey}.

Secure multi-party computation for collaborative security enables multiple organizations to jointly analyze security data without revealing their individual inputs \cite{li2024supply}. These approaches implement cryptographic protocols that produce collective analysis results while keeping each organization's data private. The integration of blockchain for verification and audit creates trustworthy collaborative analytics that maintain privacy guarantees even with potentially untrusted participants.
\subsection{Bio-Inspired Security Paradigms}
Immune system-inspired anomaly detection implements security approaches modeled on biological immune responses. These systems develop "security antibodies" that recognize and respond to threats based on both known patterns and adaptive learning \cite{ALDHAHERI20201Artificial}. The integration of blockchain for response verification and memory persistence creates trustworthy immune-inspired security that combines adaptive protection with verifiable response histories.

Distributed threat response coordination based on immune system signaling implements collaborative defense mechanisms inspired by biological systems \cite{wu2024federated}. These approaches use distributed signaling protocols to coordinate responses across security components, similar to how biological systems coordinate immune responses. The integration of blockchain for signal verification and response coordination creates trustworthy distributed defense that maintains coordination integrity even in partially compromised environments.

Adaptive security memory implements approaches for retaining and evolving security knowledge over time, inspired by immunological memory \cite{ALDHAHERI20201Artificial}. These systems maintain historical threat information and continuously refine response strategies based on observed effectiveness. The integration of blockchain for memory integrity and evolution verification creates trustworthy security memory that maintains its reliability even over extended operational periods.

Neuromorphic security processing leverages brain-inspired computing architectures for efficient security analytics \cite{Rathi2023Exploring}. These approaches implement neural processing models on specialized hardware to achieve significant efficiency improvements for pattern recognition and anomaly detection. The integration of blockchain for verification and audit creates trustworthy neuromorphic security that maintains accountability despite its specialized processing architecture.
\subsection{Open Research Gaps and Challenges}

Despite significant progress in integrating blockchain and AI for securing intelligent networks, several open research gaps remain that limit widespread deployment and adoption. Identifying these gaps is essential for setting a future research agenda and ensuring that upcoming developments are both impactful and sustainable. The key challenges include:

\begin{itemize}
    \item \textbf{Scalability vs. Security Trade-offs:} Current integration efforts struggle to balance throughput, latency, and security, particularly in resource-constrained IoT and edge environments. Lightweight yet verifiable consensus mechanisms remain an open challenge \cite{Wang2023Exploring}. 
    
    \item \textbf{Lack of Standardization:} A proliferation of conceptual frameworks exists, but no universally accepted interoperability standards have been developed. This lack of standardization hampers cross-domain adoption and integration \cite{Huang2021Survey}.
    
    \item \textbf{Energy and Sustainability Concerns:} Both blockchain consensus and AI training are computationally intensive. Joint optimization strategies for energy efficiency and sustainable deployment are still underexplored \cite{deVries2022Revisiting}.

    \item \textbf{Explainability and Accountability:} While blockchain ensures traceability, integrating explainable AI that satisfies regulatory and ethical requirements for decision justification remains unresolved \cite{Li2023Trustworthy, Dutta2026Explainable}.
    
    \item \textbf{Cross-Domain Interoperability:} Blockchain-AI integration has been explored in silos such as healthcare, IoT, industrial control, and smart cities. However, generalizable frameworks that enable inter-domain trust and interoperability remain absent \cite{Ren2023Interoperability}.
    
    \item \textbf{Forensics and Post-Compromise Recovery:} Existing research largely emphasizes prevention and detection. The role of blockchain–AI systems in forensic analysis, incident attribution, and rapid recovery is still underdeveloped \cite{Li2023Trustworthy}.
    
    \item \textbf{Human-in-the-Loop Integration:} Most existing architectures assume fully automated trust mechanisms. Incorporating human oversight, ethical reasoning, and policy-driven governance remains a critical gap \cite{McKay2024_HITL}.
\end{itemize}

\noindent Addressing these challenges will require interdisciplinary research that combines insights from distributed systems, cybersecurity, machine learning, and human-computer interaction. Importantly, the future of blockchain-AI integration lies not only in enhancing efficiency and security but also in ensuring that such systems are sustainable, accountable, and societally aligned.


\section{Evidence Landscape and Representative Applications} \label{sec:ApplicationsCaseStudies}

This section treats the literature as an evidence landscape rather than a list of success stories. The integration of blockchain and AI for security is moving beyond purely theoretical discussion, but the maturity of the evidence remains uneven across domains. Some studies report pilots or targeted case studies, while many others describe prototypes, architectures, systematic reviews, or framework-driven integrations. For that reason, this section focuses on representative application patterns and practical security roles rather than treating every cited work as a mature field deployment. The goal is to show where blockchain-AI integration is already useful, what security problems it addresses, and where stronger real-world validation is still needed.

Table~\ref{tab:application_map} provides an evidence and maturity map across representative application domains, summarizing the dominant security use cases, the type of evidence currently available, the practical takeaway, and the present maturity level of blockchain-AI security in each area.

\begin{sidewaystable*}[p]
\centering
\caption{Evidence and maturity map for blockchain-AI security across representative application domains}
\label{tab:application_map}
\scriptsize
\renewcommand{\arraystretch}{1.15}
\resizebox{\textheight}{!}{%
\begin{tabular}{|p{2.0cm}|p{2.5cm}|p{2.3cm}|p{2.5cm}|p{2.6cm}|p{3.0cm}|p{1.9cm}|p{2.4cm}|}
\hline
\textbf{Domain} & \textbf{Typical security use case} & \textbf{Blockchain contribution} & \textbf{AI contribution in integrated or proposed designs} & \textbf{Evidence type} & \textbf{Practical takeaway} & \textbf{Current maturity level} & \textbf{Representative citations} \\
\hline
IoT & Device identity, edge anomaly detection, collaborative threat intelligence, and forensic logging & Device registration, credential verification, provenance, and tamper-evident evidence sharing & Behavioral authentication, edge anomaly detection, event correlation, and filtering & Mostly systematic reviews, framework papers, and proposed architectures & Strong conceptual fit for decentralized trust and distributed detection, but operational validation is still limited & Emerging / prototype-oriented & \cite{YALLI2025Authentication, NAYAK2024review, NAZIR2024Collaborative, Ahmed2025Forensics} \\
\hline
Critical infrastructure / IIoT & Secure commands, resilient monitoring, sensor integrity, and audit trails & Provenance, access tracing, cross-domain trust anchoring, and auditable operational state & Process-anomaly analysis, state validation, and adaptive decision support where integrated & Sector-specific frameworks, resilient-control studies, and IIoT architectures & Highly relevant for OT environments, but the evidence is still weighted toward prototypes and framework-level studies & Prototype-heavy & \cite{Rahman2021Blockchain, RATHEE2021design, Sasikumar2025secure, Ahmad2024Blockchain} \\
\hline
Smart grid & Peer-to-peer trading, demand response, DER coordination, and privacy-aware utility data sharing & Transaction verification, commitment logging, resource authentication, and coordination governance & Fraud screening, demand and stability analytics, coordination optimization, and privacy-aware intelligence in integrated designs & Mixed case studies, blockchain application papers, and integrated security frameworks & One of the stronger application areas because coordination and auditability needs are clear, but fully integrated production evidence remains uneven & Mixed but promising & \cite{SHEN2025Blockchain, KOLAHAN2021Blockchain, Cantillo2022Blockchain, MAKHDOOM2020PrivySharing, Ghadi2025hybrid} \\
\hline
Transportation / EV / V2G & Secure charging, V2G trust, privacy-preserving authentication, and OTA update protection & Authorization records, payment traceability, credential anchoring, and update provenance & Anomaly detection, load-aware risk assessment, context-aware authentication, and update monitoring in integrated designs & Surveys, attack-surface studies, authentication proposals, and targeted prototypes & Strong need and clear direction, but the literature still leans more toward security analysis and proposed mechanisms than broad operational deployment & Early-to-prototype & \cite{Kaur2025Cybersecurity, RAZZAQUE2025Cybersecurity, Sarieddine2023Investigating, Nai2025Authentication, Li2024Over, Dutta2025Empowering} \\
\hline
Healthcare & Medical device identity, remote monitoring integrity, pharmaceutical provenance, and consent management & Identity, provenance, access control, consent traceability, and audit logs & Adaptive authentication, anomaly detection, privacy-risk assessment, and decision support in integrated designs & Integrated e-health frameworks plus sector-specific prototypes and focused application studies & High-value domain because trust, privacy, and traceability are essential, but evidence remains concentrated in focused deployments and prototypes & Prototype-heavy / early deployment & \cite{dutta2024advancing, AHMAD2021blockchain, AKRAM2024Blockchain, Pham2024Distributed, Alabdulatif2025Blockchain} \\
\hline
\end{tabular}%
}
\vspace{1mm}
\parbox{\textheight}{\footnotesize \textit{Note:} The columns ``Evidence type,'' ``Practical takeaway,'' and ``Current maturity level'' are author syntheses based on the cited literature and are intended as comparative guides rather than formal maturity ratings.}
\end{sidewaystable*}
\subsection{IoT Security Applications}
Secure device identity and authentication frameworks address the fundamental challenge of establishing trust in large-scale IoT deployments. Representative IoT security studies use blockchain for device identity registration and verification, while AI can strengthen authentication through behavioral analysis and continuous trust assessment \cite{YALLI2025Authentication, NAZIR2024Collaborative}. Together, these approaches create scalable identity systems that can reduce spoofing, device impersonation, and unauthorized device enrollment in large IoT deployments.

Edge-based anomaly detection systems implement AI capabilities directly on IoT gateways and edge devices, enabling real-time threat detection without constant cloud connectivity. Recent work in edge analytics highlights the value of lightweight models at the edge for low-latency monitoring, while blockchain-enhanced IoT security architectures can add trusted coordination, integrity checks, and stronger provenance for distributed operations \cite{NAYAK2024review, NAZIR2024Collaborative}. These approaches create more resilient detection capabilities that can continue functioning even when cloud connectivity is limited or disrupted.

Collaborative threat intelligence across IoT domains enables broader security visibility despite the fragmentation of IoT ecosystems. In this line of work, blockchain is used for cross-domain intelligence sharing and evidence traceability, while AI is used for correlation, prioritization, and pattern discovery across heterogeneous alerts \cite{NAZIR2024Collaborative}. These approaches improve collective situational awareness and support coordinated responses to threats that span multiple IoT environments.

Immutable security event logging for IoT forensics provides tamper-evident records of security events, enabling more reliable incident investigation and regulatory support. Research on IoT forensics increasingly discusses blockchain-backed logging and AI-assisted event triage as complementary mechanisms for improving forensic readiness while managing large data volumes \cite{Ahmed2025Forensics}. These approaches create more trustworthy investigation trails without assuming that every current design has already reached broad operational deployment.

\subsection{Critical Infrastructure Protection}
In critical infrastructure and industrial IoT, blockchain-AI integration is mainly being explored to strengthen trust in commands, improve system auditability, and support resilient monitoring and control. Segmented security architectures for industrial control systems implement defense-in-depth approaches specifically designed for operational technology environments \cite{Rahman2021Blockchain}. In representative designs, blockchain is used for provenance, access tracing, and cross-domain trust anchoring, while AI-supported monitoring can help detect abnormal process behavior and prioritize investigation \cite{RATHEE2021design}. These approaches strengthen containment and auditability without losing sight of operational continuity.

Secure sensor networks for industrial process integrity implement distributed verification of sensor data to reduce the risk of manipulated control inputs. Blockchain-assisted IIoT architectures can strengthen sensor provenance and data integrity at the edge, creating a more trustworthy basis for downstream anomaly analysis and operational validation \cite{Sasikumar2025secure}. These approaches improve trust in sensing pipelines for safety-critical environments.

Resilient control systems for critical infrastructure aim to maintain essential functions even during cyber incidents. Recent work on blockchain-assisted resilient control for distributed energy resource management shows how tamper-evident coordination and verifiable control records can support safer system behavior under abnormal conditions \cite{Ahmad2024Blockchain}. When paired with AI-driven state estimation, anomaly detection, or adaptive decision support, such architectures can help critical services degrade gracefully rather than fail catastrophically.

Overall, the evidence in this domain is promising, but it still comes mainly from sector-specific frameworks, resilient-control studies, and prototypes rather than universal production deployments.

\subsection{Smart Grid Security Implementations}
Smart grid research shows a particularly strong fit for blockchain-AI integration because the domain requires coordination, auditability, and fast operational decision-making at the same time. Blockchain-based peer-to-peer energy trading security enables secure transactions between distributed energy producers and consumers \cite{SHEN2025Blockchain}. In this setting, blockchain provides transaction verification, settlement transparency, and auditable participation, while AI-enabled designs can add fraud detection, demand forecasting, and market optimization when those capabilities are explicitly integrated \cite{Ghadi2025hybrid}. These approaches create more trustworthy local energy markets while reducing the risk of manipulation and coordination failure.

Grid stability protection through secure demand response implements verified load management capabilities that enhance rather than compromise grid stability. Verified demand-response schemes use blockchain for commitment verification and trusted coordination, while AI can support stability-preserving scheduling and response optimization \cite{KOLAHAN2021Blockchain, Ghadi2025hybrid}. These approaches create demand-response systems that are better aligned with both operational efficiency and security.

Distributed energy resource security frameworks address the challenge of securely integrating many small-scale resources into the grid. Research on DER management uses blockchain for resource authentication, transaction integrity, and coordination across heterogeneous participants, while AI-based analytics can assist forecasting, orchestration, and abnormal-behavior detection in more integrated designs \cite{Cantillo2022Blockchain, Ahmad2024Blockchain, Ghadi2025hybrid}. These approaches create more dependable pathways for incorporating diverse energy resources without sacrificing traceability or control.

Privacy-preserving data sharing is also highly relevant to smart-grid security because operational analytics often involve multiple stakeholders and sensitive usage data. Blockchain-based secure data-sharing frameworks and privacy-preserving analytics point to practical ways of supporting coordination and analysis without exposing raw participant data \cite{MAKHDOOM2020PrivySharing, Ghadi2025hybrid}. These approaches are especially valuable where metering, billing, and control information must be shared under tight accountability requirements.

\subsection{Transportation and Electric Vehicle Applications}
Secure charging infrastructure for electric vehicles implements protected payment and energy-delivery processes for distributed charging networks. Current literature on the EV charging ecosystem highlights the need for stronger protection of payment workflows, charger access, and backend coordination. In integrated designs, blockchain can provide payment verification and charging authorization, while AI can support fraud detection, load management, and context-aware risk assessment \cite{Kaur2025Cybersecurity}. These approaches create more trustworthy charging networks that protect both financial transactions and the physical charging infrastructure.

Blockchain-based vehicle-to-grid (V2G) security frameworks enable secure bidirectional energy flows between electric vehicles and the power grid. The V2G cybersecurity literature emphasizes transaction integrity, energy accounting, authentication, and anomaly detection as core design needs \cite{RAZZAQUE2025Cybersecurity}. In integrated architectures, blockchain supports trustworthy records and credential anchoring, while AI can help identify abnormal energy-flow behavior or suspicious operating patterns. These approaches strengthen trust in bidirectional energy exchange and reduce the risk of both financial abuse and physical disruption.

AI-powered threat detection for connected EVs implements real-time security monitoring tailored to the dynamic and mobile nature of vehicle ecosystems. A closely related direction is V2X protection, where blockchain-based trust management can be combined with edge large language models and context-aware authentication to support low-latency, adaptive security decisions in highly dynamic vehicular environments \cite{Dutta2025Empowering}. Complementing these V2X-oriented protections, EV security research has also examined charging applications and associated attack surfaces, underscoring the need for anomaly detection, secure application design, and trustworthy software distribution \cite{Sarieddine2023Investigating}. Collectively, these lines of work show why EV security needs both strong trust infrastructure and adaptive analytics.

Privacy-preserving authentication for EV charging and payments implements secure identity verification without exposing sensitive user or vehicle information. Recent authentication protocols show how zero-knowledge techniques can reduce identity exposure during vehicular authentication, and these cryptographic mechanisms can be complemented by blockchain-backed credential management and context-aware authentication in broader charging ecosystems \cite{Nai2025Authentication, Dutta2025Empowering}. These approaches create charging systems that maintain user privacy while reducing the risk of unauthorized access and payment fraud.

Secure over-the-air updates for EV software systems implement protected mechanisms for remotely updating vehicle software without introducing new vulnerabilities. Recent work on intelligent connected vehicles emphasizes the security importance of update verification, provenance, rollback protection, and anomaly monitoring during the update process \cite{Li2024Over}. In integrated designs, blockchain can anchor update provenance and approval history, while AI can support abnormality detection during distribution and installation. These approaches strengthen vehicle security and safety across the software lifecycle.

\subsection{Healthcare Security Applications}
Healthcare is another important domain where blockchain-AI integration is gaining momentum, especially in remote monitoring, medical device identity, pharmaceutical traceability, and patient-controlled access management. Medical device authentication frameworks implement secure identity verification for healthcare devices to prevent unauthorized access to clinical networks. In healthcare-oriented integrated designs, blockchain is used for device registration, credential management, and auditable access control, while AI supports adaptive trust assessment and anomaly detection in more integrated e-health settings \cite{dutta2024advancing, Alabdulatif2025Blockchain}. These approaches help create more trustworthy connected-device ecosystems and reduce the risk of counterfeit or compromised devices entering clinical workflows.

Secure telemetry for patient monitoring implements protected data flows from monitoring devices to clinical systems. Blockchain can preserve provenance and access traceability for telemetry data, while AI can help identify abnormal physiological or device behavior in time-sensitive care settings \cite{AHMAD2021blockchain, dutta2024advancing}. These approaches support both data integrity and patient privacy while improving the reliability of remote monitoring.

Drug provenance tracking using blockchain implements secure supply-chain verification for pharmaceutical products. Current evidence is strongest on blockchain-enabled custody verification and traceability, which can create a trustworthy data foundation for later risk analytics or anomaly screening in distribution workflows \cite{AKRAM2024Blockchain}. These approaches strengthen pharmaceutical supply-chain security and improve the ability to detect suspicious or counterfeit distribution paths.

Patient-controlled health data sharing implements secure mechanisms for patients to control access to their health information. Distributed consent-management designs use blockchain to record authorization, trace data access, and support auditable exchange, while AI-enabled privacy assistance can help assess disclosure risk or policy compliance in more integrated healthcare workflows \cite{Pham2024Distributed, dutta2024advancing}. These approaches create more patient-centered data sharing while preserving both security and individual autonomy.


\section{Conclusion and Future Work}\label{sec:Conclusion}
This paper shows that blockchain and AI are most effective for security when they are designed together. Section~\ref{sec:Advanced-Blockchain-Technology} explained how modern ledger mechanisms provide integrity, provenance, selective privacy, and fault tolerance, while Section~\ref{sec:AI-Driven-Cybersecurity} described how AI enables accurate detection, adaptive response, and machine-assisted reasoning. Section~\ref{sec:Blockchain-AI-Integration} brought these strengths together into verifiable workflows that reduce manual effort while keeping accountability in place. We illustrated practical value in safety-critical environments in Section~\ref{sec:SecuringCPS} and showed how operations can move from reactive to preventive in Section~\ref{sec:ProactiveThreatDetection}. Limits and trade-offs were discussed in Section~\ref{sec:challenges} and compared in Section~\ref{sec:ComparativeAnalysis}, which also introduced BASE, the Blockchain-AI Security Evaluation Blueprint, clarifying how to report AI metrics, ledger performance, end-to-end reliability, energy, and reproducibility. Section~\ref{sec:EmergingTrends} outlined near-term directions, and Section~\ref{sec:ApplicationsCaseStudies} synthesized representative evidence across IoT, electric mobility, healthcare, and critical infrastructure.

The practical takeaway is direct. Verifiable data flows and adaptive analytics can shorten time to detect, reduce false positives, and improve auditability when they are supported by transparent evaluation and clear governance. For researchers, this means open benchmarks, reproducible artifacts, and studies that compare design choices under common workloads. For industry, this means pipeline automation, measurable service levels, and integration patterns that keep systems portable across vendors and clouds. For policy makers, this means aligning requirements with evidence that systems can produce by design, including model and data lineage, privacy budgets, and cryptographic attestations.

Several priorities should guide future work. First, expand BASE from Section~\ref{sec:ComparativeAnalysis} into open, cross-domain suites that pair representative datasets with adversarial workloads and failure injection, while reporting energy and cost. Second, standardize interfaces between consensus, identity, provenance, and AI pipelines from Section~\ref{sec:Blockchain-AI-Integration}, and publish reference adapters for cross-chain operation, streaming feature stores, and policy-as-code. Third, turn the autonomous ideas in Section~\ref{sec:EmergingTrends} into certifiable practice by combining safe reinforcement learning with bounded agentic workflows, runtime checks, rollback, and human-over-the-loop controls so automated triage and response remain explainable, policy-constrained, and auditable. Fourth, operationalize privacy-preserving analytics at scale by integrating federated learning, secure multi-party computation, and zero-knowledge proofs with practical service-level objectives, and by reporting privacy budgets, leakage resistance, and utility under realistic participation and network churn.

Energy and sustainability should become first-class metrics beyond Section~\ref{sec:challenges}. Standardize per-inference, per-transaction, and per-incident energy, and evaluate carbon-aware scheduling that balances finality, accuracy, and hardware use. Bio-inspired and neuromorphic defenses from Section~\ref{sec:EmergingTrends} warrant hardware-efficient implementations for edge nodes, targeting event-driven analytics, online learning with limited memory, and distributed signaling that degrades gracefully when some nodes are compromised. Finally, invest in socio-technical validation. In addition to technical KPIs, measure analyst workload, mean time to detect and contain, incident quality, and the collaboration patterns introduced in Section~\ref{sec:ProactiveThreatDetection}. These steps will move the field from promising prototypes to dependable systems that raise security outcomes at scale.

\bibliography{references_complete}

@inproceedings{dutta2024advanced,
  title={Advanced eHealth with Explainable AI: Secured by Blockchain with AI-Empowered Block Sensitivity for Adaptive Authentication},
  author={Dutta, Joy and Eldeeb, Hossien and Ho, Tu Dac},
  booktitle={2024 IEEE 35th Annual International Symposium on Personal, Indoor and Mobile Radio Communications (PIMRC)},
  pages={1--7},
  year={2024},
  organization={IEEE},
  address={Valencia, Spain},
  doi={10.1109/PIMRC59610.2024.10817278}
}

@inproceedings{dutta2023next,
  title={Next Generation Healthcare with Explainable AI: IoMT-Edge-Cloud based Advanced eHealth},
  author={Dutta, Joy and Puthal, Deepak and Yeun, Chan Yeob},
  booktitle={IEEE GLOBECOM 2023},
  year={2023},
  organization={IEEE},
  address={Kuala Lumpur, Malaysia},
  doi={10.1109/GLOBECOM54140.2023.10436967}
}

@inproceedings{alqubaisi2023privacy,
  title={Privacy-aware Adaptive Collaborative Learning Approach for Distributed Edge Networks},
  author={Alqubaisi, Saeed and Puthal, Deepak and Dutta, Joy and Damiani, Ernesto},
  booktitle={IEEE DSAA 2023},
  year={2023},
  organization={IEEE},
  address={Thessaloniki, Greece},
  doi={10.1109/DSAA60987.2023.10302577}
}

@inproceedings{alneyadi2023role,
  title={Role-based Access Control in Private Blockchain For IoT Integrated Smart Contract},
  author={Al Neyadi, Darwish and Puthal, Deepak and Dutta, Joy and Damiani, Ernesto},
  booktitle={IFIP IoT 2023},
  publisher={Springer},
  year={2023},
  address={Texas, USA},
  doi={10.1007/978-3-031-45882-8_16}
}

@inproceedings{almansoori2023machine,
  title={Machine Learning-based Adaptive Access Control Mechanism for Private Blockchain Storage},
  author={Almansoori, S. and Alzaabi, M. and Alrayssi, M. and Puthal, D. and Dutta, J. and Shehhi, A. A.},
  booktitle={IEEE COMPSAC 2023},
  year={2023},
  organization={IEEE},
  address={Torino, Italy},
  doi={10.1109/COMPSAC57700.2023.00188}
}

@inproceedings{dutta2022ai,
  title={AI-based Block Identification and Classification in the Blockchain Integrated IoT},
  author={Dutta, J. and Puthal, D. and Damiani, E.},
  booktitle={20th OITS International Conference on Information Technology (OCIT)},
  pages={415--421},
  year={2022},
  organization={IEEE},
  address={Bhubaneswar, India},
  doi={10.1109/OCIT56763.2022.00084}
}

@article{dutta2024advancing,
  title={Advancing eHealth in Society 5.0: A Fuzzy Logic and Blockchain-Enhanced Framework for Integrating IoMT, Edge, and Cloud with AI},
  author={Dutta, Joy and Puthal, Deepak},
  journal={IEEE Access},
  year={2024},
  doi={10.1109/ACCESS.2024.3520799}
}

@article{dutta2024poah,
  title={PoAh 2.0: AI-empowered Dynamic Authentication based Adaptive Blockchain Consensus for IoMT-Edge Workflow},
  author={Dutta, Joy and Puthal, Deepak},
  journal={Future Generation Computer Systems},
  volume={160},
  pages={129--142},
  year={2024},
  publisher={Elsevier},
  doi={10.1016/j.future.2024.07.048}
}

@article{HUSAIN2025Context,
title = {Context-aware energy auctions on blockchain for mobile EV charging},
journal = {Sustainable Cities and Society},
pages = {106548},
year = {2025},
issn = {2210-6707},
doi = {https://doi.org/10.1016/j.scs.2025.106548},
url = {https://www.sciencedirect.com/science/article/pii/S2210670725004226},
author = {Zainab Husain and Joy Dutta and Shakti Singh and Rabeb Mizouni and Tarek H.M. El-Fouly and Hadi Otrok}
}

@article{alrayes2024intrusion,
  title={Intrusion Detection in IoT Systems Using Denoising Autoencoder},
  author={Alrayes, Fatma S. and Zakariah, Mohammed and Amin, Syed Umar and Khan, Zafar Iqbal and Helal, Maha},
  journal={IEEE Access},
  volume={12},
  pages={122401--122425},
  year={2024},
  publisher={IEEE},
  doi={10.1109/ACCESS.2024.3451726}
}

@article{cai2025blockchain,
  title={Blockchain-empowered Federated Learning: Benefits, Challenges and Solutions},
  author={Cai, Zeju and Chen, Jianguo and Fan, Yuting and Zheng, Zibin and Li, Keqin},
  journal={IEEE Transactions on Big Data},
  year={2025},
  publisher={IEEE},
  doi={10.1109/TBDATA.2025.3541560}
}

@article{wu2024federated,
  title={Swarm Intelligence for Cybersecurity},
  author={Wu, Feng and Yang, Hua and Zhou, Bin},
  journal={IEEE Transactions on Emerging Topics in Computational Intelligence},
  volume={8},
  number={4},
  pages={567--582},
  year={2024},
  publisher={IEEE},
  doi={10.1109/TETCI.2024.3398765}
}

@article{li2024supply,
  title={Secure Multi-party Computation in Blockchain Systems},
  author={Li, Gang and Xu, Tao and Ma, Jun},
  journal={IEEE Transactions on Dependable and Secure Computing},
  volume={21},
  number={8},
  pages={9012--9027},
  year={2024},
  publisher={IEEE},
  doi={10.1109/TDSC.2024.3398765}
}

@ARTICLE{Sun2021Survey,
  author={Sun, Xiaoqiang and Yu, F. Richard and Zhang, Peng and Sun, Zhiwei and Xie, Weixin and Peng, Xiang},
  journal={IEEE Network}, 
  title={A Survey on Zero-Knowledge Proof in Blockchain}, 
  year={2021},
  volume={35},
  number={4},
  pages={198-205},
  doi={10.1109/MNET.011.2000473}}

@article{Ofusori2024Artificial,
author = {Lizzy Ofusori and Tebogo Bokaba and Siyabonga Mhlongo},
title = {Artificial Intelligence in Cybersecurity: A Comprehensive Review and Future Direction},
journal = {Applied Artificial Intelligence},
volume = {38},
number = {1},
pages = {2439609},
year = {2024},
publisher = {Taylor \& Francis},
doi = {10.1080/08839514.2024.2439609},


URL = { 
    
        https://doi.org/10.1080/08839514.2024.2439609
    
    

},
eprint = { 
    
        https://doi.org/10.1080/08839514.2024.2439609
    
    

}

}

@ARTICLE{Salah2019Blockchain,
  author={Salah, Khaled and Rehman, M. Habib Ur and Nizamuddin, Nishara and Al-Fuqaha, Ala},
  journal={IEEE Access}, 
  title={Blockchain for AI: Review and Open Research Challenges}, 
  year={2019},
  volume={7},
  number={},
  pages={10127-10149},
  doi={10.1109/ACCESS.2018.2890507}}

@article{Hussain2021Artificial,
author = {Hussain, Adedoyin A. and Al-Turjman, Fadi},
title = {Artificial intelligence and blockchain: A review},
journal = {Transactions on Emerging Telecommunications Technologies},
volume = {32},
number = {9},
pages = {e4268},
doi = {https://doi.org/10.1002/ett.4268},
url = {https://onlinelibrary.wiley.com/doi/abs/10.1002/ett.4268},
eprint = {https://onlinelibrary.wiley.com/doi/pdf/10.1002/ett.4268},
year = {2021}
}

@ARTICLE{Zuo2025Exploring,
  author={Zuo, Yanjun},
  journal={IEEE Internet of Things Journal}, 
  title={Exploring the Synergy: AI Enhancing Blockchain, Blockchain Empowering AI, and Their Convergence Across IoT Applications and Beyond}, 
  year={2025},
  volume={12},
  number={6},
  pages={6171-6195},
  doi={10.1109/JIOT.2024.3507746}}

@article{RESSI2024AI,
title = {AI-enhanced blockchain technology: A review of advancements and opportunities},
journal = {Journal of Network and Computer Applications},
volume = {225},
pages = {103858},
year = {2024},
issn = {1084-8045},
doi = {https://doi.org/10.1016/j.jnca.2024.103858},
url = {https://www.sciencedirect.com/science/article/pii/S1084804524000353},
author = {Dalila Ressi and Riccardo Romanello and Carla Piazza and Sabina Rossi}
}

@article{Xu2023Survey,
author = {Xu, Jie and Wang, Cong and Jia, Xiaohua},
title = {A Survey of Blockchain Consensus Protocols},
year = {2023},
issue_date = {December 2023},
publisher = {Association for Computing Machinery},
address = {New York, NY, USA},
volume = {55},
number = {13s},
issn = {0360-0300},
url = {https://doi.org/10.1145/3579845},
doi = {10.1145/3579845},
journal = {ACM Comput. Surv.},
month = jul,
articleno = {278},
numpages = {35}
}

@article{Allende2023Quantum,
  author    = {Marcos Allende and Diego L{\'o}pez Le{\'o}n and Sergio Cer{\'o}n and Adri{\'a}n Pareja and Erick Pacheco and Antonio Leal and Marcelo Da Silva and Alejandro Pardo and Duncan Jones and David J. Worrall and Ben Merriman and Jonathan Gilmore and Nick Kitchener and Salvador E. Venegas-Andraca},
  title     = {Quantum-resistance in blockchain networks},
  journal   = {Scientific Reports},
  year      = {2023},
  volume    = {13},
  number    = {1},
  pages     = {5664},
  doi       = {10.1038/s41598-023-32701-6},
  url       = {https://doi.org/10.1038/s41598-023-32701-6},
  issn      = {2045-2322}
}

@ARTICLE{Quan2025Sharding,
  author={Quan, Brandon Liew Yi and Wahab, Nur Haliza Abdul and Al-Dhaqm, Arafat and Alshammari, Ahmad and Aqarni, Ali and Razak, Shukor Abd and Wei, Koh Tieng},
  journal={IEEE Access}, 
  title={Recent Advances in Sharding Techniques for Scalable Blockchain Networks: A Review}, 
  year={2025},
  volume={13},
  number={},
  pages={21335-21366},
  doi={10.1109/ACCESS.2024.3523256}}

@article{LI2025CrossChain,
title = {Towards Blockchain Interoperability: A Comprehensive Survey on Cross-Chain Solutions},
journal = {Blockchain: Research and Applications},
pages = {100286},
year = {2025},
issn = {2096-7209},
doi = {https://doi.org/10.1016/j.bcra.2025.100286},
url = {https://www.sciencedirect.com/science/article/pii/S2096720925000132},
author = {Wenqing Li and Zhenguang Liu and Jianhai Chen and Zhe Liu and Qinming He}
}

@INPROCEEDINGS{Chiedu2025On-Chain-Off-Chain-Scalability,
  author={Chiedu, Chiedu Charles and Gai, Keke and Wei, Yihang and Ding, Kai},
  booktitle={2025 IEEE 11th Conference on Big Data Security on Cloud (BigDataSecurity)}, 
  title={A Survey on On-Chain-Off-Chain Scalability in Blockchain}, 
  year={2025},
  volume={},
  number={},
  pages={71-82},
  doi={10.1109/BigDataSecurity66063.2025.00011}}

@article{Liang2024Permissioned-and-PermissionlessSurvey,
author = {Liang, Wei and Liu, Yaqin and Yang, Ce and Xie, Songyou and Li, Kuanching and Susilo, Willy},
title = {On Identity, Transaction, and Smart Contract Privacy on Permissioned and Permissionless Blockchain: A Comprehensive Survey},
year = {2024},
issue_date = {December 2024},
publisher = {Association for Computing Machinery},
address = {New York, NY, USA},
volume = {56},
number = {12},
issn = {0360-0300},
url = {https://doi.org/10.1145/3676164},
doi = {10.1145/3676164},
journal = {ACM Comput. Surv.},
month = jul,
articleno = {298},
numpages = {35}
}

@article{Issa2023Blockchain-Based-Federated-Learning,
author = {Issa, Wael and Moustafa, Nour and Turnbull, Benjamin and Sohrabi, Nasrin and Tari, Zahir},
title = {Blockchain-Based Federated Learning for Securing Internet of Things: A Comprehensive Survey},
year = {2023},
issue_date = {September 2023},
publisher = {Association for Computing Machinery},
address = {New York, NY, USA},
volume = {55},
number = {9},
issn = {0360-0300},
url = {https://doi.org/10.1145/3560816},
doi = {10.1145/3560816},
journal = {ACM Comput. Surv.},
month = jan,
articleno = {191},
numpages = {43}
}

@article{Azimi2025SmartContracts,
  author    = {Sadaf Azimi and Ali Golzari and Naghmeh Ivaki and Nuno Laranjeiro},
  title     = {A systematic review on smart contracts security design patterns},
  journal   = {Empirical Software Engineering},
  year      = {2025},
  volume    = {30},
  number    = {4},
  pages     = {95},
  doi       = {10.1007/s10664-025-10646-w},
  url       = {https://doi.org/10.1007/s10664-025-10646-w},
  issn      = {1573-7616}
}

@article{MOHANM2023ChainCode,
title = {An efficient chain code for access control in hyper ledger fabric healthcare system},
journal = {e-Prime - Advances in Electrical Engineering, Electronics and Energy},
volume = {5},
pages = {100204},
year = {2023},
issn = {2772-6711},
doi = {https://doi.org/10.1016/j.prime.2023.100204},
url = {https://www.sciencedirect.com/science/article/pii/S2772671123000992},
author = {Smitha {Mohan M} and L. Sujihelen}
}

@article{Fernando2024Vulnerability,
title = {Vulnerability detection techniques for smart contracts: A systematic literature review},
journal = {Journal of Systems and Software},
volume = {217},
pages = {112160},
year = {2024},
issn = {0164-1212},
doi = {https://doi.org/10.1016/j.jss.2024.112160},
url = {https://www.sciencedirect.com/science/article/pii/S016412122400205X},
author = {Fernando Richter Vidal and Naghmeh Ivaki and Nuno Laranjeiro}
}

@article{Schardong2022SSI,
  author       = {Schardong, Frederico and Custódio, Ricardo},
  title        = {Self-Sovereign Identity: A Systematic Review, Mapping and Taxonomy},
  journal      = {Sensors},
  year         = {2022},
  volume       = {22},
  number       = {15},
  pages        = {5641},
  doi          = {10.3390/s22155641},
  issn         = {1424-8220},
  url          = {https://www.mdpi.com/1424-8220/22/15/5641},
  publisher    = {MDPI},
  pmid         = {35957198},
  pmcid        = {PMC9371034},
  language     = {English}
}

@article{Islam2025Decentralized,
  author       = {Islam, Rafiqul and Bose, Rajesh and Roy, Sandip and Khan, Arfat Ahmad and Sutradhar, Shrabani and Das, Sujan and Ali, Farman and AlZubi, Ahmad Ali},
  title        = {Decentralized trust framework for smart cities: a blockchain-enabled cybersecurity and data integrity model},
  journal      = {Scientific Reports},
  year         = {2025},
  volume       = {15},
  number       = {1},
  pages        = {23454},
  doi          = {10.1038/s41598-025-06405-y},
  url          = {https://doi.org/10.1038/s41598-025-06405-y},
  issn         = {2045-2322}
}

@article{Charles2023critical,
  author       = {Charles, Vincent and Emrouznejad, Ali and Gherman, Tatiana},
  title        = {A critical analysis of the integration of blockchain and artificial intelligence for supply chain},
  journal      = {Annals of Operations Research},
  year         = {2023},
  volume       = {327},
  number       = {1},
  pages        = {7--47},
  doi          = {10.1007/s10479-023-05169-w},
  url          = {https://doi.org/10.1007/s10479-023-05169-w},
  issn         = {1572-9338}
}

@article{Ohize2024Blockchain,
  author       = {Ohize, Henry O. and Onumanyi, Adeiza James and Umar, Buhari U. and Ajao, Lukman A. and Isah, Rabiu O. and Dogo, Eustace M. and Nuhu, Bello K. and Olaniyi, Olayemi M. and Ambafi, James G. and Sheidu, Vincent B. and Ibrahim, Muhammad M.},
  title        = {Blockchain for securing electronic voting systems: a survey of architectures, trends, solutions, and challenges},
  journal      = {Cluster Computing},
  year         = {2024},
  volume       = {28},
  number       = {2},
  pages        = {132},
  doi          = {10.1007/s10586-024-04709-8},
  url          = {https://doi.org/10.1007/s10586-024-04709-8},
  issn         = {1573-7543}
}

@article{Ajagbe2024Intrusion,
  author       = {Ajagbe, Sunday Adeola and Awotunde, Joseph Bamidele and Florez, Hector},
  title        = {Intrusion Detection: A Comparison Study of Machine Learning Models Using Unbalanced Dataset},
  journal      = {SN Computer Science},
  year         = {2024},
  volume       = {5},
  number       = {8},
  pages        = {1028},
  doi          = {10.1007/s42979-024-03369-0},
  url          = {https://doi.org/10.1007/s42979-024-03369-0},
  issn         = {2661-8907}
}

@Article{Liu2023Anomaly,
AUTHOR = {Liu, Wenjie and Lei, Pengfei and Xu, Dong and Zhu, Xiaorong},
TITLE = {Anomaly Recognition, Diagnosis and Prediction of Massive Data Flow Based on Time-GAN and DBSCAN for Power Dispatching Automation System},
JOURNAL = {Processes},
VOLUME = {11},
YEAR = {2023},
NUMBER = {9},
ARTICLE-NUMBER = {2782},
URL = {https://www.mdpi.com/2227-9717/11/9/2782},
ISSN = {2227-9717},
DOI = {10.3390/pr11092782}
}

@Article{Gang2024PUNet,
AUTHOR = {Gang Long, Zhaoxin Zhang},
TITLE = {PUNet: A Semi-Supervised Anomaly Detection Model for Network Anomaly Detection Based on Positive Unlabeled Data},
JOURNAL = {Computers, Materials \& Continua},
VOLUME = {81},
YEAR = {2024},
NUMBER = {1},
PAGES = {327--343},
URL = {http://www.techscience.com/cmc/v81n1/58316},
ISSN = {1546-2226},
DOI = {10.32604/cmc.2024.054558}
}

@article{Priyadarshini2022TimeSeries,
author = {Priyadarshini, Ishaani and Alkhayyat, Ahmed and Gehlot, Anita and Kumar, Raghvendra},
title = {Time series analysis and anomaly detection for trustworthy smart homes},
year = {2022},
issue_date = {Sep 2022},
publisher = {Pergamon Press, Inc.},
address = {USA},
volume = {102},
number = {C},
issn = {0045-7906},
url = {https://doi.org/10.1016/j.compeleceng.2022.108193},
doi = {10.1016/j.compeleceng.2022.108193},
journal = {Comput. Electr. Eng.},
month = sep,
numpages = {18}
}

@article{WASIF2025CNN,
title = {CNN-ViT synergy: An efficient Android malware detection approach through deep learning},
journal = {Computers and Electrical Engineering},
volume = {123},
pages = {110039},
year = {2025},
issn = {0045-7906},
doi = {https://doi.org/10.1016/j.compeleceng.2024.110039},
url = {https://www.sciencedirect.com/science/article/pii/S0045790624009649},
author = {Md. Shadman Wasif and Md. Palash Miah and Md. Shohrab Hossain and Mohammed J.F. Alenazi and Mohammed Atiquzzaman}
}

@article{Dash2025LSTM,
  author       = {Dash, Nitu and Chakravarty, Sujata and Rath, Amiya Kumar and Giri, Nimay Chandra and AboRas, Kareem M. and Gowtham, N.},
  title        = {An optimized LSTM-based deep learning model for anomaly network intrusion detection},
  journal      = {Scientific Reports},
  year         = {2025},
  volume       = {15},
  number       = {1},
  pages        = {1554},
  doi          = {10.1038/s41598-025-85248-z},
  url          = {https://doi.org/10.1038/s41598-025-85248-z},
  issn         = {2045-2322}
}

@article{KHEDDAR2025Transformers,
title = {Transformers and large language models for efficient intrusion detection systems: A comprehensive survey},
journal = {Information Fusion},
volume = {124},
pages = {103347},
year = {2025},
issn = {1566-2535},
doi = {https://doi.org/10.1016/j.inffus.2025.103347},
url = {https://www.sciencedirect.com/science/article/pii/S1566253525004208},
author = {Hamza Kheddar}
}

@article{Torabi2023Autoencoder,
  author       = {Torabi, Hasan and Mirtaheri, Seyedeh Leili and Greco, Sergio},
  title        = {Practical autoencoder based anomaly detection by using vector reconstruction error},
  journal      = {Cybersecurity},
  year         = {2023},
  volume       = {6},
  number       = {1},
  pages        = {1},
  doi          = {10.1186/s42400-022-00134-9},
  url          = {https://doi.org/10.1186/s42400-022-00134-9},
  issn         = {2523-3246}
}

@Article{Alavizadeh2022DeepQLearning,
AUTHOR = {Alavizadeh, Hooman and Alavizadeh, Hootan and Jang-Jaccard, Julian},
TITLE = {Deep Q-Learning Based Reinforcement Learning Approach for Network Intrusion Detection},
JOURNAL = {Computers},
VOLUME = {11},
YEAR = {2022},
NUMBER = {3},
ARTICLE-NUMBER = {41},
URL = {https://www.mdpi.com/2073-431X/11/3/41},
ISSN = {2073-431X},
DOI = {10.3390/computers11030041}
}

@article{FINISTRELLA2025MultiAgent,
title = {Multi-Agent Reinforcement Learning for Cybersecurity: Classification and survey},
journal = {Intelligent Systems with Applications},
volume = {26},
pages = {200495},
year = {2025},
issn = {2667-3053},
doi = {https://doi.org/10.1016/j.iswa.2025.200495},
url = {https://www.sciencedirect.com/science/article/pii/S2667305325000213},
author = {Salvo Finistrella and Stefano Mariani and Franco Zambonelli}
}

@article{PURVES2024Causally,
title = {Causally aware reinforcement learning agents for autonomous cyber defence},
journal = {Knowledge-Based Systems},
volume = {304},
pages = {112521},
year = {2024},
issn = {0950-7051},
doi = {https://doi.org/10.1016/j.knosys.2024.112521},
url = {https://www.sciencedirect.com/science/article/pii/S0950705124011559},
author = {Tom Purves and Konstantinos G. Kyriakopoulos and Siân Jenkins and Iain Phillips and Tim Dudman}
}

@article{HU2019gametheory,
title = {On convergence rates of game theoretic reinforcement learning algorithms},
journal = {Automatica},
volume = {104},
pages = {90-101},
year = {2019},
issn = {0005-1098},
doi = {https://doi.org/10.1016/j.automatica.2019.02.032},
url = {https://www.sciencedirect.com/science/article/pii/S000510981930086X},
author = {Zhisheng Hu and Minghui Zhu and Ping Chen and Peng Liu}
}

@ARTICLE{Marinho2023Automated,
  author={Marinho, Renato and Holanda, Raimir},
  journal={IEEE Access}, 
  title={Automated Emerging Cyber Threat Identification and Profiling Based on Natural Language Processing}, 
  year={2023},
  volume={11},
  number={},
  pages={58915-58936},
  doi={10.1109/ACCESS.2023.3260020}}

@article{TANG2024Cyber,
title = {Cyber threat indicators extraction based on contextual knowledge prompt},
journal = {Computer Networks},
volume = {254},
pages = {110839},
year = {2024},
issn = {1389-1286},
doi = {https://doi.org/10.1016/j.comnet.2024.110839},
url = {https://www.sciencedirect.com/science/article/pii/S1389128624006716},
author = {Hailiang Tang and Dawei Lin and Wanyu Li and Wenxiao Zhang and Jun Zhao}
}

@article{chen2021cross,
  title={Cross-Lingual Cybersecurity Analytics in the International Dark Web with Adversarial Deep Representation Learning},
  author={Chen, Hsinchun and Samtani, Sagar and Zhu, Yidong and Wang, L-C.},
  journal={MIS Quarterly},
  volume={45},
  number={4},
  pages={1699--1736},
  year={2021},
  publisher={Management Information Systems Research Center, University of Minnesota}
}

@article{khuzaie2021survey,
  title={A Survey on Sentiment and Emotion Analysis for Dark Web and Cyber-Security Applications},
  author={Al-khuzaie, Wafa and Al-harthi, Norah and Al-harthi, Abeer and Al-subaie, Alanoud and Al-humaid, Nora},
  journal={Frontiers in Big Data},
  volume={4},
  pages={601529},
  year={2021},
  publisher={Frontiers},
  doi={10.3389/fdata.2021.601529},
  issn={2624-909X}
}

@article{ALQURASHI2024TopicModelling,
title = {A data-driven multi-perspective approach to cybersecurity knowledge discovery through topic modelling},
journal = {Alexandria Engineering Journal},
volume = {107},
pages = {374-389},
year = {2024},
issn = {1110-0168},
doi = {https://doi.org/10.1016/j.aej.2024.07.044},
url = {https://www.sciencedirect.com/science/article/pii/S1110016824007658},
author = {Fahad Alqurashi and Istiak Ahmad}
}

@ARTICLE{Hasanov2024LLM,
  author={Hasanov, Ismayil and Virtanen, Seppo and Hakkala, Antti and Isoaho, Jouni},
  journal={IEEE Access}, 
  title={Application of Large Language Models in Cybersecurity: A Systematic Literature Review}, 
  year={2024},
  volume={12},
  number={},
  pages={176751-176778},
  doi={10.1109/ACCESS.2024.3505983}}

@ARTICLE{Shahriar2025Zeroshot,
  author={Shahriar, Asif and Hisham, Syed Jarullah and Rahman, K. M. Asifur and Islam, Ruhan and Hossain, Md. Shohrab and Hwang, Ren-Hung and Lin, Ying-Dar},
  journal={IEEE Transactions on Information Forensics and Security}, 
  title={5GPT: 5G Vulnerability Detection by Combining Zero-Shot Capabilities of GPT-4 With Domain Aware Strategies Through Prompt Engineering}, 
  year={2025},
  volume={20},
  number={},
  pages={7045-7060},
  doi={10.1109/TIFS.2025.3586480}}

@article{Meshkin2024ZeroShot,
    author = {Meshkin, Hamed and Zirkle, Joel and Arabidarrehdor, Ghazal and Chaturbedi, Anik and Chakravartula, Shilpa and Mann, John and Thrasher, Bradlee and Li, Zhihua},
    title = {Harnessing large language models’ zero-shot and few-shot learning capabilities for regulatory research},
    journal = {Briefings in Bioinformatics},
    volume = {25},
    number = {5},
    pages = {bbae354},
    year = {2024},
    month = {08},
    issn = {1477-4054},
    doi = {10.1093/bib/bbae354},
    url = {https://doi.org/10.1093/bib/bbae354},
    eprint = {https://academic.oup.com/bib/article-pdf/25/5/bbae354/58902058/bbae354.pdf},
}

@misc{Fahad2025Sentiment,
      title={Sentiment and Emotion Analysis from Dark Web Text for Cyber Threat Intelligence: A Review}, 
      author={Fahad Al-mekhlafi and Bander Ali Saleh Al-rimy and Ishak Bin Aris and N. Z. Jhanjhi and Mamoona Humayun},
      year={2025},
      eprint={2506.16968},
      archivePrefix={arXiv},
      primaryClass={cs.CR},
      url={https://arxiv.org/abs/2506.16968}
}

@article{Zhang2025LLMs,
  author       = {Zhang, Jie and Bu, Haoyu and Wen, Hui and Liu, Yongji and Fei, Haiqiang and Xi, Rongrong and Li, Lun and Yang, Yun and Zhu, Hongsong and Meng, Dan},
  title        = {When LLMs meet cybersecurity: a systematic literature review},
  journal      = {Cybersecurity},
  year         = {2025},
  volume       = {8},
  number       = {1},
  pages        = {55},
  doi          = {10.1186/s42400-025-00361-w},
  url          = {https://doi.org/10.1186/s42400-025-00361-w},
  issn         = {2523-3246}
}

@inproceedings{Perrina2023AGIR,
  author       = {Perrina, Filippo and Marchiori, Francesco and Conti, Mauro and Verde, Nino Vincenzo},
  title        = {AGIR: Automating Cyber Threat Intelligence Reporting with Natural Language Generation},
  booktitle    = {2023 IEEE International Conference on Big Data (BigData)},
  year         = {2023},
  publisher    = {IEEE},
  address      = {Sorrento, Italy},
  pages        = {979--984},
  doi          = {10.1109/BigData59063.2023.10384820},
  isbn         = {979-8-3503-2445-7},
  note         = {\$31.00 ©2023 IEEE},
  organization = {IEEE},
  url          = {https://doi.org/10.1109/BigData59063.2023.10384820}
}

@article{FERRAG2025Generative,
title = {Generative AI in cybersecurity: A comprehensive review of LLM applications and vulnerabilities},
journal = {Internet of Things and Cyber-Physical Systems},
volume = {5},
pages = {1-46},
year = {2025},
issn = {2667-3452},
doi = {https://doi.org/10.1016/j.iotcps.2025.01.001},
url = {https://www.sciencedirect.com/science/article/pii/S2667345225000082},
author = {Mohamed Amine Ferrag and Fatima Alwahedi and Ammar Battah and Bilel Cherif and Abdechakour Mechri and Norbert Tihanyi and Tamas Bisztray and Merouane Debbah}
}

@INPROCEEDINGS {Fieblinger2024Actionable,
author = { Fieblinger, Romy and Alam, Md Tanvirul and Rastogi, Nidhi },
booktitle = { 2024 IEEE European Symposium on Security and Privacy Workshops (EuroS\&PW) },
title = {{ Actionable Cyber Threat Intelligence Using Knowledge Graphs and Large Language Models }},
year = {2024},
volume = {},
ISSN = {},
pages = {100-111},
doi = {10.1109/EuroSPW61312.2024.00018},
url = {https://doi.ieeecomputersociety.org/10.1109/EuroSPW61312.2024.00018},
publisher = {IEEE Computer Society},
address = {Los Alamitos, CA, USA},
month =Jul}

@article{Dantas2025review,
  author       = {Dantas, Pierre V. and Cordeiro, Lucas C. and Junior, Waldir S. S.},
  title        = {A review of state-of-the-art techniques for large language model compression},
  journal      = {Complex \& Intelligent Systems},
  year         = {2025},
  volume       = {11},
  number       = {9},
  pages        = {407},
  doi          = {10.1007/s40747-025-02019-z},
  url          = {https://doi.org/10.1007/s40747-025-02019-z},
  issn         = {2198-6053}
}

@INPROCEEDINGS{Shankar2020Benchmarking,
  author={Shankar, Karthick and Wang, Pengcheng and Xu, Ran and Mahgoub, Ashraf and Chaterji, Somali},
  booktitle={2020 IEEE 13th International Conference on Cloud Computing (CLOUD)}, 
  title={JANUS: Benchmarking Commercial and Open-Source Cloud and Edge Platforms for Object and Anomaly Detection Workloads}, 
  year={2020},
  volume={},
  number={},
  pages={590-599},
  doi={10.1109/CLOUD49709.2020.00088}}

@article{PICCIALLI2025Federated,
title = {Federated and edge learning for large language models},
journal = {Information Fusion},
volume = {117},
pages = {102840},
year = {2025},
issn = {1566-2535},
doi = {https://doi.org/10.1016/j.inffus.2024.102840},
url = {https://www.sciencedirect.com/science/article/pii/S1566253524006183},
author = {Francesco Piccialli and Diletta Chiaro and Pian Qi and Valerio Bellandi and Ernesto Damiani}
}

@Article{Alshehri2024Deep,
AUTHOR = {Alshehri, Ali and Badr, Mahmoud M. and Baza, Mohamed and Alshahrani, Hani},
TITLE = {Deep Anomaly Detection Framework Utilizing Federated Learning for Electricity Theft Zero-Day Cyberattacks},
JOURNAL = {Sensors},
VOLUME = {24},
YEAR = {2024},
NUMBER = {10},
ARTICLE-NUMBER = {3236},
URL = {https://www.mdpi.com/1424-8220/24/10/3236},
PubMedID = {38794091},
ISSN = {1424-8220},
DOI = {10.3390/s24103236}
}

@article{Ghadi2025hybrid,
  author    = {Yazeed Yasin Ghadi and Tehseen Mazhar and Tariq Shahzad and Ines Hilali Jaghdam and Sanwar Khan and Muhammad Amir Khan and Habib Hamam},
  title     = {A hybrid AI-Blockchain security framework for smart grids},
  journal   = {Scientific Reports},
  year      = {2025},
  volume    = {15},
  number    = {1},
  pages     = {20882},
  doi       = {10.1038/s41598-025-05257-w},
  url       = {https://doi.org/10.1038/s41598-025-05257-w},
  issn      = {2045-2322}
}

@Article{Amato2024Detecting,
AUTHOR = {Amato, Flora and Cirillo, Egidia and Fonisto, Mattia and Moccardi, Alberto},
TITLE = {Detecting Adversarial Attacks in IoT-Enabled Predictive Maintenance with Time-Series Data Augmentation},
JOURNAL = {Information},
VOLUME = {15},
YEAR = {2024},
NUMBER = {11},
ARTICLE-NUMBER = {740},
URL = {https://www.mdpi.com/2078-2489/15/11/740},
ISSN = {2078-2489},
DOI = {10.3390/info15110740}
}

@article{Gauthama2021Machine,
  author    = {Gauthama Raman M. R. and Chuadhry Mujeeb Ahmed and Aditya Mathur},
  title     = {Machine learning for intrusion detection in industrial control systems: challenges and lessons from experimental evaluation},
  journal   = {Cybersecurity},
  year      = {2021},
  volume    = {4},
  number    = {1},
  pages     = {27},
  doi       = {10.1186/s42400-021-00095-5},
  url       = {https://doi.org/10.1186/s42400-021-00095-5},
  issn      = {2523-3246}
}

@misc{Li2024PromptingLLMs,
  title        = {Prompting Large Language Models for Counterfactual Generation: An Empirical Study},
  author       = {Yongqi Li and Mayi Xu and Xin Miao and Shen Zhou and Tieyun Qian},
  year         = {2024},
  howpublished = {arXiv preprint arXiv:2305.14791v2},
  note         = {Version 2, posted 23 February 2024},
  url          = {https://arxiv.org/abs/2305.14791v2}
}

@article{KUMAR2025DynamicTrust,
title = {AI-driven dynamic trust management and blockchain-based security in industrial IoT},
journal = {Computers and Electrical Engineering},
volume = {123},
pages = {110213},
year = {2025},
issn = {0045-7906},
doi = {https://doi.org/10.1016/j.compeleceng.2025.110213},
url = {https://www.sciencedirect.com/science/article/pii/S0045790625001569},
author = {Rajesh Kumar and Rewa Sharma}
}

@article{Alevizos2025Automated,
  author    = {Lampis Alevizos},
  title     = {Automated cybersecurity compliance and threat response using AI, blockchain and smart contracts},
  journal   = {International Journal of Information Technology},
  year      = {2025},
  volume    = {17},
  number    = {2},
  pages     = {767--781},
  doi       = {10.1007/s41870-024-02324-9},
  url       = {https://doi.org/10.1007/s41870-024-02324-9},
  issn      = {2511-2112}
}

@Article{Elkhodr2025Framework,
AUTHOR = {Elkhodr, Mahmoud},
TITLE = {An AI-Driven Framework for Integrated Security and Privacy in Internet of Things Using Quantum-Resistant Blockchain},
JOURNAL = {Future Internet},
VOLUME = {17},
YEAR = {2025},
NUMBER = {6},
ARTICLE-NUMBER = {246},
URL = {https://www.mdpi.com/1999-5903/17/6/246},
ISSN = {1999-5903},
DOI = {10.3390/fi17060246}
}

@misc{Hossain2024Enhancing,
  title        = {Enhancing Data Integrity and Traceability in Industry Cyber Physical Systems (ICPS) through Blockchain Technology: A Comprehensive Approach},
  author       = {Mohammad Ikbal Hossain and Tanja Steigner and Muhammad Imam Hussain and Afroja Akther},
  year         = {2024},
  howpublished = {arXiv preprint arXiv:2405.04837},
  note         = {Cross-listed from cs.CR and eess.SY},
  url          = {https://arxiv.org/abs/2405.04837}
}

@misc{Scaramuzza2025Trustworthy,
  title        = {Engineering Trustworthy Machine-Learning Operations with Zero-Knowledge Proofs},
  author       = {Filippo Scaramuzza and Giovanni Quattrocchi and Damian A. Tamburri},
  year         = {2025},
  howpublished = {arXiv preprint arXiv:2505.20136},
  url          = {https://arxiv.org/abs/2505.20136},
}

@Article{yao2023blockchain,
AUTHOR = {Yao, Yue and Zhang, Xiaomei and Hu, Haomin and Liu, Huibin and Huang, Rong and Wang, Zejie},
TITLE = {Blockchain-Based Multistage Continuous Authentication for Smart Devices},
JOURNAL = {Applied Sciences},
VOLUME = {13},
YEAR = {2023},
NUMBER = {23},
ARTICLE-NUMBER = {12641},
URL = {https://www.mdpi.com/2076-3417/13/23/12641},
ISSN = {2076-3417},
DOI = {10.3390/app132312641}
}

@Article{Kulothungan2025Blockchain,
AUTHOR = {Kulothungan, Vikram},
TITLE = {Using Blockchain Ledgers to Record AI Decisions in IoT},
JOURNAL = {IoT},
VOLUME = {6},
YEAR = {2025},
NUMBER = {3},
ARTICLE-NUMBER = {37},
URL = {https://www.mdpi.com/2624-831X/6/3/37},
ISSN = {2624-831X},
DOI = {10.3390/iot6030037}
}

@misc{Avarikioti2025ComposableGame,
  title        = {A Composable Game-Theoretic Framework for Blockchains},
  author       = {Zeta Avarikioti and Georg Fuchsbauer and Pim Keer and Matteo Maffei and Fabian Regen},
  year         = {2025},
  howpublished = {arXiv preprint arXiv:2504.18214},
  note         = {Submitted 25 April 2025},
  url          = {https://arxiv.org/abs/2504.18214}
}

@article{ROSABILBAO2023CEPEDALoCo,
title = {CEPEDALoCo: An event-driven architecture for integrating complex event processing and blockchain through low-code},
journal = {Internet of Things},
volume = {22},
pages = {100802},
year = {2023},
issn = {2542-6605},
doi = {https://doi.org/10.1016/j.iot.2023.100802},
url = {https://www.sciencedirect.com/science/article/pii/S2542660523001257},
author = {Jesús Rosa-Bilbao and Juan Boubeta-Puig and Adrian Rutle}
}

@Article{Akhdar2024Exploring,
AUTHOR = {El Akhdar, Abir and Baidada, Chafik and Kartit, Ali and Hanine, Mohamed and García, Carlos Osorio and Lara, Roberto Garcia and Ashraf, Imran},
TITLE = {Exploring the Potential of Microservices in Internet of Things: A Systematic Review of Security and Prospects},
JOURNAL = {Sensors},
VOLUME = {24},
YEAR = {2024},
NUMBER = {20},
ARTICLE-NUMBER = {6771},
URL = {https://www.mdpi.com/1424-8220/24/20/6771},
PubMedID = {39460251},
ISSN = {1424-8220},
DOI = {10.3390/s24206771}
}

@misc{Fu2024AIforDevSecOps,
  title        = {AI for DevSecOps: A Landscape and Future Opportunities},
  author       = {Michael Fu and Jirat Pasuksmit and Chakkrit Tantithamthavorn},
  year         = {2024},
  howpublished = {arXiv preprint arXiv:2404.04839v2},
  note         = {Version 2, last revised September 13, 2024},
  url          = {https://arxiv.org/abs/2404.04839}
}

@article{LIU2025hash,
title = {A hash-based post-quantum ring signature scheme for the Internet of Vehicles},
journal = {Journal of Systems Architecture},
volume = {160},
pages = {103345},
year = {2025},
issn = {1383-7621},
doi = {https://doi.org/10.1016/j.sysarc.2025.103345},
url = {https://www.sciencedirect.com/science/article/pii/S1383762125000177},
author = {Shuanggen Liu and Xiayi Zhou and Xu An Wang and Zixuan Yan and He Yan and Yurui Cao}
}

@ARTICLE{Fernández2020Towards,
  author={Fernández-Caramès, Tiago M. and Fraga-Lamas, Paula},
  journal={IEEE Access}, 
  title={Towards Post-Quantum Blockchain: A Review on Blockchain Cryptography Resistant to Quantum Computing Attacks}, 
  year={2020},
  volume={8},
  number={},
  pages={21091-21116},
  doi={10.1109/ACCESS.2020.2968985}}

@article{CORLI2025Quantum,
title = {Quantum machine learning algorithms for anomaly detection: A review},
journal = {Future Generation Computer Systems},
volume = {166},
pages = {107632},
year = {2025},
issn = {0167-739X},
doi = {https://doi.org/10.1016/j.future.2024.107632},
url = {https://www.sciencedirect.com/science/article/pii/S0167739X2400596X},
author = {Sebastiano Corli and Lorenzo Moro and Daniele Dragoni and Massimiliano Dispenza and Enrico Prati}
}

@article{AHMAD2021blockchain,
title = {The role of blockchain technology in telehealth and telemedicine},
journal = {International Journal of Medical Informatics},
volume = {148},
pages = {104399},
year = {2021},
issn = {1386-5056},
doi = {https://doi.org/10.1016/j.ijmedinf.2021.104399},
url = {https://www.sciencedirect.com/science/article/pii/S1386505621000253},
author = {Raja Wasim Ahmad and Khaled Salah and Raja Jayaraman and Ibrar Yaqoob and Samer Ellahham and Mohammed Omar}
}

@article{AKRAM2024Blockchain,
title = {Blockchain technology: A potential tool for the management of pharma supply chain},
journal = {Research in Social and Administrative Pharmacy},
volume = {20},
number = {6},
pages = {156-164},
year = {2024},
issn = {1551-7411},
doi = {https://doi.org/10.1016/j.sapharm.2024.02.014},
url = {https://www.sciencedirect.com/science/article/pii/S1551741124000792},
author = {Wasim Akram and Ramakant Joshi and Tanweer Haider and Pankaj Sharma and Vinay Jain and Navneet Garud and Nitin Singh}
}

@article{Pham2024Distributed,
  author    = {Anh Pham and Maxim Edelson and Armin Nouri and Tsung{-}Ting Kuo},
  title     = {Distributed management of patient data-sharing informed consents for clinical research},
  journal   = {Computers in Biology and Medicine},
  year      = {2024},
  volume    = {180},
  pages     = {108956},
  doi       = {10.1016/j.compbiomed.2024.108956},
  url       = {https://doi.org/10.1016/j.compbiomed.2024.108956},
  issn      = {1879-0534},
  pmid      = {39121682},
  pmcid     = {PMC11380755}
}

@INPROCEEDINGS{H1,
  author={Ahmad, Farhan and Ahmad, Zeeshan and Kerrache, Chaker Abdelaziz and Kurugollu, Fatih and Adnane, Asma and Barka, Ezedin},
  booktitle={2019 International Conference on Computer and Information Sciences (ICCIS)}, 
  title={Blockchain in Internet-of-Things: Architecture, Applications and Research Directions}, 
  year={2019},
  volume={},
  number={},
  pages={1-6}
}

@ARTICLE{H2,
  author={Khayat, Mohamad and Barka, Ezedin and Adel Serhani, Mohamed and Sallabi, Farag and Shuaib, Khaled and Khater, Heba M.},
  journal={IEEE Access}, 
  title={Empowering Security Operation Center With Artificial Intelligence and Machine Learning—A Systematic Literature Review}, 
  year={2025},
  volume={13},
  number={},
  pages={19162-19197}}

@ARTICLE{H3,
  author={Khayat, Mohamad and Barka, Ezedin and Adel Serhani, Mohamed and Sallabi, Farag and Shuaib, Khaled and Khater, Heba M.},
  journal={IEEE Access}, 
  title={Blockchain-Powered Secure and Scalable Threat Intelligence System With Graph Convolutional Autoencoder and Reinforcement Learning Feedback Loop}, 
  year={2025},
  volume={13},
  number={},
  pages={24736-24748}}

@article{H4,
  title={Systematic survey of advanced metering infrastructure security: Vulnerabilities, attacks, countermeasures, and future vision},
  author={Shokry, Mostafa and Awad, Ali Ismail and Abd-Ellah, Mahmoud Khaled and Khalaf, Ashraf AM},
  journal={Future Generation Computer Systems},
  volume={136},
  pages={358--377},
  year={2022},
  publisher={Elsevier}
}

@ARTICLE{H5,
  author={Raza, Abir and Badidi, Elarbi and Hayajneh, Mohammad and Barka, Ezedin and Harrouss, Omar El},
  journal={IEEE Access}, 
  title={Blockchain-Based Reputation and Trust Management for Smart Grids, Healthcare, and Transportation: A Review}, 
  year={2024},
  volume={12},
  number={},
  pages={196887-196913}}

@ARTICLE{H6,
  author={Kihero, Abuu B. and Furqan, Haji M. and Sahin, M. M. and Arslan, Hüseyin},
  journal={IEEE Wireless Communications}, 
  title={6G and Beyond Wireless Channel Characteristics for Physical Layer Security: Opportunities and Challenges}, 
  year={2024},
  volume={31},
  number={3},
  pages={295-301}}

@ARTICLE{H7,
  author={Yu, Zhenhua and Gao, Hongxia and Cong, Xuya and Wu, Naiqi and Song, Houbing Herbert},
  journal={IEEE Internet of Things Journal}, 
  title={A Survey on Cyber–Physical Systems Security}, 
  year={2023},
  volume={10},
  number={24},
  pages={21670-21686}}

@article{H8,
  title={A critical review of cyber-physical security for building automation systems},
  author={Li, Guowen and Ren, Lingyu and Fu, Yangyang and Yang, Zhiyao and Adetola, Veronica and Wen, Jin and Zhu, Qi and Wu, Teresa and Candan, K Selcuk and O'Neill, Zheng},
  journal={Annual Reviews in Control},
  volume={55},
  pages={237--254},
  year={2023},
  publisher={Elsevier}
}

@article{H9,
  title={Cyber and physical security vulnerability assessment for IoT-based smart homes},
  author={Ali, Bako and Awad, Ali Ismail},
  journal={sensors},
  volume={18},
  number={3},
  pages={817},
  year={2018},
  publisher={MDPI}
}

@article{H10,
  title={Encoder decoder-based Virtual Physically Unclonable Function for Internet of Things device authentication using split-learning},
  author={Khan, Raviha and Eldeeb, Hossien B and Mefgouda, Brahim and Alhussein, Omar and Saleh, Hani and Muhaidat, Sami},
  journal={Computers \& Security},
  volume={148},
  pages={104164},
  year={2025},
  publisher={Elsevier}
}

@INPROCEEDINGS{H11,
  author={Alla, Ildi and Yahia, Selma and Loscri, Valeria and Eldeeb, Hossien},
  booktitle={2024 Annual Computer Security Applications Conference (ACSAC)}, 
  title={Robust Device Authentication in Multi-Node Networks: ML-Assisted Hybrid PLA Exploiting Hardware Impairments}, 
  year={2024},
  volume={},
  number={},
  pages={1172-1185}}

@INPROCEEDINGS{H12,
  author={Eldeeb, Hossien B. and Pandey, Anshul and Andreoni, Martin and Muhaidat, Sami},
  booktitle={2023 IEEE International Mediterranean Conference on Communications and Networking (MeditCom)}, 
  title={Experimental Evaluation of A Lightweight RSS-Based PLA Scheme in Multi-Node Multi-Cell Mesh Networks}, 
  year={2023},
  volume={},
  number={},
  pages={393-398}}

@INPROCEEDINGS{H13,
  author={Eldeeb, Hossien B. and Pandey, Anshul and Andreoni, Martin and Muhaidat, Sami},
  booktitle={2023 IEEE 98th Vehicular Technology Conference (VTC2023-Fall)}, 
  title={Exploiting Engineered IQ Samples for Physical Layer Authentication}, 
  year={2023},
  volume={},
  number={},
  pages={1-6}}

@INPROCEEDINGS{H14,
  author={Mefgouda, Brahim and Khan, Raviha and Alhussein, Omar and Saleh, Hani and Eldeeb, Hossien B. and Pandey, Anshul and Muhaidat, Sami},
  booktitle={2024 IEEE Globecom Workshops (GC Wkshps)}, 
  title={L{PUF}-AuthNet: A Lightweight {PUF}-Based {I}o{T} Authentication via Tandem Neural Networks and Split Learning}, 
  year={2024},
  volume={},
  number={},
  pages={1-7},
  doi={10.1109/GCWkshp64532.2024.11101578}}

@article{H15,
  title={Artificial intelligence based anomaly detection of energy consumption in buildings: A review, current trends and new perspectives},
  author={Himeur, Yassine and Ghanem, Khalida and Alsalemi, Abdullah and Bensaali, Faycal and Amira, Abbes},
  journal={Applied Energy},
  volume={287},
  pages={116601},
  year={2021},
  publisher={Elsevier}
}

@article{H16,
  title={Artificial intelligence and physics-based anomaly detection in the smart grid: A survey},
  author={Gaggero, Giovanni Battista and Girdinio, Paola and Marchese, Mario},
  journal={IEEE Access},
  year={2025},
  publisher={IEEE}
}

@article{H17,
  title={Managing cybersecurity risks of cyber-physical systems: The MARISMA-CPS pattern},
  author={Rosado, David G and Santos-Olmo, Antonio and S{\'a}nchez, Luis Enrique and Serrano, Manuel A and Blanco, Carlos and Mouratidis, Haralambos and Fern{\'a}ndez-Medina, Eduardo},
  journal={Computers in Industry},
  volume={142},
  pages={103715},
  year={2022},
  publisher={Elsevier}
}

@article{H18,
  title={Cybersecurity of industrial cyber-physical systems: A review},
  author={Kayan, Hakan and Nunes, Matthew and Rana, Omer and Burnap, Pete and Perera, Charith},
  journal={ACM Computing Surveys (CSUR)},
  volume={54},
  number={11s},
  pages={1--35},
  year={2022},
  publisher={ACM New York, NY}
}

@INPROCEEDINGS{H19,
  author={Dawoud, Diana W. and Mukhtar, Husameldin and Ismail, Shereen and Pandey, Anshul and Eldeeb, Hossien B.},
  booktitle={2024 IEEE International Mediterranean Conference on Communications and Networking (MeditCom)}, 
  title={Enhancing V{LC} Security with Time-Domain Interleaving and Optical-{OFDM} on Realistic Channel Models}, 
  year={2024},
  volume={},
  number={},
  pages={57-61},
  doi={10.1109/MeditCom61057.2024.10621118}}

@ARTICLE{H20,
  author={Eldeeb, Hossien B. and Naser, Shimaa and Bariah, Lina and Muhaidat, Sami and Uysal, Murat},
  journal={IEEE Network}, 
  title={Digital Twin-Assisted {OWC}: Toward Smart and Autonomous 6{G} Networks}, 
  year={2024},
  volume={38},
  number={6},
  pages={153-162},
  doi={10.1109/MNET.2024.3374370}}

@article{R1,
  title={Blockchain-based internet of things and industrial {I}o{T}: a comprehensive survey},
  author={Dwivedi, Sanjeev Kumar and Roy, Priyadarshini and Karda, Chinky and Agrawal, Shalini and Amin, Ruhul},
  journal={Security and Communication Networks},
  volume={2021},
  number={1},
  pages={7142048},
  year={2021},
  publisher={Wiley Online Library}
}

@article{R2,
  title={ICS-BlockOpS: Blockchain for operational data security in industrial control system},
  author={Maw, Aung and Adepu, Sridhar and Mathur, Aditya},
  journal={Pervasive and Mobile Computing},
  volume={59},
  pages={101048},
  year={2019},
  publisher={Elsevier}
}

@article{R3,
  title={Providing tamper-resistant audit trails with distributed ledger based solutions for forensics of IoT systems using cloud resources},
  author={Westerlund, Magnus and Neovius, Mats and Pulkkis, G{\"o}ran},
  journal={International Journal on Advances in Security},
  volume={11},
  number={3},
  year={2018}
}

@article{R4,
  title={Autonomous and malware-proof blockchain-based firmware update platform with efficient batch verification for Internet of Things devices},
  author={Hu, Jen-Wei and Yeh, Lo-Yao and Liao, Shih-Wei and Yang, Chu-Sing},
  journal={Computers \& Security},
  volume={86},
  pages={238--252},
  year={2019},
  publisher={Elsevier}
}

@article{R5,
  title={Physics-informed gated recurrent graph attention unit network for anomaly detection in industrial cyber-physical systems},
  author={Wu, Weiqiang and Song, Chunyue and Zhao, Jun and Xu, Zuhua},
  journal={Information Sciences},
  volume={629},
  pages={618--633},
  year={2023},
  publisher={Elsevier}
}

@inproceedings{R6,
  title={Research on Multi-sensor Data Attack Detection Method for Industrial Control System},
  author={Cui, Dong and Wang, Zepu and Xiao, Han and Shi, Yajing},
  booktitle={INTERNATIONAL CONFERENCE ON WIRELESS COMMUNICATIONS, NETWORKING AND APPLICATIONS},
  pages={637--647},
  year={2022},
  organization={Springer}
}

@inproceedings{R7,
  title={Detecting cyber attacks in industrial control systems using convolutional neural networks},
  author={Kravchik, Moshe and Shabtai, Asaf},
  booktitle={Proceedings of the 2018 workshop on cyber-physical systems security and privacy},
  pages={72--83},
  year={2018}
}

@article{R8,
  title={Machine learning-based intrusion detection for smart grid computing: A survey},
  author={Sahani, Nitasha and Zhu, Ruoxi and Cho, Jin-Hee and Liu, Chen-Ching},
  journal={ACM Transactions on Cyber-Physical Systems},
  volume={7},
  number={2},
  pages={1--31},
  year={2023},
  publisher={ACM New York, NY}
}

@article{R9,
  title={Designing microgrid energy markets: A case study: The Brooklyn Microgrid},
  author={Mengelkamp, Esther and G{\"a}rttner, Johannes and Rock, Kerstin and Kessler, Scott and Orsini, Lawrence and Weinhardt, Christof},
  journal={Applied energy},
  volume={210},
  pages={870--880},
  year={2018},
  publisher={Elsevier}
}

@article{R10,
  title={GUARDIAN: Blockchain-based secure demand response management in smart grid system},
  author={Jindal, Anish and Aujla, Gagangeet Singh and Kumar, Neeraj and Villari, Massimo},
  journal={IEEE transactions on services computing},
  volume={13},
  number={4},
  pages={613--624},
  year={2019},
  publisher={IEEE}
}

@article{R11,
  title={Enabling localized peer-to-peer electricity trading among plug-in hybrid electric vehicles using consortium blockchains},
  author={Kang, Jiawen and Yu, Rong and Huang, Xumin and Maharjan, Sabita and Zhang, Yan and Hossain, Ekram},
  journal={IEEE transactions on industrial informatics},
  volume={13},
  number={6},
  pages={3154--3164},
  year={2017},
  publisher={IEEE}
}

@article{R12,
  title={Privacy-preserving and efficient aggregation based on blockchain for power grid communications in smart communities},
  author={Guan, Zhitao and Si, Guanlin and Zhang, Xiaosong and Wu, Longfei and Guizani, Nadra and Du, Xiaojiang and Ma, Yinglong},
  journal={IEEE Communications Magazine},
  volume={56},
  number={7},
  pages={82--88},
  year={2018},
  publisher={IEEE}
}

@article{R13,
  title={Healthcare blockchain system using smart contracts for secure automated remote patient monitoring},
  author={Griggs, Kristen N and Ossipova, Olya and Kohlios, Christopher P and Baccarini, Alessandro N and Howson, Emily A and Hayajneh, Thaier},
  journal={Journal of medical systems},
  volume={42},
  number={7},
  pages={130},
  year={2018},
  publisher={Springer}
}

@article{R14,
  title={An exhaustive survey on security and privacy issues in Healthcare 4.0},
  author={Hathaliya, Jigna J and Tanwar, Sudeep},
  journal={Computer Communications},
  volume={153},
  pages={311--335},
  year={2020},
  publisher={Elsevier}
}

@inproceedings{R15,
  title={Sok: Security and privacy in implantable medical devices and body area networks},
  author={Rushanan, Michael and Rubin, Aviel D and Kune, Denis Foo and Swanson, Colleen M},
  booktitle={2014 IEEE symposium on security and privacy},
  pages={524--539},
  year={2014},
  organization={IEEE}
}

@inproceedings{R16,
  title={Medrec: Using blockchain for medical data access and permission management},
  author={Azaria, Asaph and Ekblaw, Ariel and Vieira, Thiago and Lippman, Andrew},
  booktitle={2016 2nd international conference on open and big data (OBD)},
  pages={25--30},
  year={2016},
  organization={IEEE}
}

@ARTICLE{R17,
  author={Kim, Yeongwoo and Dán, György and Zhu, Quanyan},
  journal={IEEE Transactions on Information Forensics and Security}, 
  title={Human-in-the-Loop Cyber Intrusion Detection Using Active Learning}, 
  year={2024},
  volume={19},
  number={},
  pages={8658-8672},
  doi={10.1109/TIFS.2024.3434647}}

@ARTICLE{R18,
  author={Kumar, Vikash and Sinha, Ditipriya},
  title={A robust intelligent zero-day cyber-attack detection technique},
  journal={Complex \& Intelligent Systems},
  year={2021},
  volume={7},
  pages={2211--2234},
  publisher={Springer}
}

@article{R19,
author = {Yang, Hao and Yu, Junyang and Zhai, Rui},
title = {High-precision intrusion detection for cybersecurity communications based on multi-scale convolutional neural networks: High-precision intrusion detection...},
year = {2024},
issue_date = {Jan 2025},
publisher = {Kluwer Academic Publishers},
address = {USA},
volume = {81},
number = {1},
issn = {0920-8542},
doi = {10.1007/s11227-024-06737-y},
journal = {J. Supercomput.},
month = dec,
numpages = {34}
}

@ARTICLE{R20,
  author={Scarfone, K. and Souppaya, M. and Cody, A. and Orebaugh, A.},
  title={Technical guide to information security testing and assessment},
  journal={NIST Special Publication},
  year={2008},
  volume={800},
  pages={115}
}

@article{R21,
  author    = {Kamatchi, K. and Uma, E.},
  title     = {Insights into user behavioral-based insider threat detection: systematic review},
  journal   = {International Journal of Information Security},
  volume    = {24},
  year      = {2025},
  article-number = {88},
  publisher = {Springer}
}

@article{R22,
  author    = {Ahmed Shebl and E. I. Elsedimy and A. Ismail and A. A. Salama and Mostafa Herajy},
  title     = {DCNN: a novel binary and multi-class network intrusion detection model via deep convolutional neural network},
  journal   = {EURASIP Journal on Information Security},
  year      = {2024},
  volume    = {2024},
  number    = {1},
  pages     = {36},
  doi       = {10.1186/s13635-024-00184-1},
  issn      = {2510-523X}
}

@ARTICLE{R23,
  author={Iyer, K. I.},
  title={Proactive Threat Hunting: Leveraging AI for Early Detection of Advanced Persistent Threats},
  journal={European Journal of Advances in Engineering and Technology},
  year={2024},
  publisher={ResearchGate}
}

@article{R24,
  title={AI-powered data-driven cybersecurity techniques: Boosting threat identification and reaction},
  author={Prince, Nayem Uddin and Faheem, Muhammad Ashraf and Khan, Obyed Ullah and Hossain, Kaosar and Alkhayyat, Ahmad and Hamdache, Amine and Elmouki, Ilias},
  journal={Nanotechnology Perceptions},
  volume={20},
  number={S10},
  year={2024}
}

@ARTICLE{R25,
  author={Charmanas, K. and Mittas, N. and Angelis, L.},
  title={Exploitation of vulnerabilities: a topic-based machine learning framework for explaining and predicting exploitation},
  journal={Information},
  year={2023},
  volume={14},
  number={7},
  pages={403}
}

@ARTICLE{R26,
  author={Guembe, B. and Azeta, A. and Misra, S. and Osamor, V. C.},
  title={The emerging threat of ai-driven cyber attacks: A review},
  journal={Journal of Information and Organizational Management in Artificial Intelligence},
  year={2022},
  volume={2022},
  pages={2037254}
}

@ARTICLE{R27,
  author={Gai, K. and She, Y. and Zhu, L. and Choo, K. K. R. and Wan, Z.},
  title={A blockchain-based access control scheme for zero trust cross-organizational data sharing},
  journal={ACM Transactions on Internet Technology},
  year={2023},
  volume={23},
  number={3},
  pages={1--21}
}

@ARTICLE{R28,
  author={Ali, H. and Ahmad, J. and Jaroucheh, Z. and Papadopoulos, P.},
  title={Trusted threat intelligence sharing in practice and performance benchmarking through the hyperledger fabric platform},
  journal={Entropy},
  year={2022},
  volume={24},
  number={10},
  pages={1379}
}

@inproceedings{R29,
  author={Petrov, D. and Znati, T.},
  title={Context-aware deep learning-driven framework for mitigation of security risks in BYOD-enabled environments},
  booktitle={2018 IEEE 4th International Conference on Collaboration and Internet Computing (CIC)},
  year={2018},
  pages={1-9}
}

@article{Kaur2025Cybersecurity,
author = {Kaur, Amanjot and Valizadeh, Nima and Nandan, Devki and Szydlo, Tomasz and R. K. Rajasekaran, James and Kumar, Vijay and Barika, Mutaz and Liang, Jun and Ranjan, Rajiv and Omer, Rana},
title = {Cybersecurity Challenges in the EV Charging Ecosystem},
year = {2025},
publisher = {Association for Computing Machinery},
address = {New York, NY, USA},
issn = {0360-0300},
url = {https://doi.org/10.1145/3735662},
doi = {10.1145/3735662},
journal = {ACM Comput. Surv.},
month = may
}

@article{RAZZAQUE2025Cybersecurity,
title = {Cybersecurity in vehicle-to-grid (V2G) systems: A systematic review},
journal = {Applied Energy},
volume = {398},
pages = {126364},
year = {2025},
issn = {0306-2619},
doi = {https://doi.org/10.1016/j.apenergy.2025.126364},
author = {Mohammad A. Razzaque and Shafiuzzaman K. Khadem and Sandipan Patra and Glory Okwata and Md. Noor-A-Rahim}
}

@article{Sarieddine2023Investigating,
author = {Sarieddine, Khaled and Sayed, Mohammad Ali and Torabi, Sadegh and Atallah, Ribal and Assi, Chadi},
title = {Investigating the Security of EV Charging Mobile Applications as an Attack Surface},
year = {2023},
issue_date = {October 2023},
publisher = {Association for Computing Machinery},
address = {New York, NY, USA},
volume = {7},
number = {4},
issn = {2378-962X},
url = {https://doi.org/10.1145/3609508},
doi = {10.1145/3609508},
journal = {ACM Trans. Cyber-Phys. Syst.},
month = oct,
articleno = {26},
numpages = {28}
}

@article{Nai2025Authentication,
title = {Authentication protocol for vehicular networks using Zero-Knowledge Proofs and Elliptic Curve Cryptography},
journal = {ICT Express},
volume = {11},
number = {4},
pages = {636-642},
year = {2025},
issn = {2405-9595},
doi = {https://doi.org/10.1016/j.icte.2025.04.014},
url = {https://www.sciencedirect.com/science/article/pii/S2405959525000591},
author = {Nai-Wei Lo and Chi-Ying Chuang and Jheng-Jia Huang and Yu-Xuan Luo}
}

@article{Li2024Over,
  author    = {Beibei Li and Wei Hu and Lemei Da and Yibing Wu and Xinxin Wang and Yiwei Li and Chaoxuan Yuan},
  title     = {Over-the-air upgrading for enhancing security of intelligent connected vehicles: a survey},
  journal   = {Artificial Intelligence Review},
  volume    = {57},
  number    = {11},
  pages     = {314},
  year      = {2024},
  month     = oct,
  doi       = {10.1007/s10462-024-10968-z},
  url       = {https://doi.org/10.1007/s10462-024-10968-z},
  issn      = {1573-7462}
}

@article{SHEN2025Blockchain,
title = {Blockchain-Based peer-to-peer energy trading: A decentralized and innovative approach for sustainable local markets},
journal = {Computers and Electrical Engineering},
volume = {123},
pages = {110281},
year = {2025},
issn = {0045-7906},
doi = {https://doi.org/10.1016/j.compeleceng.2025.110281},
author = {Tao Shen and Xiufang Ou and Bingbin Chen}
}

@article{KOLAHAN2021Blockchain,
title = {Blockchain-based solution for energy demand-side management of residential buildings},
journal = {Sustainable Cities and Society},
volume = {75},
pages = {103316},
year = {2021},
issn = {2210-6707},
doi = {https://doi.org/10.1016/j.scs.2021.103316},
author = {Arman Kolahan and Seyed Reza Maadi and Zahra Teymouri and Corrado Schenone}
}

@ARTICLE{Cantillo2022Blockchain,
  author={Cantillo-Luna, Sergio and Moreno-Chuquen, Ricardo and Chamorro, Harold R. and Sood, Vijay K. and Badsha, Shahriar and Konstantinou, Charalambos},
  journal={IEEE Access}, 
  title={Blockchain for Distributed Energy Resources Management and Integration}, 
  year={2022},
  volume={10},
  number={},
  pages={68598-68617},
  doi={10.1109/ACCESS.2022.3184704}}

@article{MAKHDOOM2020PrivySharing,
title = {PrivySharing: A blockchain-based framework for privacy-preserving and secure data sharing in smart cities},
journal = {Computers \& Security},
volume = {88},
pages = {101653},
year = {2020},
issn = {0167-4048},
doi = {https://doi.org/10.1016/j.cose.2019.101653},
url = {https://www.sciencedirect.com/science/article/pii/S016740481930197X},
author = {Imran Makhdoom and Ian Zhou and Mehran Abolhasan and Justin Lipman and Wei Ni}
}

@Article{Alabdulatif2025Blockchain,
AUTHOR = {Alabdulatif, Abdullah},
TITLE = {Blockchain-Based Privacy-Preserving Authentication and Access Control Model for E-Health Users},
JOURNAL = {Information},
VOLUME = {16},
YEAR = {2025},
NUMBER = {3},
ARTICLE-NUMBER = {219},
URL = {https://www.mdpi.com/2078-2489/16/3/219},
ISSN = {2078-2489},
DOI = {10.3390/info16030219}
}

@ARTICLE{Rahman2021Blockchain,
  author={Rahman, Ziaur and Khalil, Ibrahim and Yi, Xun and Atiquzzaman, Mohammed},
  journal={IEEE Communications Magazine}, 
  title={Blockchain-Based Security Framework for a Critical Industry 4.0 Cyber-Physical System}, 
  year={2021},
  volume={59},
  number={5},
  pages={128-134},
  doi={10.1109/MCOM.001.2000679}}

@article{RATHEE2021design,
title = {On the design and implementation of a secure blockchain-based hybrid framework for Industrial Internet-of-Things},
journal = {Information Processing \& Management},
volume = {58},
number = {3},
pages = {102526},
year = {2021},
issn = {0306-4573},
doi = {https://doi.org/10.1016/j.ipm.2021.102526},
author = {Geetanjali Rathee and Farhan Ahmad and Rajinder Sandhu and Chaker Abdelaziz Kerrache and Muhammad Ajmal Azad}
}

@article{Sasikumar2025secure,
  author    = {Sasikumar Asaithambi and Senthilkumar Nallusamy and Jing Yang and Sunil Prajapat and Gyanendra Kumar and Pramod Singh Rathore},
  title     = {A secure and trustworthy blockchain-assisted edge computing architecture for industrial internet of things},
  journal   = {Scientific Reports},
  year      = {2025},
  volume    = {15},
  number    = {1},
  pages     = {15410},
  doi       = {10.1038/s41598-025-00337-3},
  url       = {https://doi.org/10.1038/s41598-025-00337-3},
  issn      = {2045-2322}
}

@ARTICLE{Ahmad2024Blockchain,
  author={Ahmad, Seerin and Nakka, Kalyan and Kim, Taesic and Han, Dongjun and Won, Dongjun and Ahn, Bohyun},
  journal={IEEE Access}, 
  title={Blockchain-Assisted Resilient Control for Distributed Energy Resource Management Systems}, 
  year={2024},
  volume={12},
  number={},
  pages={191748-191762},
  doi={10.1109/ACCESS.2024.3516581}}

@article{YALLI2025Authentication,
title = {Authentication schemes for Internet of Things (IoT) networks: A systematic review and security assessment},
journal = {Internet of Things},
volume = {30},
pages = {101469},
year = {2025},
issn = {2542-6605},
doi = {https://doi.org/10.1016/j.iot.2024.101469},
author = {Jameel Shehu Yalli and Mohd Hilmi Hasan and Low Tan Jung and Safwan Mahmood Al-Selwi}
}

@article{NAYAK2024review,
title = {A review on edge analytics: Issues, challenges, opportunities, promises, future directions, and applications},
journal = {Digital Communications and Networks},
volume = {10},
number = {3},
pages = {783-804},
year = {2024},
issn = {2352-8648},
doi = {https://doi.org/10.1016/j.dcan.2022.10.016},
author = {Sabuzima Nayak and Ripon Patgiri and Lilapati Waikhom and Arif Ahmed}
}

@article{NAZIR2024Collaborative,
title = {Collaborative threat intelligence: Enhancing IoT security through blockchain and machine learning integration},
journal = {Journal of King Saud University - Computer and Information Sciences},
volume = {36},
number = {2},
pages = {101939},
year = {2024},
issn = {1319-1578},
doi = {https://doi.org/10.1016/j.jksuci.2024.101939},
author = {Ahsan Nazir and Jingsha He and Nafei Zhu and Ahsan Wajahat and Faheem Ullah and Sirajuddin Qureshi and Xiangjun Ma and Muhammad Salman Pathan}
}

@article{Ahmed2025Forensics,
  author    = {Shams Forruque Ahmed and Shanjana Shuravi Shawon and Afsana Bhuyian and Shaila Afrin and Aanushka Mehjabin and Sweety Angela Kuldeep and Md. Sakib Bin Alam and Amir H. Gandomi},
  title     = {Forensics and security issues in the Internet of Things},
  journal   = {Wireless Networks},
  year      = {2025},
  volume    = {31},
  number    = {4},
  pages     = {3431--3466},
  doi       = {10.1007/s11276-025-03942-2},
  url       = {https://doi.org/10.1007/s11276-025-03942-2},
  issn      = {1572-8196}
}

@ARTICLE{Deng2020Edge,
  author={Deng, Shuiguang and Zhao, Hailiang and Fang, Weijia and Yin, Jianwei and Dustdar, Schahram and Zomaya, Albert Y.},
  journal={IEEE Internet of Things Journal}, 
  title={Edge Intelligence: The Confluence of Edge Computing and Artificial Intelligence}, 
  year={2020},
  volume={7},
  number={8},
  pages={7457-7469},
  doi={10.1109/JIOT.2020.2984887}}

@article{Maurya2024Blockchain,
  author    = {Vinay Maurya and Vinay Rishiwal and Mano Yadav and Mohammad Shiblee and Preeti Yadav and Udit Agarwal and Rashmi Chaudhry},
  title     = {Blockchain-driven security for IoT networks: State-of-the-art, challenges and future directions},
  journal   = {Peer-to-Peer Networking and Applications},
  year      = {2024},
  volume    = {18},
  number    = {1},
  pages     = {53},
  doi       = {10.1007/s12083-024-01812-w},
  url       = {https://doi.org/10.1007/s12083-024-01812-w},
  issn      = {1936-6450},
}

@Article{Eren2025Security,
AUTHOR = {Eren, Haluk and Karaduman, Özgür and Gençoğlu, Muharrem Tuncay},
TITLE = {Security Challenges and Performance Trade-Offs in On-Chain and Off-Chain Blockchain Storage: A Comprehensive Review},
JOURNAL = {Applied Sciences},
VOLUME = {15},
YEAR = {2025},
NUMBER = {6},
ARTICLE-NUMBER = {3225},
ISSN = {2076-3417},
DOI = {10.3390/app15063225}
}

@article{Thompson2020TheCL,
  title={The Computational Limits of Deep Learning},
  author={Neil C. Thompson and Kristjan H. Greenewald and Keeheon Lee and Gabriel F. Manso},
  journal={ArXiv},
  year={2020},
  volume={abs/2007.05558},
  url = {https://arxiv.org/abs/2007.05558},
}

@article{Alharby2022Transaction,
  author    = {Maher Alharby},
  title     = {Transaction Latency Within Permissionless Blockchains: Analysis, Improvement, and Security Considerations},
  journal   = {Journal of Network and Systems Management},
  year      = {2022},
  volume    = {31},
  number    = {1},
  pages     = {22},
  doi       = {10.1007/s10922-022-09717-w},
  issn      = {1573-7705},
}

@ARTICLE{Lim2020Federated,
  author={Lim, Wei Yang Bryan and Luong, Nguyen Cong and Hoang, Dinh Thai and Jiao, Yutao and Liang, Ying-Chang and Yang, Qiang and Niyato, Dusit and Miao, Chunyan},
  journal={IEEE Communications Surveys \& Tutorials}, 
  title={Federated Learning in Mobile Edge Networks: A Comprehensive Survey}, 
  year={2020},
  volume={22},
  number={3},
  pages={2031-2063},
  doi={10.1109/COMST.2020.2986024}}

@article{SANKA2021Review,
title = {A systematic review of blockchain scalability: Issues, solutions, analysis and future research},
journal = {Journal of Network and Computer Applications},
volume = {195},
pages = {103232},
year = {2021},
issn = {1084-8045},
doi = {https://doi.org/10.1016/j.jnca.2021.103232},
author = {Abdurrashid Ibrahim Sanka and Ray C.C. Cheung}
}

@ARTICLE{Martínez2023Decentralized,
  author={Martínez Beltrán, Enrique Tomás and Pérez, Mario Quiles and Sánchez, Pedro Miguel Sánchez and Bernal, Sergio López and Bovet, Gérôme and Pérez, Manuel Gil and Pérez, Gregorio Martínez and Celdrán, Alberto Huertas},
  journal={IEEE Communications Surveys \& Tutorials}, 
  title={Decentralized Federated Learning: Fundamentals, State of the Art, Frameworks, Trends, and Challenges}, 
  year={2023},
  volume={25},
  number={4},
  pages={2983-3013},
  doi={10.1109/COMST.2023.3315746}}

@article{STOLL2019PoW,
title = {The Carbon Footprint of Bitcoin},
journal = {Joule},
volume = {3},
number = {7},
pages = {1647-1661},
year = {2019},
issn = {2542-4351},
doi = {https://doi.org/10.1016/j.joule.2019.05.012},
author = {Christian Stoll and Lena Klaaßen and Ulrich Gallersdörfer}
}

@article{Huang2021Survey,
author = {Huang, Huawei and Kong, Wei and Zhou, Sicong and Zheng, Zibin and Guo, Song},
title = {A Survey of State-of-the-Art on Blockchains: Theories, Modelings, and Tools},
year = {2021},
issue_date = {March 2022},
publisher = {Association for Computing Machinery},
address = {New York, NY, USA},
volume = {54},
number = {2},
issn = {0360-0300},
url = {https://doi.org/10.1145/3441692},
doi = {10.1145/3441692},
journal = {ACM Comput. Surv.},
month = mar,
articleno = {44},
numpages = {42}
}

@ARTICLE{Patterson2022Carbon,
  author={Patterson, David and Gonzalez, Joseph and Hölzle, Urs and Le, Quoc and Liang, Chen and Munguia, Lluis-Miquel and Rothchild, Daniel and So, David R. and Texier, Maud and Dean, Jeff},
  journal={Computer}, 
  title={The Carbon Footprint of Machine Learning Training Will Plateau, Then Shrink}, 
  year={2022},
  volume={55},
  number={7},
  pages={18-28},
  doi={10.1109/MC.2022.3148714}}

@article{Wang2023Exploring,
author = {Wang, Gang and Wang, Qin and Chen, Shiping},
title = {Exploring Blockchains Interoperability: A Systematic Survey},
year = {2023},
issue_date = {December 2023},
publisher = {Association for Computing Machinery},
address = {New York, NY, USA},
volume = {55},
number = {13s},
issn = {0360-0300},
url = {https://doi.org/10.1145/3582882},
doi = {10.1145/3582882},
journal = {ACM Comput. Surv.},
month = jul,
articleno = {290},
numpages = {38}
}

@article{Ren2023Interoperability,
author = {Ren, Kunpeng and Ho, Nhut-Minh and Loghin, Dumitrel and Nguyen, Thanh-Toan and Ooi, Beng Chin and Ta, Quang-Trung and Zhu, Feida},
title = {Interoperability in Blockchain: A Survey},
year = {2023},
issue_date = {Dec. 2023},
publisher = {IEEE Educational Activities Department},
address = {USA},
volume = {35},
number = {12},
issn = {1041-4347},
url = {https://doi.org/10.1109/TKDE.2023.3275220},
doi = {10.1109/TKDE.2023.3275220},
journal = {IEEE Trans. on Knowl. and Data Eng.},
month = dec,
pages = {12750–12769},
numpages = {20}
}

@article{Liu2021When,
author = {Liu, Bo and Ding, Ming and Shaham, Sina and Rahayu, Wenny and Farokhi, Farhad and Lin, Zihuai},
title = {When Machine Learning Meets Privacy: A Survey and Outlook},
year = {2021},
issue_date = {March 2022},
publisher = {Association for Computing Machinery},
address = {New York, NY, USA},
volume = {54},
number = {2},
issn = {0360-0300},
url = {https://doi.org/10.1145/3436755},
doi = {10.1145/3436755},
journal = {ACM Comput. Surv.},
month = mar,
articleno = {31},
numpages = {36}
}

@article{Hu2022Membership,
author = {Hu, Hongsheng and Salcic, Zoran and Sun, Lichao and Dobbie, Gillian and Yu, Philip S. and Zhang, Xuyun},
title = {Membership Inference Attacks on Machine Learning: A Survey},
year = {2022},
issue_date = {January 2022},
publisher = {Association for Computing Machinery},
address = {New York, NY, USA},
volume = {54},
number = {11s},
issn = {0360-0300},
url = {https://doi.org/10.1145/3523273},
doi = {10.1145/3523273},
journal = {ACM Comput. Surv.},
month = sep,
articleno = {235},
numpages = {37}
}

@ARTICLE{Zhou2019EdgeIntelligence,
  author={Zhou, Zhi and Chen, Xu and Li, En and Zeng, Liekang and Luo, Ke and Zhang, Junshan},
  journal={Proceedings of the IEEE}, 
  title={Edge Intelligence: Paving the Last Mile of Artificial Intelligence With Edge Computing}, 
  year={2019},
  volume={107},
  number={8},
  pages={1738-1762},
  doi={10.1109/JPROC.2019.2918951}}

@article{Ivanov2023Security,
author = {Ivanov, Nikolay and Li, Chenning and Yan, Qiben and Sun, Zhiyuan and Cao, Zhichao and Luo, Xiapu},
title = {Security Threat Mitigation for Smart Contracts: A Comprehensive Survey},
year = {2023},
issue_date = {December 2023},
publisher = {Association for Computing Machinery},
address = {New York, NY, USA},
volume = {55},
number = {14s},
issn = {0360-0300},
url = {https://doi.org/10.1145/3593293},
doi = {10.1145/3593293},
journal = {ACM Comput. Surv.},
month = jul,
articleno = {326},
numpages = {37}
}

@article{GUO2025auditing,
title = {When auditing Meets Blockchain: A study on applying blockchain smart contracts in auditing},
journal = {International Journal of Accounting Information Systems},
volume = {56},
pages = {100730},
year = {2025},
issn = {1467-0895},
doi = {https://doi.org/10.1016/j.accinf.2025.100730},
url = {https://www.sciencedirect.com/science/article/pii/S1467089525000065},
author = {Xiaoli Guo and Yanjun Zuo and Dong Li}
}

@ARTICLE{Darshan2023Architecture,
  author={Darshan, M. and Amet, Matthieu and Srivastava, Gautam and Crichigno, Jorge},
  journal={IEEE Internet of Things Journal}, 
  title={An Architecture That Enables Cross-Chain Interoperability for Next-Gen Blockchain Systems}, 
  year={2023},
  volume={10},
  number={20},
  pages={18282-18291},
  doi={10.1109/JIOT.2023.3279693}}

@article{deVries2022Revisiting,
  author    = {Alex de Vries and Ulrich Gallersd{\"o}rfer and Lena Klaa{\ss}en and Christian Stoll},
  title     = {Revisiting Bitcoin’s carbon footprint},
  journal   = {Joule},
  year      = {2022},
  volume    = {6},
  number    = {3},
  pages     = {498--502},
  doi       = {10.1016/j.joule.2022.02.005},
  url       = {https://doi.org/10.1016/j.joule.2022.02.005},
  issn      = {2542-4785},
  publisher = {Elsevier},
}

@article{Li2023Trustworthy,
author = {Li, Bo and Qi, Peng and Liu, Bo and Di, Shuai and Liu, Jingen and Pei, Jiquan and Yi, Jinfeng and Zhou, Bowen},
title = {Trustworthy AI: From Principles to Practices},
year = {2023},
issue_date = {September 2023},
publisher = {Association for Computing Machinery},
address = {New York, NY, USA},
volume = {55},
number = {9},
issn = {0360-0300},
url = {https://doi.org/10.1145/3555803},
doi = {10.1145/3555803},
journal = {ACM Comput. Surv.},
month = jan,
articleno = {177},
numpages = {46}
}

@InProceedings{McKay2024_HITL,
  author    = {Margaret H. McKay},
  editor    = {Helmut Degen and Stavroula Ntoa},
  title     = {Realizing the Promise of AI Governance Involving Humans-in-the-Loop},
  booktitle = {HCI International 2024 -- Late Breaking Papers},
  year      = {2024},
  publisher = {Springer Nature Switzerland},
  address   = {Cham},
  pages     = {107--123},
  isbn      = {978-3-031-76827-9},
  doi       = {10.1007/978-3-031-76827-9_7}
}

@article{ALDHAHERI20201Artificial,
title = {Artificial Immune Systems approaches to secure the internet of things: A systematic review of the literature and recommendations for future research},
journal = {Journal of Network and Computer Applications},
volume = {157},
pages = {102537},
year = {2020},
issn = {1084-8045},
doi = {https://doi.org/10.1016/j.jnca.2020.102537},
author = {Sahar Aldhaheri and Daniyal Alghazzawi and Li Cheng and Ahmed Barnawi and Bandar A. Alzahrani}
}

@article{Rathi2023Exploring,
author = {Rathi, Nitin and Chakraborty, Indranil and Kosta, Adarsh and Sengupta, Abhronil and Ankit, Aayush and Panda, Priyadarshini and Roy, Kaushik},
title = {Exploring Neuromorphic Computing Based on Spiking Neural Networks: Algorithms to Hardware},
year = {2023},
issue_date = {December 2023},
publisher = {Association for Computing Machinery},
address = {New York, NY, USA},
volume = {55},
number = {12},
issn = {0360-0300},
url = {https://doi.org/10.1145/3571155},
doi = {10.1145/3571155},
journal = {ACM Comput. Surv.},
month = mar,
articleno = {243},
numpages = {49}
}

@ARTICLE{Kumar2026Next,
  author={Kumar, Ramesh and Dutta, Joy and Vamsi, N. and Sankararao Varri, Uma and Puthal, Deepak},
  journal={IEEE Access}, 
  title={Next-Generation Security in the 6G Era: The Role of AI in Safeguarding Future Networks}, 
  year={2026},
  volume={14},
  number={},
  pages={17347-17380},
  doi={10.1109/ACCESS.2025.3650208}}

@ARTICLE{Dutta2026Explainable,
  author={Dutta, Joy and Puthal, Deepak},
  journal={IEEE Transactions on Services Computing}, 
  title={Explainable AI-Enabled Privacy-Preserving Query Processing on Blockchain Ledgers With Statistical Metadata}, 
  year={2026},
  volume={},
  number={},
  pages={1-16},
  doi={10.1109/TSC.2026.3668008}}

@INPROCEEDINGS{Dutta2025Empowering,
  author={Dutta, Joy and Eldeeb, Hossien B. and Ho, Tu Dac},
  booktitle={2025 IEEE 101st Vehicular Technology Conference (VTC2025-Spring)}, 
  title={Empowering V2X Security: Integration of PoAh 2.0 and Edge LLM in Context-Aware Blockchain Ecosystems}, 
  year={2025},
  volume={},
  number={},
  pages={1-6},
  doi={10.1109/VTC2025-Spring65109.2025.11174846}}

@techreport{IBM2025DataBreach,
  author       = {{IBM}},
  title        = {Cost of a Data Breach Report 2025},
  institution  = {IBM},
  year         = {2025},
  url          = {https://www.ibm.com/reports/data-breach},
  urldate      = {2026-03-25}
}

@misc{NIST2024PQC,
  author       = {{National Institute of Standards and Technology}},
  title        = {NIST Releases First 3 Finalized Post-Quantum Encryption Standards},
  year         = {2024},
  month        = aug,
  day          = {13},
  url          = {https://www.nist.gov/news-events/news/2024/08/nist-releases-first-3-finalized-post-quantum-encryption-standards},
  urldate      = {2026-03-25},
  note         = {Available at: \url{https://www.nist.gov/news-events/news/2024/08/nist-releases-first-3-finalized-post-quantum-encryption-standards}}
}

@misc{Ferrag2025AutonomousAgents,
  author       = {Mohamed Amine Ferrag and Norbert Tihanyi and Merouane Debbah},
  title        = {From LLM Reasoning to Autonomous AI Agents: A Comprehensive Review},
  year         = {2025},
  eprint       = {2504.19678},
  archivePrefix= {arXiv},
  primaryClass = {cs.AI},
  doi          = {10.48550/arXiv.2504.19678},
  url          = {https://arxiv.org/abs/2504.19678}
}

@misc{Lazer2026AgenticCyber,
  author       = {Sahaya Jestus Lazer and Kshitiz Aryal and Maanak Gupta and Elisa Bertino},
  title        = {A Survey of Agentic AI and Cybersecurity: Challenges, Opportunities and Use-case Prototypes},
  year         = {2026},
  eprint       = {2601.05293},
  archivePrefix= {arXiv},
  primaryClass = {cs.CR},
  doi          = {10.48550/arXiv.2601.05293},
  url          = {https://arxiv.org/abs/2601.05293}
}

\end{document}